\documentclass[12pt]{article}
\usepackage[
 linktoc=all,
 colorlinks=true,
 linkcolor=blue,
 urlcolor=blue,
 citecolor=red,
 ]{hyperref}
 
\usepackage{mathtools,amssymb,mathrsfs,microtype,slashed,tikz,ytableau,cite}

\DeclareMathOperator{\tr}{tr}
\DeclareMathOperator{\sign}{sign}
\ytableausetup{centertableaux,boxsize=.4em}
\usetikzlibrary{decorations.markings,decorations.pathmorphing,arrows.meta}

\renewcommand{\title}[1]{\vbox{\center\LARGE{#1}}\vspace{5mm}}
\renewcommand{\author}[1]{\vbox{\center\large#1}\vspace{5mm}}
\newcommand{\address}[1]{\vbox{\center\em#1}}

\makeatletter\@addtoreset{equation}{section}\makeatother

\tikzset{
  fermion/.style={
    thick,
    postaction={decorate},
    decoration={markings, mark=at position 0.5 with {\arrow{Stealth}}}
  }
}

\tikzset{
  photon/.style={
    thick,
    decorate,
    decoration={
      snake,
      segment length=3mm,
      amplitude=.5mm,
    }
  }
}

\tikzset{
  scalar/.style={
    thick,
    dashed
  }
}

\begin{document}

\begin{titlepage}
\begin{center}
\vskip-10pt
\title{\makebox[\textwidth]{\Large{
Does QFT make sense in non-integer dimensions?
}}}
\vspace{10mm}
Diego Delmastro,\footnote{\href{mailto:ddelmastro@physics.ucla.edu}{\tt ddelmastro@physics.ucla.edu}}
Yan-Yan Li,\footnote{\href{mailto:yanyanli@ucla.edu}{\tt yanyanli@ucla.edu}}
Dikshant Rathore\footnote{\href{mailto:drathore@physics.ucla.edu}{\tt drathore@physics.ucla.edu}}
\vskip 7mm
\address{
Mani L. Bhaumik Institute for Theoretical Physics, Department of Physics and Astronomy,
University of California Los Angeles, Los Angeles, CA 90095, USA.
}
\end{center}

\abstract{
We revisit the old problem of analytically continuing QFTs to fractional dimension $d\in\mathbb C$. We observe that common theories like QCD and QED have branch cuts in the complex $d$ plane. In particular, many operators in their low-energy CFT have OPEs and scaling dimensions that jump as a function of $d$.
}

{
\thispagestyle{empty}
}

\end{titlepage}

{\hypersetup{linkcolor=black}
\tableofcontents
}


\setcounter{tocdepth}{3}

\section{Introduction and summary}\label{sec:intro}\setcounter{footnote}{0}

\paragraph{The goal.} Chiral gauge theories are famously difficult to analytically continue away from integer dimensions, an obstacle first raised in the context of dimensional regularization~\cite{Bollini:1972ui,THOOFT1972189}. There, the motivation for extending $d\in\mathbb N$ to $d\in\mathbb C$ was that such a continuation is an extremely convenient regulator, massively simplifying computations. For vector-like theories, this continuation was promptly solved, but chiral theories, which involve the matrix $\gamma^\star$, required significant extra work. Arguably, a completely satisfactory treatment, especially in the context of supersymmetric theories~\cite{SIEGEL1979193,SIEGEL198037}, is still lacking. That being said, various working prescriptions are known: for example, one can use the standard 't Hooft-Veltman trick, where we split our spacetime into two submanifolds, the ``transverse'' and ``parallel'' directions, and define $\gamma^\star$ only with respect to the latter. For instance, if we were working in $d=2+\epsilon$, we would typically declare that
\begin{equation}
\begin{aligned}
\{\gamma^\star,\gamma^\mu\}&=0\qquad \mu=1,2\\
[\gamma^\star,\gamma^\mu]&=0\qquad \mu\neq 1,2\\
\end{aligned}
\end{equation}

This is all well and good if we are interested in eventually taking $d\to2$, but not otherwise: if we use this continuation in $d$, and take $d\to4$ instead, the answer we would get would differ from the correct one as computed directly in $d=4$. In other words, 't Hooft-Veltman-like prescriptions are only useful for an infinitesimal neighborhood around a given $d\in\mathbb N$. 

Here we are more ambitious and would like to extend various theories to a \emph{non-infinitesimal} subset of the complex $d$ plane, in the sense that we want a family of QFTs, labeled by $d\in\mathbb C$, which exactly matches the corresponding QFT when we take \emph{several distinct} $d\in\mathbb N$. In other words, a small patch around some specific $d\in\mathbb N$ is not enough for us: we want some open set in $\mathbb C$ that contains at least a couple of distinct integer dimensions $d$, with agreement when restricted to those. 't Hooft-Veltman-like prescriptions like the one above are insufficient for this.

\paragraph{Motivation.} There are various motivations for this endeavor. First, a better understanding of this continuation for generic $d$ instead of $d$ close to some specified integer might help us in coming up with better prescriptions for dim-reg, perhaps eventually leading to a completely satisfactory treatment of SUSY theories.

But, beyond practical reasons like dim-reg, a proper continuation of QFT to fractional $d$ has attracted significant attention in recent years, see e.g.~\cite{Giombi:2016fct,Diab:2016spb,Cappelli:2018vir,Giombi:2019upv,Henriksson:2022rnm,Giombi:2022vnz,Sirois:2022vth,Zhou:2022pah,Henriksson:2022gpa,Bonanno:2022ztf,Chester:2022hzt,Jin:2023cce,Jin:2023fbz,Reehorst:2024vyq,Giombi:2025evu,deSabbata:2024xwn,Henriksson:2025kws,Zan:2026oyb,Fraser-Taliente:2026iuj,Chiang:2026nmd,DeCesare:2025ukl,DeCesare:2026dwm,Guo:2026cjw} for a very incomplete list of examples. Before we try to define this continuation more precisely, let us spend a few paragraphs explaining some of the reasons why this operation is interesting in the first place.

For us, the fundamental reason is structural: analytic continuation of discrete parameters allows us to meaningfully connect different QFTs in the space of theories. For example, we intuitively tend to think of $2d$ Ising and $3d$ Ising as ``the same theory'', just at different values of $d$. On the other hand, one would never say that $2d$ Ising is the same theory as, say, $3d$ ABJM. The reason the former QFTs are ``the same'' but the latter are not is that the parameter $d$ in Ising can be continued to $d\in\mathbb C$, and observables become analytic functions of this variable\footnote{Throughout this paper we will use ``analytic'' to actually mean ``meromorphic''.}. This is the same reason we think of the $O(N)$ model as ``the same theory'', just at different values of $N$: this makes sense only because the parameter $N$ can be analytically continued\footnote{See e.g.~\cite{Cardy:2013rqg,Gorbenko:2020xya,Nekrasov:2023xzm,Budzik:2023vtr,Cao:2023psi,Jacobsen:2023vdf,Roux:2024ubh,Padayasi:2021sik,Binder:2019zqc} for some recent literature on analytic continuation of $N$.}. This is not just a question of academic interest, but has important practical implications:
\begin{itemize}
\item The $\epsilon$ expansion~\cite{Wilson:1971dc} explicitly assumes that the $d=3$ and $d=4$ theories are connected in the space of QFTs. The fact that this technique works at all relies on analytic continuation in $d$. Indeed, unless there is ``something'' in between Ising$_4$ and Ising$_3$, this expansion would be entirely meaningless. For example, one would never expect to be able to solve graphene by expanding, say, Banks-Zaks around $d=4$.

\item The large $N$ expansion explicitly assumes that the $O(N)$ theory and the $O(N+1)$ theory are connected in the space of QFTs. As before, one would never expect a $1/N$ expansion to work for a random sequence of QFTs, say, something like $T(1)=\text{Ising}$, $T(2)=SU(2)+\psi_\text{adj}$, $T(3)=\text{abelian Higgs}$, etc. This expansion works for $O(N)$, but not for $T(N)$, precisely because the former make up a sequence that is analytic in $N$.

\end{itemize}

The goal of the analytic continuation program is to make this intuition rigorous: in what precise sense are these theories connected to each other in the space of QFTs? Is this continuation unique? And what does the interpolation teach us about the individual QFTs?

Another common motivation is that analytic continuation in $d$ leads to controlled violations of unitarity~\cite{Hogervorst:2015akt}, making them a very nice playground for studying non-unitary QFT. For instance,~\cite{Giombi:2015haa} examines the free energy in QED in continuous dimension in order to scrutinize the $F$-theorem~\cite{Jafferis:2011zi,Klebanov:2011gs}, which in principle is only proven to hold in unitary theories~\cite{Casini:2012ei}. More generally, the fundamental principles of the conformal bootstrap program are rather agnostic to the concrete value of $d$, and the formalism itself appears particularly well adapted to treating this integer as a continuous parameter~\cite{El-Showk:2013nia}.

Yet another reason it is worth our while to explore this continuation is that much of the emergent phenomena in QCD-like theories that we usually care about, such as non-perturbative symmetry breaking, infrared dualities, confinement, etc., are present in many examples in different integer dimensions~\cite{PhysRev.128.2425,COLEMAN1976239,THOOFT1974461,PhysRevD.13.1649,PhysRevD.11.2088,Witten:1983ar,PhysRevD.29.2423,PhysRevLett.60.2575,PhysRevD.33.3704,PhysRevLett.64.721,POLYAKOV1977429,BANKS1977493,PhysRevB.16.1217,PhysRevLett.47.1556,Affleck:1983mk,Seiberg:1996nz,Intriligator:1996ex,Aharony:2015mjs,PhysRev.122.345,PhysRevLett.37.8,PhysRevLett.45.100,VAFA1984173,PhysRevLett.53.535,PhysRevD.10.2445,MANDELSTAM1976245,SEIBERG199419,SEIBERG1994484,MONTONEN1977117,GODDARD19771,Seiberg:1994pq,Seiberg:1996bd,Intriligator:1997pq,Witten:1995zh,Witten:1997sc,Vafa:1994tf,DeCesare:2021pfb}. If we were able to extend these theories to fractional $d$, we could try to track these phenomena as we go from one integer dimension to another, which would likely teach us quite a bit about the concrete mechanisms behind these features in our original, integer-dimensional gauge theories.

As a concrete example, there are in the literature various instances of gauge theories that confine in $d=4$ (see e.g.~\cite{Intriligator:1995au}), but deconfine in $d=3$ (see e.g.~\cite{Bashmakov:2018wts,Gomis:2017ixy}). In this case, one could guess that there is a confinement/deconfinement transition at some $d^\star\in[3,4]$, which could shed some light on the dynamics of the analogous transition in real-world QCD.

A related example is the fact that the critical number of flavors for which there is symmetry breaking, $N_f^\star$, is meaningful in both $d=3$ and $d=4$, but in principle these two numbers are independent~\cite{Komargodski:2017keh}. By extending QCD to fractional $d$ we get an interpolation between $N_f^\star(d=4)$ and $N_f^\star(d=3)$, potentially connecting these two quantities. With some wishful thinking, it might be possible that connecting these two objects could lead to a better understanding of both\footnote{As a simple example,~\cite{DiPietro:2015taa} used the known value $N^\star_f(d=2)=\infty$ to improve their estimate of $N_f^\star(d=3)$.}.

So, with these motivations in mind, here we reconsider the problem of extending a given QFT to $d\in\mathbb C$. As mentioned above, this problem is particularly tricky for chiral theories but, at least naively, one would expect that vector-like theories can be continued without significant complications. In this work we point out that, already for vector-like theories, this continuation is very subtle and, despite what some partial results might suggest, still an open problem.

\paragraph{Complications.} In section~\ref{sec:generalities} we will discuss much more carefully the various options for how this analytic continuation is defined in the first place. For the purposes of this summary, let us settle for a vague, aspirational prescription: compute things for general $d\in\mathbb N$ and, if the answer is ``uniform in $d$'', declare that the result is also true when $d\in\mathbb C$.

The main obstruction to naive prescriptions like the one above is the existence of \emph{low-rank exceptions}: situations where observables are not in fact ``uniform in $d$''. The simplest example where this shows up is the free, massless Dirac fermion. Consider the scalar $\bar\psi\psi$; its first few correlation functions are all unremarkable: for example, the two-point function equals $\langle\bar\psi\psi (x)\bar\psi\psi(0)\rangle\propto 1/x^{2(d-1)}$. This result has an obvious continuation to $d\in\mathbb C$.

The three- and four-point functions of $\bar\psi\psi$ are also both uniform in $d$, but something interesting happens for the five-point function. By brute-force computation one can check that it vanishes for all integers $d\ge 4$, but it is non-zero in three dimensions:
\begin{equation}\label{eq:5pt_z}
\langle\bar\psi\psi(x_1)\cdots\bar\psi\psi(x_5)\rangle \propto \delta_{3,d}
\end{equation}
(see~\eqref{eq:5pt_3d} for the actual $d=3$ expression). What should its continuation to $d\in\mathbb C$ be? The answer is far from obvious, at least to us: the integer sequence $\delta_{3,d}$ has no canonical continuation to $d\in\mathbb C$. (There are infinitely many functions that interpolate this sequence, but there is no criterion that singles out a specific choice.) 

The phenomenon above is in fact quite generic in QFT: most correlation functions have a qualitative, discrete change in behavior at some finite $d\in\mathbb N$, which complicates any naive attempts at making $d\in\mathbb C$. We next explain the main mechanism for this discontinuity in $d$.

Consider the $n$-point function of some scalar operator $\mathcal O(x)$:
\begin{equation}
\langle\mathcal O(x_1)\cdots\mathcal O(x_n)\rangle
\end{equation}
We assume that our family of QFTs is invariant under translations and rotations, i.e., the special euclidean group $\mathrm{SE}(d)=SO(d)\ltimes \mathbb R^d$. This symmetry then implies that this correlator is defined on $\mathrm{Conf}_n(\mathbb R^d)/\mathrm{SE}(d)$, i.e., it can be expressed in terms of scalars constructed out of $v_i=x_i-x_{i+1}$.

What are these scalars, specifically? If the number of insertions is small enough, the ring of $SO(d)$-invariants is generated by the matrix of inner products $g_{ij}=v_i\cdot v_j$. But if the number of $v_i$'s is at least as large as the dimension of $\mathbb R^d$, there is one more piece of $SO(d)$-invariant information that is not captured by $g_{ij}$, namely the orientation defined by the basis $v_1,\dots,v_d$. In other words, in this case the ring of $SO(d)$-invariants is generated by $g_{ij}$ together with $\varepsilon=\sign(\det(v_1,\dots,v_d))$. Therefore, one can generically write
\begin{equation}
\langle\mathcal O(x_1)\cdots\mathcal O(x_n)\rangle=\begin{cases}
\displaystyle \mathcal S_{d,n}^\text{even}(g_{ij})+\varepsilon \mathcal S_{d,n}^\text{odd}(g_{ij}) & d\le n-1\\[+2ex]
\displaystyle \mathcal S_{d,n}^\text{even}(g_{ij}) & d>n-1
\end{cases}
\end{equation}
for some functions $\mathcal S$ (even/odd refer to the intrinsic parity, see below). This is an example of a small-$d$ exception we mentioned above. For fixed $n$, there is a discontinuous change between $d\le n-1$ and $d>n-1$, which means that the correlation function cannot be uniform in $d$: to the extent that there is a meaningful continuation to $d\in\mathbb C$, there should be some sort of jump around $d\approx n-1$.

Things are a little nicer if we assume that the theory also has reflection symmetry, namely that it is invariant under the full euclidean group $\mathrm{ISO}(d)=O(d)\ltimes \mathbb R^d$. In this case, correlation functions of \emph{parity-even} scalars are defined on $\mathrm{Conf}_n(\mathbb R^d)/\mathrm{ISO}(d)$. Importantly, $\varepsilon$ is parity-odd, and therefore it cannot appear in their correlation functions. In other words, $\mathrm{Conf}_n(\mathbb R^d)/\mathrm{ISO}(d)$ is coordinatized by the Gram matrix $g_{ij}$ alone. In the notation above, the $n$-point function equals $\mathcal S_{d,n}^\text{even}(g_{ij})$ for all $d$. As there is no fundamental change in behavior as we vary the number of spacetime dimensions, there is hope that a nice continuation to $d\in\mathbb C$ might exist.

Parity-odd operators, on the other hand, are less constrained than parity-even ones. Under a parity transformation, the $n$-point function picks up the sign $(-1)^n$. Therefore, if $n$ is even, this \emph{forbids} the presence of $\varepsilon$, but if $n$ is odd, it \emph{forces} its presence instead. Therefore, the case of $n$ even should be uniform in $d$, but the case of odd $n$ is again discontinuous: the correlation function must vanish if $d>n-1$. This explains why the two-point function of the parity-odd operator $\bar\psi\psi$ was well-defined for $d\in\mathbb C$, but the five-point function was $\propto \delta_{3,d}$. 

The final conclusion is, then, that generic operators in $SO(d)$-invariant theories, and parity-odd operators in $O(d)$-invariant theories, present discrete, qualitative changes in behavior as we vary $d\in\mathbb N$, and uniform answers only emerge for sufficiently large $d$. In the naive approach to analytic continuation, observables become analytic in $d$ only after we ignore by hand low-rank exceptions. In an intuitive sense, the low $d$ behavior is non-perturbative in $1/d$. Needless to say, at the end of the day we only care about QFTs in low dimension anyway, which makes the problem both hard and interesting.

Let us mention for completeness that, for more general QFTs where the symmetry group is neither $SO(d)$ nor $O(d)$, but some more complicated global form of $\mathfrak{so}(d)$, the question is whether the representation theory of this global group is stable or not, i.e., whether there are low-rank isomorphisms in $\operatorname{Rep}(G)$. For example, in the case of $G=SO(d)$ one has the low-rank isomorphism $\wedge^d\ydiagram1\cong\boldsymbol 1$, which allows for the exceptional scalar $\det(v_1,\dots,v_d)$. We will discuss this in more detail in section~\ref{sec:kinematics}.

An important incarnation of the subtleties associated with small $d$ exceptions is the following. Say, for example, that we are interested in a perturbative theory defined for general $d$, whose spectrum contains a pair of scalar operators $\mathcal O_1$ and $\mathcal O_2$. If their scaling dimensions are different, then these operators cannot mix, and we can compute loop corrections independently. But if the scaling dimensions depend on $d$, and there is some $d^\star$ for which there is level crossing, then perturbation theory breaks down around $d\approx d^\star$, and we must switch to degenerate perturbation theory (see e.g.~\cite{Korchemsky:2015cyx}). This typically gives rise to branch points close to $|d|=d^\star$. So far, this is standard CFT phenomena.

The striking fact is that this phenomenon can occur even if $\mathcal O_1$ and $\mathcal O_2$ have different spin. For example, a rank-2 anti-symmetric field $f_{\mu\nu}$ is different from a vector $v_\mu$ for generic $d$, but the isomorphism $\ydiagram{1,1}\cong \ydiagram1$ in $SO(3)$ means that these two fields have the same spin in three dimensions, and therefore they can once again mix. Low-$d$ accidental spin degeneracy is rather common, and our general expectation is that analytically continued QFTs will develop branch cuts all over the complex $d$ plane.

We can see this explicitly in the main example we will be looking at in this paper: QCD with gauge group $G$ and fermions $\psi$ in some representation $R\in\operatorname{Rep}(G)$. We will take the limit of large $R$ in order to be able to access CFT data reliably.

In section~\ref{sec:results} we will compute the scaling dimension of the defect operator $\psi$ to leading order in $1/\dim(R)$. To tree level $\Delta^{(0)}(\psi)=(d-1)/2$, and the first correction is
\begin{equation}
\Delta^{(1)}(\psi)=\frac{4(d-1)^2 \Gamma (d-2)}{d \Gamma(d/2)^3 \Gamma(1-d/2)}\frac{\dim(G)}{\dim(R)}
\end{equation}
This quantity admits a natural continuation in $d$.

On the other hand, for the scaling dimension of the local operator $\bar\psi\psi$, the first correction turns out to be
\begin{equation}\label{eq:Delta_1_exact}
\Delta^{(1)}(\bar\psi\psi)=\frac{4(d-1)^2 \Gamma (d-2)}{d \Gamma(d/2)^3 \Gamma(1-d/2)}\biggl(
2-\frac{d-4}{d-2}\frac{\tr[\gamma^{\alpha\mu\nu}]\tr[\gamma_{\alpha\mu\nu}]}{\tr[1]^2}
\biggr)\frac{\dim(G)}{\dim(R)}
\end{equation}
We find the structure $\tr[\gamma^{\alpha\mu\nu}]\tr[\gamma_{\alpha\mu\nu}]$, which vanishes for all integers $d\neq3$, but is non-zero for $d=3$. Unlike $\Delta(\psi)$, this scaling dimension doesn't have an obvious analytic continuation in $d$.

As emphasized above, the obstruction to analyticity in $d$ is the existence of low rank exceptions, in this case the exception being the $3d$ isomorphism $\ydiagram{1,1,1}\cong\boldsymbol 1$, which allows $\tr[\gamma_{\mu\nu\rho}]$ to be non-zero.

The usual way to recover analyticity in the literature is to declare that $\tr[\gamma_{\mu_1\cdots\mu_n}]\equiv 0$ for all $d\in\mathbb C$. If we do this, the scaling dimension above becomes
\begin{equation}\label{eq:analytic_wrong}
\widehat\Delta^{(1)}(\bar\psi\psi)=\frac{8(d-1)^2 \Gamma (d-2)}{d \Gamma(d/2)^3 \Gamma(1-d/2)}\frac{\dim(G)}{\dim(R)}
\end{equation}
where the hat signifies that this holds only for \emph{generic} $d\in\mathbb N$, but not necessarily for \emph{all} $d\in\mathbb N$. This formula is now analytic in $d$, and it agrees with $\Delta^{(1)}$ for all integers $d>3$, but it gives the wrong answer when $d=3$. Therefore, this prescription fails to provide a good analytic continuation: $\widehat{\Delta}^{(1)}$ is not a good interpolation of QCD$_d$ for $d\in\mathbb C$, since it disagrees with the honest answer as computed directly in $d=3$.

Does this mean analytic continuation is not possible? Is there any way to work with analytic objects $\widehat{\Delta}^{(1)}$, while at the same time to recover the correct answer $\Delta^{(1)}$ when $d\in\mathbb N$?

In fact, analytic continuation is possible, but the mechanism is not immediately obvious. To simplify the notation, let us omit the $\dim(G)/\dim(R)$ factors (or, equivalently, let us take $G=U(1)$). Then, the correct answer in $d=3$ is $\Delta^{(1)}=256/3\pi^2$, as given by~\eqref{eq:Delta_1_exact} with $\tr[\gamma^{\alpha\mu\nu}]\tr[\gamma_{\alpha\mu\nu}]/\tr[1]^2=-3!$. On the other hand, the analytic expression~\eqref{eq:analytic_wrong} evaluated at $d=3$ reads $\widehat{\Delta}^{(1)}=-128/3\pi^2$, which is of course wrong.

The resolution is the following. Close to $d=3$, the operator $\bar\psi\psi$ becomes degenerate with $\bar\psi\gamma_{\mu\nu\rho}\psi$. This is impossible for generic $d$, since these operators have different spin, but in three dimensions one has the low-rank exception $\gamma_{\mu\nu\rho}\propto\epsilon_{\mu\nu\rho}$, and therefore these become the exact same operator. Hence, intuitively speaking, these operators can mix for $d$ close to $3$.

If we compute the correction $\Delta^{(1)}$ for $\bar\psi\gamma_{\mu\nu\rho}\psi$, and force it to be analytic by dropping by hand the odd traces $\tr[\gamma_{\mu_1\cdots\mu_n}]$, we find
\begin{equation}\label{eq:stable_psi3psi}
\widehat{\Delta}^{(1)}(\bar\psi\gamma_{\mu\nu\rho}\psi)=\frac{8 (d-4) \Gamma (d+2)}{d^2(d-2)^2 \Gamma(d/2)^3 \Gamma(1-d/2)}
\end{equation}
If we now take $d=3$, we find $\widehat{\Delta}^{(1)}(\bar\psi\gamma_{\mu\nu\rho}\psi)=256/3\pi^2$, which is the correct answer for $\Delta^{(1)}(\bar\psi\psi)$. In this way, we recover the correct scaling dimension, as computed honestly in $d=3$, from an object $\widehat{\Delta}^{(1)}$ that is analytic. But do note that~\eqref{eq:stable_psi3psi} disagrees with $\Delta^{(1)}(\bar\psi\psi)$ at $d=2$, so one should not think of this expression as the correct analytic continuation of $\Delta^{(1)}(\bar\psi\psi)$ for all $d\in\mathbb C$: this branch is only correct for $d$ close to $3$.

This last statement will perhaps be more clear if we plot the analytic dimensions $\widehat{\Delta}^{(1)}(\bar\psi\gamma_{\mu_1\cdots\mu_n}\psi)$ for $d\in\mathbb R$: 
\begin{center}
\scalebox{0.9}{
\tikz{
\node[anchor=south west,inner sep=0] (image) at (0,0) {\includegraphics[width=0.8\textwidth]{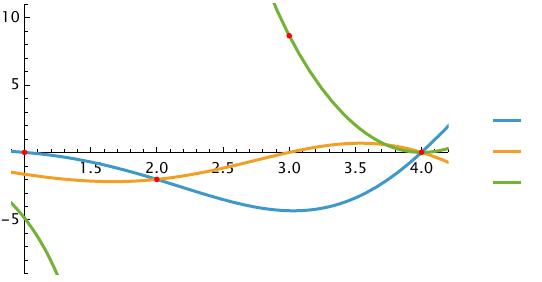}};
\node at (11.2, 3.2)  {$d$};
\node at (1.3, 6.5)  {$\widehat\Delta^{(1)}$};
\node[anchor=west] at (12.7, 4.)  {$\bar\psi\psi$};
\node[anchor=west] at (12.7, 3.25)  {$\bar\psi\gamma_{\mu_1\mu_2}\psi$};
\node[anchor=west] at (12.7, 2.45)  {$\bar\psi\gamma_{\mu_1\mu_2\mu_3}\psi$};
}}\label{fig:plot_delta}
\end{center}
where the red dots correspond to the honest value of $\Delta^{(1)}(\bar\psi\psi)$ as computed in integer $d$, using the rigorous gamma matrix traces in that dimension, without throwing away odd traces. The solid lines, on the other hand, correspond to $\widehat{\Delta}^{(1)}$, i.e., they are the analytic scaling dimensions that arise after setting $\tr[\gamma_{\mu\nu\rho}]\equiv0$ by hand. This picture makes it clear that all red dots agree with \emph{some} analytic branch, but no branch is correct for \emph{all} dots.

To order $1/\dim(R)$ the expression for $\Delta(\bar\psi\psi)$ only includes the trace $\tr[\gamma_{\mu\nu\rho}]$, which is why this quantity requires only one new branch, around $d=3$. To order in $1/\dim(R)^2$ one finds a new trace $\tr[\gamma_{\mu\nu\rho\sigma\tau}]\tr[\gamma^{\mu\nu\rho\sigma\tau}]$ in the honest, general $d$ expression for $\Delta(\bar\psi\psi)$. As before, we can make $\Delta^{(2)}$ analytic in $d$ by dropping these odd traces, producing an expression $\widehat{\Delta}^{(2)}(\bar\psi\psi)$ that agrees with $\Delta^{(2)}(\bar\psi\psi)$ for all $d>5$, but fails otherwise. One can anyway recover the correct answer at $d=5$ by considering the operator $\bar\psi\gamma_{\mu\nu\rho\sigma\tau}\psi$, which becomes isomorphic to $\bar\psi\psi$ when $d=5$; this gives rise to a new branch, defined around this integer. The ``all orders'' exact answer $\Delta(\bar\psi\psi)$ involves infinitely many branches, one adapted to each odd $d$, as a result of the infinitely many exceptional traces $\tr[\gamma_{\mu_1\cdots\mu_n}]$.

To the extent that a continuation of chiral quantities is well-defined, we expect operators like $\bar\psi\gamma^\star\psi$ to present discontinuities at the \emph{even} integers, as a result of the exceptional traces $\tr[\gamma^\star\gamma_{\mu_1\cdots\mu_n}]$. We do not have much to say about this in this work (however, see appendix~\ref{app:chiral} for some superficial comments).

In a general CFT, there is no single expression for $\Delta(\mathcal O)$ that is both analytic in $d$ and that agrees with the correct, honest answer for all $d\in\mathbb N$. Instead, there is an infinite family of expressions $\widehat{\Delta}(\mathcal O_i)$, where $\mathcal O_i$ are all the operators that become isomorphic to $\mathcal O$ at some $d\in\mathbb N$. These $\widehat{\Delta}$'s are defined by ignoring by hand low-rank exceptions and, as such, they are all analytic in $d$. But at the same time, they also capture the correct value of $\Delta(\mathcal O)$ for those integer dimensions $d$ in which $\mathcal O_i=\mathcal O$. This allows us to have our cake and eat it too: we can work with analytic objects $\widehat{\Delta}$ which, at the same time, contain all the same information as the honest $\Delta$'s. The only price one has to pay is that one has to choose the correct branch adapted to each $d\in\mathbb N$.

Presumably, a similar branched structure exists for all CFT data, and not just scaling dimensions. We leave a more thorough analysis of this for future work; the confusing five-point function of $\bar\psi\psi$ (cf.~\eqref{eq:5pt_z}) should be a good place to start.

If the flavor symmetry of the theory has a stable representation theory (e.g., if we consider parity-even operators in an $O(d)$-invariant QFT), then there are no low-rank isomorphisms, and hence no spin degeneracy. In this case, quantities like $\Delta(\mathcal O)$ are single-valued; this is why ``simple theories'' like Wilson-Fisher have scaling dimensions that are analytic in $d$.

The plan for the rest of this paper is as follows. In section~\ref{sec:generalities} we make an attempt at defining the problem more precisely, and we also describe some of the complications that arise in a general QFT. In section~\ref{sec:results} we illustrate these observations by computing a few observables in a specific theory, viz.~QCD. In section~\ref{section_QEDGNY} we study analogous observables in the QED Gross--Neveu--Yukawa model, including the effects of four-fermion interactions. In section~\ref{sec:conclusions} we finish with some open problems and general speculations.

\section{Generalities}\label{sec:generalities}

In this section we try to set up the problem of analytic continuation more systematically and precisely.

\subsection{Definition (or lack thereof)}

Let us begin with the important clarification that we do not claim every QFT admits analytic continuation, nor that this continuation is necessarily unique. For example, it might be the case that Chern-Simons theories are stuck to $d=3$, but it might also be the case that some clever re-writing allows us to define them for general $d\in\mathbb C$. However unlikely the second one might sound, either option is logically possible, and we make no claim either way.

As for uniqueness, this depends on how one formulates the question. For example, $3d$ Maxwell should admit several natural extensions away from $d=3$: it is connected to $4d$ Maxwell in the obvious way, but also to a $4d$ compact scalar, since $f_{\mu\nu}\leftrightarrow \epsilon_{\mu\nu\sigma}\partial^\sigma\phi$ in $d=3$. More generally, analytic continuation of a given QFT$_d$ is always relative to a choice of target at $d\pm1$, it is not an intrinsic operation one can perform on a specific QFT. Whenever we talk about analytic continuation in $d$, we must choose a sequence of QFTs defined for multiple integers $d\in\mathbb N$, and then ask whether they can be interpolated via $d\in\mathbb C$.

In fact, if uniqueness is to hold at all, it can only do so if we have a target for \emph{all} $d\in\mathbb N$, and not just for a few consecutive integers. Indeed, assume that a certain computation produces the expression $\delta^{\mu\nu}\delta_{\mu\nu}$. The obvious analytic continuation is that $\delta^{\mu\nu}\delta_{\mu\nu}=d$. But if we only require the continuation to give the right answer for $d=2$ and $d=3$, then we can write $\delta^{\mu\nu}\delta_{\mu\nu}=f(d)$ where $f(d)$ is any analytic function that satisfies $f(2)=2$, $f(3)=3$. There are infinitely many such functions, and therefore this construction is not unique. Only if we impose $f(d)=d$ for all $d\in\mathbb N$ (together with some further constraints on the growth of $f$ in the complex plane) is the equality $f(d)=d$ for all $d\in\mathbb C$ unique.\footnote{The precise statement is the following. Say we have a quantity $f(d)$ defined for multiple integers $S\subsetneq\mathbb N$, and we wish to continue it to some analytic function on some open, connected set $D\subset \mathbb C$ that contains these integers. Is this continuation unique? The answer is clearly \emph{no}: we can always shift $f(d)$ by, say, $\sin(\pi d)$. The only situation where we can guarantee uniqueness is when $S$ has an accumulation point in $D$. And the only accumulation point in $\mathbb N$ (as embedded in the Riemann sphere $\mathbb C\cup\{\infty\}$) is $d=\infty$. Therefore, the only way to ensure that $f(d)$ has a unique continuation is to assume that it is defined for all sufficiently large $d\in\mathbb N$.} This is of course a simplified version of the well-known theorem~\cite{Collins:1984xc} that the usual rules of dim-reg lead to a unique analytic continuation only if we require them to be correct for \emph{all} $d\in\mathbb N$.

An analogous statement holds for continuation of flavor symmetries: we need to specify an \emph{infinite} sequence of $O(N)$-symmetric QFTs to analytically continue $N$. Any finite collection of QFTs can always be made part of infinitely many sequences, and they can all be used to define continuations. As will hopefully become clear as we go on, analytic continuation in $N$ and in $d$ are not entirely independent operations, and there are many conceptual similarities. As a very simple example, note that upon compactification (on a sufficiently symmetric manifold), the Lorentz group becomes a flavor symmetry.

With this in mind, a very rough, intuitive definition could look something like this. Given an infinite sequence of QFTs defined for all $d\in\mathbb N$, one computes a certain observable, such as a scaling dimension $\Delta_d$, for all $d\in\mathbb N$. Then, if this sequence is sufficiently well-behaved in some sense (such as having bounded forward differences), then it has a unique analytic continuation to complex $d$, and we declare that this is the observable the QFT computes at fractional $d$.

Of course this is too vague to be an actual definition. For starters, a QFT is not just a collection of numbers $\{\Delta\}$, but a much richer structure that implies infinitely many coherence relations between them. Unfortunately, the honest answer is that we don't have a general definition. We can mention a few special cases here where a working definition exists, but we will spend the rest of this paper explaining some of the obstructions associated to this operation and why the problem is still very much open.

One situation where rigorous analytic continuation to fractional $d$ may be possible is for QFTs that admit a lattice discretization~\cite{10.1143/PTP.69.65,YGefen_1983,YGefen_1984,YGefen_19842,PhysRevB.34.2060,LEGUILLOU1990559,Eyink:1989dv,PhysRevB.63.184420,Kar:2011vt,Yi_2015,Yoshida:2014szs}. These can put on a lattice whose continuum limit has fractional Hausdorff dimension, and this gives a non-perturbative definition of the field theory at that $d\in\mathbb R$. A natural question is whether this continuum limit always exists, whether it is unique, and whether the result is actually analytic in $d$. We do not know the answer to any of these questions. Another important issue is how to define this continuation in theories that do not admit a nice discretization; as we shall see, fermions are particularly subtle in fractional dimension, precisely for the same reason these objects refuse to sit peacefully on a lattice.

As for analytic continuation directly in the continuum, we only know how to define this in perturbative settings. For example, the $O(N)$ model admits a $1/N$ expansion, and this perturbative series happens to be analytic in $d$. Indeed, the large $N$ expansion is organized in terms of Feynman integrals, and these admit a general definition of analytic continuation~\cite{Collins:1984xc}. 

So, to summarize:
\begin{itemize}
\item We do not have a general definition of analytic continuation in $d$, nor in $N$. There are some cases where one can give a working prescription, but these are very non-generic and it is not clear if the result is always well-defined and unique.
\item If a general definition exists at all, it will only lead to a unique extension if we choose an \emph{infinite} sequence of QFTs as our starting point. This holds both for analytic continuation of $d$ and $N$.
\item Not every infinite sequence admits a continuation. Indeed, there is no reason to expect a random sequence like $T(1)=\text{Ising}$, $T(2)=SU(2)+\psi_\text{adj}$, $T(3)=\text{abelian Higgs}$, etc.~to admit a meaningful interpolation. 

\end{itemize}

\subsection{Kinematics}\label{sec:kinematics}

As above, there is no general definition of analytic continuation of a QFT to $d\in\mathbb C$. We can try to make some progress by disentangling the dynamics from the kinematics. Namely, here we would like to understand the representation theory of the Lorentz group at fractional $d$. We begin with the Lorentz group being $O(d)$, and then explain why other global forms such as $SO(d)$ and $Spin(d)$ do not really have a satisfactory continuation.

\paragraph{Orthogonal group $\boldsymbol{O(d)}$.} We first recall some basic facts about $\operatorname{Rep}(O(d))$~\cite{Goodman2009-ge}. When $d$ is odd, any irreducible representation can be specified by an $SO(d)$ irreducible representation $\lambda\in\mathrm{Rep}(SO(d))$, together with a choice of intrinsic parity $\pm$. When $d$ is even, $SO(d)$ has a $\mathbb Z_2$ outer automorphism $\mathsf C$ that permutes the last two Dynkin labels. Then, if $\lambda$ is fixed by this action, it induces two representations of $O(d)$ which take the same form as before, $(\lambda,\pm)$. If $\lambda\in\mathrm{Rep}(SO(d))$ is not fixed by $\mathsf C$, only the direct sum $\lambda+\lambda^{\mathsf C}$ induces an $O(d)$ representation. We call $(\lambda,+)$ a true tensor and $(\lambda,-)$ a pseudo-tensor. For representations of type $\lambda+\lambda^{\mathsf C},$ there is no intrinsic parity (the two choices for this sign are exchanged by electric-magnetic duality).

We now explain what the analogue of $\operatorname{Rep}(O(d))$ in the case of fractional $d$ is. The precise mathematical framework to understand groups of continuous rank is that of Deligne categories~\cite{Deligne:rep}; see~\cite{Binder:2019zqc} for a physics-oriented presentation. Here we simply summarize some facts about Deligne categories without providing precise definitions; see the references for more details.

In short, an irreducible representation of $O(d)$ at fractional $d$ is a simple object in the category $\widetilde{\operatorname{Rep}}(O(d))$. Such objects are labeled by terminating sequences of non-negative integers, which play the role of the usual Dynkin labels for integer $d$.

For convenience we will denote these sequences via the associated Young diagram. For example, the vector of $\widetilde{\operatorname{Rep}}(O(d))$ corresponds to the sequence $\ydiagram1=(1,0,\dots)$. It becomes a true vector of $O(d)$ when $d\in\mathbb N$; we will explain how to recover pseudo-tensors below.

The dimension of a given object in $\widetilde{\operatorname{Rep}}(O(d))$ can be obtained by plugging its Dynkin labels into the regular Weyl formula of $O(d)$ for increasing values of $d\in\mathbb N$; this produces an integer sequence of polynomial growth which, by standard arguments, has a unique analytic continuation to $d\in\mathbb C$. For example, the vector representation of $O(d)$ has dimension $1,2,3,\dots$ for $d=1,2,3,\dots$, and this has a unique analytic continuation to the Riemann sphere, namely $f(d)=d$. Then, the dimension of the object labeled by $(1,0,\dots)$ is $\dim(\ydiagram1)=d$ for all $d\in\mathbb C$.

Conversely, any sequence of true-tensor representations of $O(d)$ that share the same Young diagram for all $d\in\mathbb N$ defines an object in the Deligne category above. We will call such sequences \emph{stable}. An example of an unstable sequence is $S^d\ydiagram1$: this has $d$ boxes, which increases as we increase $d$. The dimension of this sequence is $\dim(S^d\ydiagram1)=\binom{2d-1}{d}-\binom{2d-3}{d-2}$, which grows like $\sim2^{2d}$, too fast to have a good, unique continuation to $d\in\mathbb C$. The category $\widetilde{\operatorname{Rep}}(O(d))$ only describes stable sequences, precisely because these, and only these, have a unique analytic continuation to complex $d$.

Let us stress that the Deligne category above is not the ``Rep'' of anything: there is no structure $O(d),d\in\mathbb C$, whose category of representations coincides with $\widetilde{\operatorname{Rep}}(O(d))$. In order to talk about $O(d)$ at fractional $d$ we do not literally consider a Lie group of complex rank; instead, it is enough to understand $O(d)$ for all $d\in\mathbb N$. (Intuitively, $\widetilde{\operatorname{Rep}}(O(d))$ is a regularized version of the representation category of the direct limit $O(\infty)$.)

The objects in $\widetilde{\operatorname{Rep}}(O(d))$ can be composed, which generalizes the usual notion of tensor product of representations. In practice, composition looks exactly like regular tensor products as computed in any $O(d)$ for $d\in\mathbb N$ large enough to fit the representations. For example,
\begin{equation}\label{eq:tensor_frac_d}
\ydiagram{1,1}\times\ydiagram{1,1}=\boldsymbol1+\ydiagram{1,1}+\ydiagram2+\ydiagram{2,2}+\ydiagram{2,1,1}+\ydiagram{1,1,1,1}
\end{equation}
holds for all integers $d\ge11$, and therefore it is also true in $\widetilde{\operatorname{Rep}}(O(d))$. 

The category $\widetilde{\operatorname{Rep}}(O(d))$ allows us to meaningfully talk about tensor fields at fractional $d$. An important question is how it relates to $\operatorname{Rep}(O(d))$ when $d$ happens to be an integer. These are not identical on the nose since, for example, $\wedge^4\ydiagram1$ is a non-trivial object in $\widetilde{\operatorname{Rep}}(O(d))$ for any $d$, but it is the zero object in $\operatorname{Rep}(O(3))$. The short answer is that, if $d\in\mathbb N$, there is an equivalence of categories $\operatorname{Rep}(O(d))\cong \widetilde{\operatorname{Rep}}(O(d))/\mathcal I$ where $\mathcal I$ is the collection of ``negligible morphisms'' --- the quotient kills all objects in $\widetilde{\operatorname{Rep}}(O(d))$ that are trivial in $\operatorname{Rep}(O(d))$.

A followup question is what happens to the pseudo-tensors of $O(d)$, which are not naturally described by $\widetilde{\operatorname{Rep}}(O(d))$. The answer is that they ``reappear'' when we quotient by the null representations above. To see this, note that any pseudo-tensor can equivalently be written as the product of a true tensor and the Levi-Civita symbol. For example $(\ydiagram2,-)\cong (\ydiagram{2,1},+)$ when $d=3$, $(\ydiagram2,-)\cong(\ydiagram{2,1,1},+)$ when $d=4$, etc. Therefore, for any fixed $d\in\mathbb N$, the pseudo-tensor $(\ydiagram2,-)$ is described by a certain object in $\widetilde{\operatorname{Rep}}(O(d))$: $\ydiagram{2,1}$ when $d=3$, $\ydiagram{2,1,1}$ when $d=4$, etc. 

In this sense, the Deligne category is not ``missing'' any $O(d)$ representation: for any fixed $d\in\mathbb N$, there is some object in $\widetilde{\operatorname{Rep}}(O(d))$ that descends to any given pseudo-tensor. But at the same time there is no meaningful way to talk about pseudo-tensors for $d\in\mathbb C\setminus\mathbb N$. As far as this categorical framework is concerned, only true tensors make sense for general $d\in\mathbb C$; pseudo-tensors only reappear when $d\in\mathbb N$. (An  analogous statement holds for representations of type $\lambda+\lambda^{\mathsf C}$. These do not correspond to any object in $\widetilde{\operatorname{Rep}}(O(d))$ but, for any given $d\in\mathbb N$, there is some object that descends to $\lambda+\lambda^{\mathsf C}$. An example of a tensor of this type is the $(d/2)$-form. This does not correspond to an object in $\widetilde{\operatorname{Rep}}(O(d))$ since its Dynkin labels are $(0,\dots,0,2,0)\oplus (0,\dots,0,0,2)$, which does not have a meaningful direct limit, this is not a terminating sequence. But this $(d/2)$-form does have an image when $d\in\mathbb N$: $\wedge^2\ydiagram1$ when $d=4$, $\wedge^3\ydiagram1$ when $d=6$, etc.)

In conclusion, the category $\widetilde{\operatorname{Rep}}(O(d))$ captures all representations of $O(d)$ and, for stable, true tensors, it gives their canonical continuation to $d\in\mathbb C$. We can anticipate that, in QFT, pseudo-tensors and tensors whose spin depends on $d$ (say, some field that transforms as $S^d\ydiagram1$) will both fight back if we try to continue them to fractional $d$. Only fields that are true tensors and that define stable sequences as we vary $d\in\mathbb N$ should have a nice analytic continuation. So for example, the generalization of free Maxwell where the gauge field is a $p$-form should be amenable to analytic continuation for any fixed $p$, but if we want a generalization of the theta term, we should take $p=d/2$, in which case the theory will likely protest.

\paragraph{Other groups: $\boldsymbol{SO(d),Spin(d)}$, etc.} Let us now discuss what goes wrong if we try to repeat this story for other global forms of the Lorentz group. For example, say that we try to define a putative $\widetilde{\operatorname{Rep}}(SO(d))$ via stable sequences of $SO(d)$ representations, i.e., we assume that its simple objects are again labeled by terminating sequences of non-negative integers. One problem with this is that, unlike $\widetilde{\operatorname{Rep}}(O(d))$, our naive $\widetilde{\operatorname{Rep}}(SO(d))$ is actually missing some representations. In the case of $O(d)$ we explained that there is no simple object in $\widetilde{\operatorname{Rep}}(O(d))$ that corresponds to the $(d/2)$-form but, for any fixed $d\in\mathbb N$, this representation does have an image: $\wedge^2\ydiagram1$ when $d=4$, $\wedge^3\ydiagram1$ when $d=6$, etc. On the other hand, in the case of $SO(d)$, this $(d/2)$-form is actually reducible, and we can break it into the self-dual and anti-self-dual parts. These irreducible pieces do not have an image in our $\widetilde{\operatorname{Rep}}(SO(d))$ at all. (And indeed, it was already noticed in the early days of dim-reg that (anti-)self-dual fields are especially tricky to analytically continue~\cite{BOS1988177}.)

Another problem with this naive definition is that certain operations, like tensor products, are non-analytic in $d$. For example, consider the decomposition~\eqref{eq:tensor_frac_d}. This equation is true in $SO(d)$ for all $d\ge11$, and therefore we would guess that it is also true in $\widetilde{\operatorname{Rep}}(SO(d))$ for all $d\in\mathbb C$. But then, we would conclude that $\ydiagram{1,1}\times\ydiagram{1,1}$ contains only one singlet for any $d\in\mathbb C$, which is actually false: in $SO(4)$ one has $\wedge^4\ydiagram1\cong\boldsymbol 1$, and therefore there are two singlets when $d=4$. In a formal sense, objects like $\operatorname{Hom}(R_1,R_2)$ in the would-be category $\widetilde{\operatorname{Rep}}(SO(d))$ are discontinuous in $d$, due to low-rank exceptions.

In fact, the same non-analyticity would occur in $\widetilde{\operatorname{Rep}}(O(d))$ if we insisted that this category should describe pseudo-tensors as well. Indeed, one has $(\ydiagram{1,1},-)\times(\ydiagram{1,1},+)=(\boldsymbol 1,-)+(\wedge^4\ydiagram1,-)+\cdots$ in $O(d)$ for all $d\ge11$, which would imply that there are no  singlets for any $d\in\mathbb C$, which is actually false: when $d=4$, there is one singlet, since in this case $(\wedge^4\ydiagram1,-)\cong(\boldsymbol1,+)$. True tensors are stable, pseudo-tensors are not.

Finally, let us consider the $Spin(d)$ form of the Lorentz group. This group has the same issues as $SO(d)$: self-dual and anti-self-dual representations are not captured by terminating sequences of Dynkin labels at all, and low-rank exceptions make tensor products non-analytic in $d$. One could wonder if $Pin(d)$ might fare better, since $O(d)$ does admit a continuation as opposed to $SO(d)$. The answer is \emph{no, not really}: pinor representations are also not captured by terminating sequences of Dynkin labels. Indeed, this representation has Dynkin labels $(0,\dots,0,1)$ for odd $d$, and $(0,\dots,0,1,0)\oplus (0,\dots,0,0,1)$ for even $d$, and these sequences do not have a meaningful direct limit. Pinors are unstable representations. For example, their dimension $2^{\lfloor d/2\rfloor}$ is non-polynomial in $d$ and, as such, it does not have a well-defined, unique continuation to $d\in\mathbb C$\footnote{A possible formalization of these facts, suggested to us by Justin Kulp, is the following. As above, $\widetilde{\operatorname{Rep}}(O(d))$ is the correct category to talk about stable representations. On the other hand, unstable representations are not objects in this category. But, the fact that unstable representations can be tensored to produce stable ones (e.g., the square of a pinor is $1+\ydiagram1+\wedge^2\ydiagram1+\cdots$) means that these large representations should be objects in a module category of $\widetilde{\operatorname{Rep}}(O(d))$, i.e., what we are missing is some version of $\widetilde{\operatorname{Rep}}(\mathrm{Cliff}(d))$ as a module over $\widetilde{\operatorname{Rep}}(O(d))$. See also~\cite{Gaiotto:2025nrd} (and in particular the discussion around footnote 8).}.

As a well-known example of the complications associated to pinors, consider the trace $\tr[\gamma^{\mu\nu\rho}]$. If we compute this for small values of $d$, we find
\begin{equation}
\begin{aligned}
d=2&\colon \tr[\gamma^{\mu\nu\rho}]=0\\
d=3&\colon \tr[\gamma^{\mu\nu\rho}]=2i\epsilon^{\mu\nu\rho}\\
d=4&\colon \tr[\gamma^{\mu\nu\rho}]=0\\
d=5&\colon \tr[\gamma^{\mu\nu\rho}]=0\\
&\vdots
\end{aligned}
\end{equation}
whose stable limit is $\tr[\gamma^{\mu\nu\rho}]=0$. This is the standard 't Hooft-Veltman prescription for this trace; while correct around $d=4$, it gives the wrong answer for the small rank exception $d=3$. This means that if we use this continuation for $d\in\mathbb C$, we will necessarily get the wrong answer when we take $d\to3$. More generally, all prescriptions for fermions in fractional dimension found in the literature ignore by hand low-rank exceptions that occur below $d=4$, which means that they will necessarily fail for small enough $d$. If our only goal is to continue to an infinitesimal neighborhood of $d=4$, this is not the end of the world, but if we want a continuation to $d\in\mathbb C$ that gives the correct answer when $d=3$ or $d=2$, we cannot drop small $d$ exceptions, and we have to work harder.

\paragraph{Back to physics.} We should note here that the fact that the representation theory of $O(d)$ is well-defined for non-integer $d$ does not imply that analytic continuation of $O(d)$-symmetric QFTs is necessarily easy. For example, scalar fields transform according to the trivial representation of $O(d)$, which is stable. But, out of these scalars, one can construct more complicated tensors, including unstable ones. Indeed, consider the evanescent operators of~\cite{Hogervorst:2015akt}, given by $R_p:=\tr_{\wedge^p\ydiagram1}[M]$, where $M_{\mu\nu}=\partial_\mu\partial_\nu\phi$. The representations $\wedge^p\ydiagram1$ are all well-defined for $d\in\mathbb C$. But if we take $d$ to be an integer, then the operators with $p>d$ are all null. As such, their continuation to fractional $d$ is rather subtle: for example, they often have negative norm~\cite{Hogervorst:2015akt}.

For QFTs that are invariant under $SO(d)$ only, things are even less straightforward. The lack of a proper $\widetilde{\operatorname{Rep}}(SO(d))$ category means that parity-breaking QFTs cannot be continued to complex $d$, at least not if we want to use the framework of Deligne categories as above. Of course, in physics we never let the lack of a rigorous framework stop us. We should try to continue such theories anyway; if the results are consistent, then a suitable generalization of Deligne categories must exist.

In this paper we will only study vector-like theories, but with a focus on parity-odd operators. Such representations are unstable from the point of view of $\widetilde{\operatorname{Rep}}(O(d))$, and therefore they should present the same obstructions as generic operators in a chiral theory. For example, much like $\operatorname{Hom}_{SO(d)}(\ydiagram{1,1}\times\ydiagram{1,1},\boldsymbol1)=\delta_{d,4}\mathbb C$ is non-analytic in $d$ due to the $SO(4)$ low-rank isomorphism $\wedge^4\ydiagram1\cong\boldsymbol 1$, one also has $\operatorname{Hom}_{O(d)}((\ydiagram{1,1},+)\times(\ydiagram{1,1},-),\boldsymbol1)=\delta_{d,4}\mathbb C$ due to the $O(4)$ low-rank isomorphism $(\wedge^4\ydiagram1,-)\cong(\boldsymbol1,+)$. These low-rank exceptions will force us to go beyond the current framework of analytic continuation in QFT.


For completeness, let us mention that this whole discussion also goes through for analytic continuation of flavor (or gauge) symmetry groups. For example, an $O(N)$ symmetry should allow for a somewhat straightforward continuation of $N$, but an $SO(N)$ symmetry will likely be harder to define. For a general symmetry group, its Deligne interpolating category is well defined if its representation theory is stable; the best understood examples~\cite{Binder:2019zqc} are $U(N)$, $Sp(N)$ and $S_N$. For example, the fact that $\widetilde{\operatorname{Rep}}(U(N))$ exists but $\widetilde{\operatorname{Rep}}(SU(N))$ does not is related to the well-known fact that mesons in planar QCD are much easier to understand than baryons~\cite{Witten:1979kh}. The latter are unstable in the sense above: while $\ydiagram1\otimes\overline{\ydiagram1}$ contains a singlet for any number of colors, there is no finite power of $\ydiagram1$ that yields a singlet in ``$\widetilde{\operatorname{Rep}}(SU(N))$''. Unstable representations are non-perturbative in $1/N$, and they obstruct any naive attempts at analytic continuation of this parameter. (Similarly, the existence of $\widetilde{\operatorname{Rep}}(S_N)$ is related to the success of operations like the replica trick and the symmetric product orbifold).


\subsection{A simple example: free fermions}

As we reviewed above, fields transforming in unstable representations of the Lorentz group are particularly adverse to analytic continuation, (s)pinors being the most important example. Does this mean fermionic theories cannot be analytically continued to $d\in\mathbb C$? We believe a meaningful continuation is possible, although we will explain here some of the obstructions one has to overcome, and then describe our best attempts at doing so.

Here we will frame the discussion in the context of free Dirac fermions. We will perform a few computations in this QFT and then ask whether the result admits a meaningful continuation or not. This is an extended version of the toy model discussed in the introduction; readers interested in the interacting case only can safely skip to section~\ref{sec:results}.

A simple observable one can consider is the free energy of this system. This quantity is given by~\cite{Benjamin:2023qsc}
\begin{equation}
\log Z=2^{\lfloor d/2\rfloor+2-d}\sum_{n=1}^\infty\frac{(-1)^{n+1}}{n}\sinh(n/2T)^{1-d}
\end{equation}
Does this admit a meaningful continuation to complex $d$? The factor $2^{\lfloor d/2\rfloor}$ doesn't seem to like $d\not\in\mathbb N$, but this is obviously a moot point: we could just as well ask about the free energy \emph{per fermionic degree of freedom}, which simply amounts to dividing the expression above by this awkward factor. The resulting object does admit a perfectly healthy continuation to complex $d$. (And note that one can extract some interesting physics already for this extremely simple observable: for example, the limits $T\to0$ and $d\to1$ do not commute. This reflects the fact that the ground state is discontinuous at $d=1$: there is only one vacuum for $d>1$, but two vacua at $d=1$.)

Are there observables for which analytic continuation is not possible? A potential example is the following. Give the fermion a mass $m$, and couple it to a background $U(1)$ gauge field $A$. At energies below $m$ we can integrate out the fermion, producing a local effective action for $A$~\cite{PhysRevD.29.2366,PhysRevLett.48.975}:
\begin{equation}\label{eq:eff_S_A}
\begin{aligned}
S_\text{eff}(A)&\supset\operatorname{Tr}\log(\slashed\partial+m+i\slashed A)\\
&=\operatorname{Tr}\log(\slashed\partial+m)
+\operatorname{Tr}\biggl[\frac{1}{\slashed\partial+m}i\slashed A\biggr]-\frac12\operatorname{Tr}\biggl[\frac{1}{\slashed\partial+m}i\slashed A\frac{1}{\slashed\partial+m}i\slashed A\biggr]+\cdots
\end{aligned}
\end{equation}

The parity-even part of this effective action is straightforward. For example, at the quadratic level $\mathcal O(A^2)$, it reads\footnote{We denote momentum integrals by $\int_k:=\int\frac{\mathrm d^dk}{(2\pi)^d}$ and the free fermion propagator by $S(p)=\frac{1}{i\slashed p+m}$.}
\begin{equation}
S^\text{even}_\text{eff}(A)\supset \frac12\int_k A_\mu(k)A_\nu(-k)\int_p\tr[S(p)\gamma^\mu S(p+k)\gamma^\nu]_\text{even}
\end{equation}
where 
\begin{equation}\label{eq:A2_trace}
\begin{aligned}
\tr[S(p)\gamma^\mu S(p+k)\gamma^\nu]_\text{even}&=\frac{-1}{(p^2+m^2)((p+k)^2+m^2)}\bigl(\tr[\slashed p\gamma^\mu(\slashed p+\slashed k)\gamma^\nu]-m^2\tr[\gamma^\mu\gamma^\nu]\bigr)\\
\tr[S(p)\gamma^\mu S(p+k)\gamma^\nu]_\text{odd}&=\frac{-im}{(p^2+m^2)((p+k)^2+m^2)}\bigl(\tr[\slashed p\gamma^\mu\gamma^\nu]+\tr[\gamma^\mu(\slashed p+\slashed k)\gamma^\nu]\bigr)
\end{aligned}
\end{equation}
Evaluating the trace and performing the $p$ integral\footnote{Here we neglect the question of whether such momentum integrals converge in the first place. While very important in general, and even more so in the context of varying $d$, this issue is orthogonal to the one we are trying to address here. So below we just cut off the integrals at $p=\Lambda$ and consider $m\gg\Lambda\gg k$, and throw away subleading terms like $\Lambda/m$.}, we learn that the heavy fermion simply generates a renormalization of the would-be gauge coupling:
\begin{equation}
S^\text{even}_\text{eff}(A)\supset2^{\lfloor d/2\rfloor}\frac{\Gamma(2-d/2)}{3(4 \pi )^{d/2}}| m|^{d-4}\int_k \frac12A_\mu(k) (k^\mu k^\nu-k^2 \delta^{\mu\nu}) A_\nu(-k)
\end{equation}
Up to the inconsequential factor $2^{\lfloor d/2\rfloor}$, this admits a natural analytic continuation to $d\in\mathbb C$.

The parity-odd part in $S_\text{eff}(A)$ is a very different story. For example, at the linear level $\mathcal O(A)$ we find
\begin{equation}
\begin{aligned}
S^\text{odd}_\text{eff}(A)&\supset i\int_kA_\mu(k)\int_p\tr[S(p)\gamma^\mu]\\
&=i\tr[\gamma^\mu]\sign(m)| m|^{d-1}\frac{\Gamma(1-d/2)}{(4 \pi )^{d/2}}\int_kA_\mu(k) 
\end{aligned}
\end{equation}

Now, while the coefficient $\frac{\Gamma(1-d/2)}{(4 \pi )^{d/2}}$ admits a meaningful continuation to $d\in\mathbb C$, things are not so simple for the trace $\tr[\gamma^\mu]$. In $d=1$ this is just $\tr[\gamma^\mu]=\pm1$, and we obtain the well-known $1d$ Chern-Simons term that arises when integrating out a heavy fermion: $S^\text{odd}_\text{eff}(A)\supset \frac i2\sign(m)\int\! A$. If $d$ is an integer different from $1$, this trace vanishes instead. So, for general $d\in\mathbb N$, the effective action contains a Kronecker delta $\delta_{d,1}$, and there is no natural way to analytically continue such an object to fractional $d$. Similarly, at the quadratic level we find something proportional to $\tr[\gamma^{\mu\nu\rho}]\propto\delta_{d,3}\epsilon^{\mu\nu\rho}$, for which there is no obvious analytic continuation either. 
And of course the pattern continues: there is a $5d$ Chern-Simons term at the cubic level, etc.

So even for this very simple theory, which is free and vector-like, it is not entirely clear how to continue away from integer dimensions. Some questions, like the free energy, are well-defined for $d\in\mathbb C$; some others, like the free energy in presence of an external field, not quite so. And we see that the main obstruction is the existence of low-rank exceptions. Indeed, the even part of $S_\text{eff}(A)$ only involves traces of an even number of gammas, and these are the same for all $d$; but the odd part involves traces with an odd number of gammas, and those vanish unless the number of insertions equals the number of spacetime dimensions. In other words, $S_\text{eff}^\text{even}(A)$ has a stable limit (and it agrees with it) but $S_\text{eff}^\text{odd}(A)$ does not.

One could complain that the issue with this analysis is that we are trying to analytically continue an object, $S_\text{eff}(A)$, which is a function of a \emph{vector} field, and it is sort of meaningless to talk about non-trivial tensors in fractional dimension anyway; only scalars are to be continued. This is not entirely true since, as we explained above, stable tensor fields make sense in fractional dimension, and vectors are stable. But in any case, there are complications for scalar questions as well, as we explain next.

For simplicity let us turn off $A$ and let us set the fermion mass to zero, and consider the correlation functions of $\bar\psi\psi$. This operator is parity-odd\footnote{Parity here refers to reflections, namely flipping the sign of a \emph{single} spatial coordinate, say, $x_1$. This acts as $\psi\mapsto\gamma_1\psi$ and $\bar\psi\mapsto-\bar\psi\gamma_1$ (the relative sign is inherited from the Lorentzian identification $\bar\psi=\psi^\dagger\gamma^0$). The usual mass term $\bar\psi\psi$ is odd under reflections. In even dimensions, one can also consider $\bar\psi\gamma^\star\psi$, which is even.}, so we might expect the same difficulties as above; but crucially, it is a Lorentz scalar, so if we find any obstacles to analytically continuing its correlators, those will be harder to dismiss. We can begin with its expectation value, which is given by
\begin{equation}\label{eq:1pt_intro}
\begin{aligned}
\tikz[baseline=-3pt]{\draw[thick] (0,0) circle (.5cm);
\draw[fermion] (.5,.1) -- (.5,.11);
\filldraw (-.5,0) circle (1pt);}\ &=-\lim_{x\to0}\tr[S(x)]\\
&=\frac{\Gamma(d/2)}{2\pi^{d/2}}\tr[\gamma_\mu]\lim_{x\to0}\frac{x^\mu}{x^d}
\end{aligned}
\end{equation}
which contains once again the problematic $\tr[\gamma_\mu]$ factor. This trace vanishes for all integers $d>1$. But in $d=1$ the trace is non-zero and we find $\langle\bar\psi\psi\rangle=\frac12\delta_{1,d}$. Again, it is unclear how this should be analytically continued in $d$.

There is a simple explanation for this discontinuity at $d=1$. This operator has scaling dimension $\Delta(\bar\psi\psi)=d-1$. For $d>1$, this is non-zero, and its \emph{vev} is forced to vanish by the (unbroken) scaling symmetry. But in $d=1$ the scaling dimension vanishes, and this operator becomes degenerate with the identity, which allows for mixing of these operators, explaining why $\bar\psi\psi$ is allowed to acquire a \emph{vev}. This mixing is a discontinuous process and we should not be alarmed by it. In $d=1$ we could define a different operator by subtracting the contribution of the identity, namely by normal ordering ${:}\bar\psi\psi{:}=\bar\psi\psi-1/2$, which now has vanishing \emph{vev}. Apparently, the reasonable analytic continuation of this expectation value is $\langle\bar\psi\psi\rangle=0$ for all $d$, including $d=1$. This is a bit misleading: the combination $\bar\psi\psi-1/2$ doesn't make sense above $d=1$, since $\bar\psi\psi$ has non-zero $\Delta$ (more precisely, this combination is not a conformal primary). So we either work with $\bar\psi\psi$, which has a discontinuous \emph{vev}, or with the discontinuous object $\bar\psi\psi-\delta_{1,d}/2$, which has continuous \emph{vev}. Either way, this bilinear operator is simply non-analytic in $d$ around $d=1$.

Are there any further discontinuities in $\bar\psi\psi$ at higher $d$? We can answer this question by looking at its connected five-point function
\begin{equation}\label{eq:5pt_function}
\tikz[baseline=-3pt]{\draw (0,0) circle (.5cm);

\foreach \x in {0,...,4}{
\filldraw ({.5*cos(72*\x)},{.5*sin(72*\x)}) circle (1pt);
\draw[thick, postaction={decorate}, decoration={markings, mark=at position 0.5 with {\arrow[rotate=-10]{Stealth}}},rotate=72*\x] (-20:.5cm) arc(-20:120:.5cm);
}
\node[scale=.8] at ({.8*cos(72*0)},{.8*sin(72*0)}) {$x_1$};
\node[scale=.8] at ({.8*cos(72*1)},{.8*sin(72*1)}) {$x_5$};
\node[scale=.8] at ({.8*cos(72*2)},{.8*sin(72*2)}) {$x_4$};
\node[scale=.8] at ({.8*cos(72*3)},{.8*sin(72*3)}) {$x_3$};
\node[scale=.8] at ({.8*cos(72*4)},{.8*sin(72*4)}) {$x_2$};

}+\ \text{permutations}
\end{equation}
for a total of $4!=24$ permutations. The first permutation is proportional to
\begin{equation}
\biggl(\prod_{i=1}^5\frac{(x_i-x_{i+1})_{\mu_i}}{|x_i-x_{i+1}|^d}\biggr)
\tr[\gamma^{\mu_1}\gamma^{\mu_2}\gamma^{\mu_3}\gamma^{\mu_4}\gamma^{\mu_5}]
\end{equation}
We can use the Clifford relation to write the product of five gammas as the fully anti-symmetrized product $\gamma^{\mu_1\mu_2\mu_3\mu_4\mu_5}$, plus terms with three gammas, $\gamma^{\mu_1\mu_2\mu_3}\delta^{\mu_4\mu_5}$ (plus permutations), and terms with a single gamma, $\gamma^{\mu_1}\delta^{\mu_2\mu_3}\delta^{\mu_4\mu_5}$ (plus permutations). The fully anti-symmetric term $\gamma^{\mu_1\mu_2\mu_3\mu_4\mu_5}$ vanishes due to translation invariance, since we only have four linearly independent vectors to contract it with. So the final answer is that the five-point function is proportional to various combinations of the $x_i$'s contracted with $\tr[\gamma^{\mu\nu\rho}]$ and with $\tr[\gamma^\mu]$. Although not obvious from this discussion, the overall coefficient is non-zero, i.e., the various combinations do not cancel each other out\footnote{The full expression is too large to include here. If we take the simplifying configuration $x_1=(0,0,0)$, $x_2=(0,0,a)$, $x_3=(0,0,1)$, $x_4=(0,b,0)$, $x_5=(c,0,0)$, and take the the limit $a\to0,b\to\infty$ so that the result fits in a single line, we find
\begin{equation}\label{eq:5pt_3d}
b^5|a|\langle \bar\psi\psi(x_1)\cdots\bar\psi\psi(x_5)\rangle= i\operatorname{sign}(c)\biggl(\frac{1}{c^2}-\frac{|c|}{(1+c^2)^{3/2}}\biggr)
\end{equation}
up to factors of $2\pi$. This is of course odd under parity, as required, but crucially non-zero. 
}.

The conclusion of this little exercise is that this correlation function vanishes for all $d\in\mathbb N$ except for $d=3$. Then, the stable limit of this correlation function is that it vanishes for all $d\in\mathbb C$, and we again discover a non-analytic behavior of $\bar\psi\psi$, this time not only at $d=1$, but also at $d=3$. It is not hard to imagine that the seven-point function is proportional to the trace of five gammas, giving rise to a discontinuity at $d=5$, etc. Not only is the \emph{vev} of $\bar\psi\psi$ discontinuous as a function of $d$, so are its higher correlation functions.


For future reference we mention that, if we were interested in \emph{quartic} operators such as $(\bar\psi\psi)^2$, then there are well-known subtleties such as the existence of operators of the form~\cite{BONDI1990268,DUGAN1991239}
\begin{equation}\label{eq:evanes}
\mathscr O_n:=\bar\psi\gamma^{\mu_1\cdots\mu_n}\psi\,\bar\psi\gamma_{\mu_1\cdots\mu_n}\psi
\end{equation}
These are scalars and degenerate with $(\bar\psi\psi)^2$, so these operator will generically mix. In integer dimensions, the operators with $n>d$ vanish by anti-symmetry. But in fractional dimension these are independent and non-trivial, and one must consider them together with the original quartic~\cite{DiPietro:2015taa,DiPietro:2017kcd,DiPietro:2017vsp}. The quartics with $n>d$ are evanescent: non-trivial when $d\not\in \mathbb N$, but null when $d\in\mathbb N$. The difficulty in defining traces of gamma matrices in non-integer dimension also shows up in $\mathscr O_n$; for example, the OPE of $\bar\psi\psi$ with the quartics above is
\begin{equation}
\bar\psi\psi(x)\,\mathscr O_n(0)\sim\frac{1}{x^{2(d-1)}}\tr[\gamma_{\mu_1\cdots\mu_n}]\bar\psi\gamma^{\mu_1\cdots\mu_n}\psi+\cdots
\end{equation}
Without a prescription for $\tr[\gamma_{\mu_1\cdots\mu_n}]$ in fractional dimension, we are somewhat forced to admit that OPEs are discontinuous as a function of $d$, even for scalar operators as we are considering here. 

Clearly, even free theories present some difficulties in defining a meaningful continuation in $d$. In the reminder of this paper we will study interacting theories instead. But before doing that, this might be a good point to pause and discuss our conventions for gamma matrices in fractional dimension.

\subsubsection{Gamma matrices}\label{sec:gamma}

\paragraph{Integer dimension.} For integer $d$ we define the Clifford algebra as
\begin{equation}\label{eq:cliff}
\gamma^\mu\gamma^\nu+\gamma^\nu\gamma^\mu=2\delta^{\mu\nu}
\end{equation}

It is a well-known fact that, if $d$ is an integer and $\delta^{\mu\nu}$ denotes the standard metric on $\mathbb R^d$, then the Clifford algebra admits an (essentially unique) irreducible matrix representation, of dimension $2^{\lfloor d/2\rfloor}$ (see e.g.~\cite{BERG_2001}). For the discussion below we don't need to assume that the matrix representation has been chosen to be irreducible. In fact, when a QFT contains $N_f$ fermions, one might as well declare that the theory has a \emph{single} fermion that transforms under the reducible representation that consists of $N_f$ copies of the irreducible one. This latter perspective is notationally more convenient, especially in the case of fractional $d$.

The Clifford relation~\eqref{eq:cliff} implies the usual reduction identities, such as
\begin{equation}\label{eq:cliff_red}
\begin{aligned}
\gamma^\mu\gamma_\mu&=d \\
\gamma^\mu\gamma^\nu\gamma_\mu&=(2-d)\gamma^\nu
\end{aligned}
\end{equation}
etc. Similarly, one can also take the matrix trace of the Clifford relation, which yields the usual trace identities
\begin{equation}\label{eq:master_trace}
\begin{aligned}
\tr[\gamma^\mu\gamma^\nu]&=\delta^{\mu\nu}\tr[1]\\
\tr[\gamma^\mu\gamma^\nu\gamma^\rho\gamma^\sigma]&=(\delta^{\mu\nu}\delta^{\rho\sigma}-\delta^{\mu\rho}\delta^{\nu\sigma}+\delta^{\mu\sigma}\delta^{\nu\rho})\tr[1]
\end{aligned}
\end{equation}
etc.

Traces over the odd part of Clifford are a little less well-behaved. First, we can use the Clifford relation to write a product of gamma matrices as their full anti-symmetrization, plus products of fewer matrices. For example,
\begin{equation}\label{eq:cliff_reduce_3}
\gamma_\mu\gamma_\nu\gamma_\rho=\gamma_{\mu\nu\rho}+\gamma_\mu\delta_{\nu\rho}-\gamma_\nu\delta_{\mu\rho}+\gamma_\rho\delta_{\mu\nu}
\end{equation}
where $\gamma_{\mu\nu\rho}:=\frac{1}{3!}(\gamma_\mu\gamma_\nu\gamma_\rho\pm\text{permutations})$. Therefore, we only need to work out traces of fully anti-symmetrized gammas $\tr[\gamma_{\mu_1\cdots\mu_n}]$ since more general products can always be reduced to these.

As is well-known, $\tr[\gamma_{\mu_1\cdots\mu_n}]$ vanishes unless $n$ is equal to $d$ and is also odd. This follows from the basic identity
\begin{equation}
\gamma_\alpha\gamma^{\mu_1\cdots\mu_n}\gamma^\alpha=(-1)^n(d-2n)\gamma^{\mu_1\cdots\mu_n}
\end{equation}
together with cyclicity. When this trace is non-zero, it equals $\pm i^{(n-1)/2}\epsilon_{\mu_1\cdots\mu_n}\tr[1]$. The sign is arbitrary, since $\pm\gamma^\mu$ both satisfy the Clifford algebra. This sign can be understood as a choice of orientation; it does not affect the even part of the algebra, which is why both choices lead to the same representation of $Spin(d)$.

\paragraph{Fractional $\boldsymbol d$.} As explained in subsection~\ref{sec:kinematics}, fractional $d$ does not literally mean that the rank of $O(d)$ is fractional. Instead, we are instructed to look at the Deligne category $\widetilde{\operatorname{Rep}}(O(d))$. Unfortunately, the fermionic version $\widetilde{\operatorname{Rep}}(Pin(d))$ is not actually well-defined: pinor representations are unstable.

We of course do not want to admit defeat either, so let us push through anyway. Intuitively, fractional $d$ should look something like $\operatorname{Rep}(Pin(\infty))$. We will not try to make this precise, but we do ask: is there any sense in which the various formulas like~\eqref{eq:cliff_red} and~\eqref{eq:master_trace} have a direct limit? If not in the rigorous mathematical sense, at least in a more practical one?

In fact, some formulas do admit a meaningful stabilization. For example, the relation $\gamma^\mu\gamma^\nu\gamma_\mu=(2-d)\gamma^\nu$ is true for all $d\in\mathbb N$, and therefore it makes sense to claim that it is also true when $d\in\mathbb C$. In order to make this rigorous one would have to specify what type of object $\gamma^\mu$ is when $d\in\mathbb C$. We will not let these minutiae distract us too much; in practice, one can regard these matrices as elements of a formal algebra of symbols subject to the Clifford relation.

What about the identity $\tr[\gamma^\mu\gamma^\nu]=\delta^{\mu\nu}\tr[1]$? This equation is again true for all $d\in\mathbb N$, so it is tempting to once again declare that it is also true when $d\in\mathbb C$. This is not entirely immediate though:
\begin{itemize}
\item If we think of the $\gamma^\mu$'s as living in a certain algebra of symbols, then this trace identity formally follows from the Clifford relation, as long as ``$\tr$'' satisfies linearity $\tr[\lambda A+\mu B]=\lambda\tr[A]+\mu\tr[B]$ and cyclicity $\tr[AB]=\tr[BA]$. But this is not actually well-defined, since there is no sense in which such an algebra has a finite-dimensional representation, and therefore there is no reason a trace acting on it should exist.
\item Relatedly, the trace relation looks stable when written in terms of $\tr[1]$. But if we replace this factor by $2^{\lfloor d/2\rfloor}$, the identity no longer looks particularly nice, since this sequence does not have a  unique analytic continuation.
\end{itemize}

One can formally fix both problems at the same time by writing instead $\tr[\gamma^\mu\gamma^\nu]/\tr[1]=\delta^{\mu\nu}$. Now, the \emph{l.h.s.}~can be thought of as a ``regularized trace'', one that has a finite $d\to\infty$ limit, and the \emph{r.h.s.}~is a perfectly well-defined, stable tensor. In practice, this just means that the trace relations over the even part of the Clifford algebra are all correct at fractional $d$, but with the understanding that the factor $\tr[1]$ is to be left as is, instead of replacing it by $2^{\lfloor d/2\rfloor}$ or some interpolation thereof.

This last comment actually has an important corollary. When we work with Dirac fermions in a certain integer-dimensional QFT, we take $\psi$ to transform according to a pinor representation of the Lorentz group. Such a representation is always given by a certain number of copies of the irreducible, $2^{\lfloor d/2\rfloor}$-dimensional representation, and we define $N_f$ to be this number. In other words, ``number of fermions'' in a given QFT simply means how many copies of the smallest Dirac representation appear in our theory. In fractional $d$, there is no such ``smallest'' representation, and therefore no canonical definition of $N_f$. The object that plays the role of $N_f$ is $\tr[1]$, where the trace refers to a sum over both flavor and spin indices. If one wishes to specialize our general-$d$ formulas to a given $d\in\mathbb N$, the dictionary is
\begin{equation}\label{eq:gamma_d_dict}
\tr[1]=N_f\times 2^{\lfloor d/2\rfloor}
\end{equation}
If $d\not\in\mathbb N$, there is no canonical way to disentangle spin and flavor, and the only meaningful operation is tracing over both, at the same time. Therefore, in the same way $N_f$ normally labels families of fermionic QFTs, in fractional dimension it is $\tr[1]$ what labels such families.


The traces over the odd part of Clifford are again less well-behaved. Indeed, the identity $\tr[\gamma_{\mu_1\cdots\mu_n}]=\pm\delta_{n,d} i^{(n-1)/2}\epsilon_{\mu_1\cdots\mu_n}\tr[1]$ implies that these traces all vanish for sufficiently large $d$, and therefore their stable limit is that they vanish for all $d\in\mathbb C$. But of course setting them to zero does not lead to a good analytic continuation, since they miss the low-rank exceptions where these traces are occasionally non-zero.

For the time being, we will simply leave the odd traces unspecified. That is to say, in the same way the factor $\tr[1]$ is not meant to be continued to $d\in\mathbb C$, but just carried along, the odd traces $\tr[\gamma_{\mu_1\cdots\mu_n}]$ will also be left as is. As we shall see, one can get very far without ever trying to write down a specific formula for these traces.

The rules above are enough to have a working theory of fermions in fractional dimension. Our general philosophy is that, if one wants to be on the safe side when working with gamma matrices, the parameter $d$ is always an integer, there is no such a thing as $\gamma^\mu\in Pin(d)$ for $d\in\mathbb C$. Instead, one computes observables using formulas that are true for all $d\in\mathbb N$, and analytically continues the answer, not the intermediate steps. (It goes without saying, this last part is still very non-trivial, and not even necessarily well-defined; here we are just kicking the can down the road.)

Conversely, if one insists on trying to work with formal symbols defined for all $d\in\mathbb C$, one runs into many well-known paradoxes. Let us describe a few of them here for illustration purposes. In the discussion below we abandon our general philosophy that one should only talk about $\gamma^\mu$ for integer $d$, and assume that these exist at fractional $d$ as well.

If $\gamma^\mu$ is a valid object at fractional $d$, and ``$\tr$'' is a valid operation acting on it, then we must admit the existence of a symbol $\tr[\gamma^\mu]$. This is non-zero for $d=1$, and therefore we cannot just claim that $\tr[\gamma^\mu]=0$ for all $d\in\mathbb C$. For example, it might be the case that
\begin{equation}
\tr[\gamma^\mu]\tr[\gamma_\mu]\overset?=\frac{1}{\Gamma(d)\Gamma(2-d)}
\end{equation}
or some other analytic function of $d$ that equals $\delta_{1,d}$ when $d\in\mathbb N$.

One can in fact give a formal argument to determine $\tr[\gamma_\mu]$ for fractional $d$. First, note that this symbol satisfies the identity
\begin{equation}\label{eq:weird_identity}
\tr[\gamma_\rho]\delta_{\mu\nu}=\tr[\gamma_\mu]\delta_{\nu\rho}=\tr[\gamma_\nu]\delta_{\mu\rho}
\end{equation}
as can be checked by taking the trace of~\eqref{eq:cliff_reduce_3} and using cyclicity. By contracting $\mu$ and $\nu$ one finds $d\tr[\gamma_\rho]=\tr[\gamma_\rho]$ which implies that either $d=1$ or $\tr[\gamma_\rho]=0$. This is of course the correct answer for integer $d$, but also appears to follow from the Clifford relation for fractional $d$ as well. Therefore, if we take the point of view that gamma matrices for fractional $d$ are well-defined as formal symbols, and that their traces are meaningful as formal operations that satisfy linearity and cyclicity, then $\tr[\gamma_\rho]$ is identically zero at fractional $d$. In other words, the only possible value we can assign to this symbol that is compatible with the standard properties of ``$\tr$'' is $\tr[\gamma^\mu]\tr[\gamma_\mu]=\delta_{1,d}$, which is non-analytic in $d$.

Of course, another solution to $d\tr[\gamma_\rho]=\tr[\gamma_\rho]$ is $\tr[\gamma_\rho]=\infty$, which is perhaps the only logical trace for an object that is not trace class. This also nicely matches the intuition that fractional $d$ is formally equivalent to infinite $d$.

The identity~\eqref{eq:weird_identity}, while admittedly rather weird, seems to be required for consistency of momentum integrals. For example, we typically say that, under the integral sign, $k^\mu k^\mu\to\frac{1}{d}\delta^{\mu\nu}k^2$, since $\delta^{\mu\nu}$ is the only tensor that has the right index structure. But if we admit that $\tr[\gamma_\mu]$ is a valid tensor, one should aslo allow for $\tr[\gamma_\mu]\tr[\gamma_\nu]$, which technically is also a symmetric rank-2 tensor. The identity~\eqref{eq:weird_identity} implies that these are not independent structures: indeed, multiplying both sides by $\tr[\gamma_\nu]$, we find $\tr[\gamma_\mu]\tr[\gamma_\nu]=\delta_{\mu\nu}\tr[\gamma_\rho]\tr[\gamma^\rho]$. So at the very least, these formal manipulations do not interfere with the standard rules of analytic continuation of loop integrals.

While the argument above concerns the trace of a single gamma, the same pattern occurs for any odd number of gammas. For example, once again using Clifford plus cyclicity, one can check that
\begin{equation}\label{eq:tr_gamma3_0}
\begin{aligned}
&\delta_{\mu \rho}\tr[\gamma_{\nu \sigma \tau}]+\delta_{\nu \sigma }\tr[\gamma_{\mu \rho \tau}]+\delta_{\mu \tau}\tr[\gamma_{\nu \rho \sigma }]-\delta_{\nu \rho }\tr[\gamma_{\mu \sigma \tau}]-\delta_{\mu \sigma }\tr[\gamma_{\nu \rho \tau}]-\delta_{\nu \tau}\tr[\gamma_{\mu \rho \sigma }]=0
\end{aligned}
\end{equation}
Contracting $\mu$ and $\rho $ simplifies to $(d-3)\tr[\gamma_{\nu \sigma \tau}]=0$, which again implies that this trace vanishes unless $d=3$. The general statement is that the trace of $n$ gamma matrices, where $n$ is odd, is proportional to $\delta_{n,d}$, even if $d$ is fractional.

Moreover, if we insist that these traces make sense at $d\in\mathbb C$, there are some further questions one must reckon with. For example, is the operator $\tr[\gamma_\mu]\bar\psi\gamma^\mu\psi$ a Lorentz scalar? If the symbol $\tr[\gamma_\mu]$ defines a valid expression with a free Lorentz index, we should be able to contract it with other free Lorentz indices to construct scalars. More generally, one would guess that $\bar\psi\psi$ is infinitely degenerate with operators of the form
\begin{equation}\label{eq:extraneous}
\tr[\gamma_{\mu_1\cdots\mu_n}]\bar\psi\gamma^{\mu_1\cdots\mu_n}\psi,\qquad \text{$n$ odd}
\end{equation}
We will term these operators \emph{extraneous}. They are similar in spirit to the evanescent operators~\eqref{eq:evanes} we encountered in the previous subsection, and it is tempting to ascribe the various discontinuities in $\bar\psi\psi$ to their existence.

These extraneous operators appear in the OPE of the evanescent quartics. For example, consider the quadratic $\bar\psi\psi$, and compute its OPE with a generic quartic $\bar\psi\Gamma\psi\,\bar\psi\bar\Gamma\psi$ for a pair of arbitrary matrices $\Gamma,\bar\Gamma$. The double contractions give
\begin{equation}
\begin{aligned}
\bar\psi\psi(x)\,\bar\psi\Gamma\psi\bar\psi\bar\Gamma\psi(0)&\supset
\tr[S(x)^2\Gamma]\bar\psi\bar\Gamma\psi+\tr[S(x)^2\bar\Gamma] \bar\psi\Gamma\psi+\bar\psi\bar\Gamma S(x)^2\Gamma\psi-\bar\psi\Gamma S(x)^2\bar\Gamma\psi \\
&\propto \frac{1}{x^{2(d-1)}}\bigl(\tr[\Gamma]\bar\psi\bar\Gamma\psi+\tr[\bar\Gamma] \bar\psi\Gamma\psi+\bar\psi[\bar\Gamma,\Gamma]\psi\bigr)
\end{aligned}
\end{equation}
If we now take $\Gamma=\gamma_{\mu_1\cdots\mu_n}$ and $\bar\Gamma=\gamma^{\mu_1\cdots\mu_n}$, the OPE becomes
\begin{equation}
\bar\psi\psi(x)\,\mathscr O_n(0)\supset\frac{1}{x^{2(d-1)}}\tr[\gamma_{\mu_1\cdots\mu_n}]\bar\psi\gamma^{\mu_1\cdots\mu_n}\psi
\end{equation}
Therefore, if we admit the existence of the evanescent operators $\mathscr O_n$, it seems somewhat unavoidable to accept the existence of extraneous operators as well.

All of these paradoxes go away if we agree that the gamma matrices are only defined for integer $d$, and the analytic continuation is not defined by literally working in $Pin(d)$ for $d\in\mathbb C$, but rather we should study $Pin(d)$ for all $d\in\mathbb N$, and (roughly speaking) take the direct limit $Pin(\infty)$.

In this work we are only concerned with vector-like theories, and we don't have much to say about the chirality matrix $\gamma^\star$. This matrix is even more subtle than the regular gamma matrices. For example, while $\tr[\gamma_{\mu\nu\rho}]$ is hard to define for $d\in\mathbb C$, at least it has a $d\to\infty$ limit: this trace vanishes for all sufficiently large $d$. This limit misses the $d=3$ low-rank exception, but at least it exists. On the other hand, $\tr[\gamma^\star]$ oscillates between $0$ (for even $d$) and $\tr[1]$ (for odd $d$), and therefore it does not even have a limit.

It appears to us that, if an analytic continuation of $\gamma^\star$ is possible at all, it will hardly be unique. For example, if we define for integer $d$
\begin{equation}
\gamma^\star:=i^{d(d-1)/2}\gamma^1\cdots\gamma^d
\end{equation}
where the factor of $i$ is included to make $\gamma^\star$ hermitian, then it follows that
\begin{equation}
\begin{aligned}
(\gamma^\star)^2&=1\\
\gamma_\mu\gamma^\star\gamma^\mu&=(-1)^{d+1}d\gamma^\star\\
\end{aligned}
\end{equation}
etc. The factor $(-1)^{d+1}$ does not have a unique, canonical continuation to $d\in\mathbb C$. Assuming that these equations make sense in fractional $d$ for \emph{some} choice of analytic continuation, then upon taking the formal trace one finds $(-1)^{d+1}\tr[\gamma^\star]=\tr[\gamma^\star]$. Therefore, if our choice of analytic continuation for $(-1)^{d+1}$ is such that it differs from $1$ for $d\not\in\mathbb N$, then we must once again conclude that $\tr[\gamma^\star]$ vanishes when $d\not\in\mathbb N$, and this trace is again non-analytic in $d$. Of course, similar conclusions hold for traces containing $\gamma^\star$ and one or more insertions of $\gamma^\mu$'s.

One can use $\gamma^\star$ to construct further evanescent operators, such as $\bar\psi(1-(-1)^{(d^2-1)/8}\gamma^\star)\psi$. Perhaps one has to take these into account as well. We will not explore the matrix $\gamma^\star$ any further in this work (however, see appendix~\ref{app:chiral} for some final thoughts).

\section{A case study: QCD}\label{sec:results}

Here we will illustrate the general discussion from the previous section in one particularly interesting example: vector-like massless QCD. Needless to say, this theory is the prototypical example of a strongly-coupled QFT. In order to recover computational control we will be working in the large $N_f$ limit, in which the strongly-coupled, low-energy fixed point becomes tractable as a series expansion in $1/N_f$. We have no insights as to what a non-perturbative definition should look like.

We refer the reader to appendix~\ref{sec:background} for a review of this expansion, as well as a reminder of the known dynamics of this theory for $d=1,2,3,4$.

In order to simplify the notation, let us restrict ourselves to the abelian case for now; we will point out the modifications required for general gauge group below. The lagrangian of the theory in Euclidean signature is given by
\begin{equation}
L=\frac{1}{4e^2} f^{\mu \nu}f_{\mu \nu} - \bar{\psi} \gamma^\mu (\partial_\mu + i a_\mu) \psi+\text{gauge-fixing}
\end{equation}

As far as the large $N_f$ expansion is concerned, this Lagrangian describes a flow between the free CFT $e^2\to0$ and the interacting CFT $e^2\to\infty$. The engineering dimension of $e^2$ is $4-d$, and therefore the direction of the flow depends on whether $d$ is larger or smaller than $4$. The interesting case is of course $d\le4$, but in order to talk about analytic continuation in $d$ we must choose a target QFT for all $d\in\mathbb N$, and therefore we will also care about this theory for $d>4$. For simplicity we will use language adapted to $d<4$, namely we will call the $e^2\to0$ fixed point the ``high energy'' theory and the $e^2\to\infty$ fixed point the ``low energy'' one, even though these roles are exchanged for $d>4$. The low-energy theory is non-perturbative in $e^2$ but weakly-coupled in $1/N_f$.

The Feynman rules of this theory at the low-energy fixed point are
\begin{equation}
\begin{aligned}
\tikz[baseline=-3pt]{
\begin{scope}
\clip (0,-.2) rectangle ++(1.5,0.5);
\draw[fermion] (0,0)--(1.8,0);
\end{scope}
\node[scale=.8] at (.1,-.3) {\phantom{$\mu$}};
\node[scale=.8] at (1.4,-.3) {\phantom{$\nu$}};
}\ &=-i\frac{\slashed p}{p^2}\\[+1ex]
\tikz[baseline=-3pt]{
\begin{scope}
\clip (0,-.2) rectangle ++(1.5,0.5);
\draw[photon] (0,0)--(1.8,0);
\end{scope}

\node[scale=.8] at (.1,-.3) {$\mu$};
\node[scale=.8] at (1.4,-.3) {$\nu$};
}\ &=\frac{g}{p^{d-2}}(\delta^{\mu\nu}+\xi p^\mu p^\nu/p^2)
\end{aligned}
\end{equation}
together with the usual vertex $-i\gamma^\mu$. Here, $g$ is the effective coupling
\begin{equation}\label{eq:eff_coupling}
g:=\frac{(4\pi)^{d/2}}{\tr[1]}\frac{2\Gamma (d-2)}{\Gamma(2-d/2)\Gamma(d/2-1)^2}\frac{d-1}{d-2}
\end{equation}
where $\tr[1]=2^{\lfloor d/2\rfloor}N_f$. Note that, in position space, the photon propagator is
\begin{equation}\label{eq:pos_photon_xi}
\begin{aligned}
\int_k e^{ikx}&\frac{g}{k^{d-2}}(\delta_{\mu\nu}+\xi k_\mu k_\nu/k^2)\\
&=\frac{2}{(4\pi)^{d/2}\Gamma(d/2)}\frac{g}{x^2}((d-2+\xi)\delta_{\mu\nu}-2 \xi x_\mu x_\nu/x^2)
\end{aligned}
\end{equation}
which has the same $1/x^2$ scaling as its $d=4$ counterpart. This implies that $\Delta(f_{\mu\nu})=2$, which is below the unitarity bound $d-2$ for all $d>4$. The $d>4$ CFTs are all non-unitary.

We will compute the following observables to leading order in $g\sim 1/N_f$:
\begin{itemize}
\item The expectation value of $W$ for various closed contours, where $W$ denotes a Wilson line:
\begin{equation}
W(\gamma)=\exp\biggl( i\oint_\gamma a\biggr)
\end{equation}
\item The scaling dimension of $\psi$. By this we mean the exponent in the two-point function
\begin{equation}
\langle\bar\psi(x)W\psi(0)\rangle\propto x^{-2\Delta(\psi)}
\end{equation}
where $W$ is a Wilson line connecting $x$ to $0$. This $x$ dependence arises for a straight Wilson line, since we want the geometry to preserve scale invariance.
\item The scaling dimension of $\bar\psi\psi$, namely the exponent in
\begin{equation}
\langle\bar\psi\psi(x)\bar\psi\psi(0)\rangle\propto x^{-2\Delta(\bar\psi\psi)}
\end{equation}
\end{itemize}
We will then try to analytically continue these observables to $d\in\mathbb C$.

Another observable that would be particularly instructive to study in the future is the magnetic current $\star f$. As above, this satisfies $\Delta(\star f)=2$, which saturates the unitarity bound, in agreement with its being conserved. This operator is parity-odd, which means that it is an unstable representation of $O(d)$, i.e., its continuation to fractional $d$ is much more involved than $f$ itself. See for example~\cite{Chester:2015wao} for some very interesting results concerning magnetic operators in fractional dimension.

\subsection{Wilson lines}

In this section we sketch the computation of the expectation value of Wilson lines along various closed contours~\cite{POLYAKOV1980171}. This is textbook material by now so we will be superficial; the main point we want to illustrate here is that these expectation values admit a well-defined analytic continuation in $d$.

As mentioned above, we begin with the case of abelian gauge group, and generalize to arbitrary gauge group afterward.

The expectation value is given by the usual cluster expansion
\begin{equation}
\langle e^{i\oint a}\rangle=e^{-\frac12\oint\!\oint \langle aa\rangle+\frac{1}{4!}\oint\!\oint\!\oint\!\oint\langle aaaa\rangle_c+\cdots}
\end{equation}
To leading order in $1/N_f$ we simply keep the first term. So we need to line-integrate the propagator $\Delta_{\mu\nu}=\langle a_\mu a_\nu\rangle$ along the contour of interest:
\begin{equation}
\log\langle W(\gamma)\rangle=-\frac12\oint\!\!\!\oint_{\gamma\times\gamma} \mathrm dx^\mu\,\mathrm dy^\nu\Delta_{\mu\nu}(x-y)\\
\end{equation}

For the actual computations we use two different regulators: analytic regularization $\frac{1}{x^2}\to \frac{\mu^2}{(\mu^2x^2)^\alpha}$ (with $\mu$ some arbitrary scale) and point-splitting regularization, where we deform one integration contour via $\gamma\to\gamma+\epsilon$. We can gain different insights from each of these two options.

In this discussion we will write the photon propagator as $\Delta_{\mu\nu}=\frac{\delta_{\mu\nu}}{x^2}$, which accounts for multiple situations:
\begin{itemize}
\item If we multiply by $\frac{e^2}{4\pi^2}$ we obtain free Maxwell in $4d$. This will be a useful cross-check for our computations.
\item If we multiply by $\frac{4g}{(4\pi)^{d/2}\Gamma(d/2-1)}$, we obtain the low-energy fixed point of QED$_{d}$ to leading order in $1/N_f$.
\item If we replace $\frac{1}{x^2}\to \frac{\mu^2}{(\mu^2x^2)^\alpha}$ and take $\alpha\to d/2-1$ instead of $\alpha\to1$, we obtain the high-energy fixed point of QED$_{d}$ instead of the low-energy one. The associated short-distance coupling is $e^2=\mu^{4-d}\frac{4\pi^{d/2}}{\Gamma(d/2-1)}$.

\end{itemize}

Let us first explain why we are allowed to use $\Delta_{\mu\nu}\propto\frac{\delta_{\mu\nu}}{x^2}$, without the $\xi$-dependent part, instead of the more general $\Delta_{\mu\nu}\propto \frac{1}{x^2}((d-2+\xi)\delta_{\mu\nu}-2 \xi x_\mu x_\nu/x^2)$ (cf.~\eqref{eq:pos_photon_xi}). A formal proof of gauge invariance is as follows. To leading order in $1/N_f$, we need to integrate the propagator along a (flat) closed 2-manifold of the form $\gamma\times\gamma$, where $\gamma$ is the support of $W$. Then, under the integral sign, we can replace $x_\mu x_\nu\to \frac12 \delta_{\mu\nu} x^2$, which means that $(d-2+\xi)\delta_{\mu\nu}-2 \xi x_\mu x_\nu/x^2\to (d-2)\delta_{\mu\nu}$, and the answer is $\xi$-independent. A perhaps better argument is to note that, in momentum space
\begin{equation}\label{eq:wilson_k_space}
\oint_\gamma\mathrm dx^\mu\oint_\gamma\mathrm dy^\nu\Delta_{\mu\nu}(x-y)=\int_k\Delta_{\mu\nu}(k)V^\mu(\gamma,k)V^\nu(\gamma,-k)
\end{equation}
where the eikonal factor $V^\mu$ is 
\begin{equation}
V^\mu(\gamma,k):=\int_\gamma\mathrm dx^\mu\ e^{-ikx}
\end{equation}
Then, the gauge part of~\eqref{eq:wilson_k_space} is proportional to $k_\mu V^\mu=\int_\gamma \mathrm d(e^{-ikx})$, which vanishes for a closed loop. Note that either of these arguments goes through even in the presence of a regulator, such as replacing $\frac{1}{x^2}\to \frac{\mu^2}{(\mu^2x^2)^\alpha}$ or shifting one of the integration contours via $\gamma\to\gamma+\epsilon$. So this justifies the claim that we are allowed to simply use $\Delta_{\mu\nu}=\frac{\delta_{\mu\nu}}{x^2}$.

We begin with a circle geometry, $x=(R\cos\theta,R\sin\theta)$. Then, $\mathrm dx^\mu\mathrm dy_\mu=R^2\cos(\theta-\theta')\mathrm d\theta\mathrm d\theta'$ and so, in analytic regularization,
\begin{equation}\label{eq:W_alpha}
\begin{aligned}
\log\langle W(\bigcirc)\rangle&=-\frac12\frac{\mu^2R^2}{(2\mu^2R^2)^\alpha}\int_0^{2\pi}\mathrm d\theta\int_0^{2\pi}\mathrm d\theta' \frac{\cos(\theta-\theta')}{(1-\cos(\theta-\theta'))^\alpha}\\
&=-\frac12\frac{\mu^2R^2}{(2\mu^2R^2)^\alpha}2\pi\biggl(-2^{\alpha +1}\pi\frac{\Gamma (1-2 \alpha )}{\Gamma (2-\alpha ) \Gamma (-\alpha )}\biggr)
\end{aligned}
\end{equation}
Taking $\alpha\to1$ we find $\log\langle W\rangle=\pi^2$, while $\alpha\to d/2-1$ leads to $\log\langle W\rangle=(\mu  R)^{4-d}\frac{2\pi^2 \Gamma (3-d) }{\Gamma(1-d/2) \Gamma(3-d/2)}$. This implies, for our large $N_f$ QED$_d$ problem, that\footnote{The UV limit in $d=3$ is special because $R^{4-d}$ becomes $R$, which is linear and hence a counterterm. In this case one has to expand around $\alpha=d/2-1$ and keep the subleading term as well. The result is the usual $R\log(\mu R)$ behavior.}
\begin{equation}\label{eq:W_circ_QED}
\log\langle W(\bigcirc)\rangle=\begin{cases}
\displaystyle\frac{4\pi^2g}{(4\pi)^{d/2}\Gamma(d/2-1)}& R\gg e^{2/(d-4)}\\[+2ex]
\displaystyle e^2R^{4-d}\frac{\pi^{2-d/2} \Gamma (3-d) \Gamma (d/2-1)}{2 \Gamma(1-d/2) \Gamma(3-d/2)} & R\ll e^{2/(d-4)}
\end{cases}
\end{equation}
For $d<4$, which is the regime where the theory is UV free and has an interesting IR fixed point, the log of the expectation value vanishes at short distances, implying that the UV defect entropy is zero. On the other hand, the long-distance answer is positive. This is not incompatible with the $g$-theorem~\cite{Cuomo:2021rkm}, since here we are dealing with a bulk RG flow, not a defect one. 

As a check of the result above, we note that the $\alpha\to1$ limit of~\eqref{eq:W_alpha}, when multiplied by $\frac{e^2}{4\pi^2}$, yields the well-known free Maxwell answer $\langle W(\bigcirc)\rangle=\exp(e^2/4)$.

Furthermore, if we take $d=1$ in (the first line of) the general expression~\eqref{eq:W_circ_QED}, we find $\log\langle W\rangle=-2/N_f$, while if we take $d=2$ we find $\log\langle W\rangle=0$. These agree with the explicit computation performed directly in $d=1,2$, see appendix~\ref{sec:background} for the details (specifically,~\eqref{eq:W_qed_1d} and~\eqref{eq:W_qed_2d}).

We can reuse the results above to easily get the answer for general gauge group. Say we take the line to be in a representation $\rho$ of $G$; then, to tree level, the expectation value of a Wilson loop is simply $\tr_\rho[1]=\dim(\rho)$. To next order we replace $\Delta_{\mu\nu}\to\Delta_{\mu\nu}\delta^{ab}$, where $a,b$ are adjoint indices, and we also take a trace over $\rho$. In other words, we replace
\begin{equation}
\begin{aligned}
\langle aa\rangle&\to\tr_\rho[\langle aa\rangle]\\
&=\tr_\rho [t^a t^b]\delta_{ab}\langle aa\rangle_\text{QED}\\
&=T(\rho)\dim(G)\langle aa\rangle_\text{QED}
\end{aligned}
\end{equation}
Therefore, the expectation value for non-abelian $G$ reads
\begin{equation}
\log(\langle W_\rho(\bigcirc)\rangle/\dim(\rho))=\log\langle W(\bigcirc)\rangle_\text{QED}\times \frac{T(\rho)\dim(G)}{\dim(\rho)T(R)}+\mathcal O(N_f^{-2})
\end{equation}
where $T(R)$ arises because the effective coupling is $g_\text{QCD}=g_\text{QED}\times\frac{1}{T(R)}$ (cf. the discussion above~\eqref{eq:g_QCD}).

Again, if we take $d=1$ we recover the answer as derived independently in QCD$_1$, cf.~\eqref{eq:W_1d_QCD}. (If we take $d=2$ we also recover~\eqref{eq:W_2d_QCD}, although in a boring way: both answers vanish to order $1/N_f$). In any case, the main conclusion is that~\eqref{eq:W_circ_QED} admits a meaningful continuation to $d\in\mathbb C$.

For future reference let us sketch the same result but in point-splitting regularization, where we add a small shift to one of the integration contours. The easiest way to do this is to push the second integration off the plane, so that $x=(R\cos\theta,R\sin\theta,0)$ and $y=(R\cos\theta',R\sin\theta',\epsilon)$. This yields
\begin{equation}
-\frac12\int_0^{2\pi}\!\mathrm d\theta\int_0^{2\pi}\!\mathrm d\theta'\,\frac{R^2 \cos (\theta -\theta')}{2 R^2-2 R^2 \cos (\theta -\theta')+\epsilon ^2}
\end{equation}
which evaluates to $-\frac{\pi ^2 R}{|\epsilon|}+\pi ^2+\mathcal O(\epsilon)$. The divergence is just $-\frac{\pi}{|\epsilon|}$ times the perimeter $2\pi R$, so it can be canceled by a local counterterm along the line. The finite part agrees with the analytic computation from before.

A mild complaint for this regulator is that it requires a third direction, so it doesn't work if $d<3$. It is in fact possible to point-split while keeping the two contours on the same plane; the only way to do this, without having the contours intersect at all, is to shrink one of them by a small amount. In other words, we take $x=(R\cos\theta,R\sin\theta)$ and $y=((R+\epsilon)\cos\theta',(R+\epsilon)\sin\theta')$. Then,
\begin{equation}
\begin{aligned}
\log\langle W(\bigcirc)\rangle&=-\frac12\int\!\!\!\!\int_0^{2\pi}\frac{R (R+\epsilon) \cos (\theta -\theta')\,\mathrm d\theta\,\mathrm d\theta'}{(R \sin\theta-(R+\epsilon) \sin\theta')^2+(R \cos\theta-(R+\epsilon) \cos\theta')^2}\\
&=-\frac122\pi\times\begin{cases}
\frac{2 \pi  R^2}{2R\epsilon+\epsilon^2} & \epsilon>0\\
-\frac{2 \pi  (R+\epsilon )^2}{2R\epsilon+\epsilon^2} & \epsilon<0
\end{cases}\\
&=-\frac{\pi^2 R}{|\epsilon|}+\begin{cases}
\frac12\pi^2&\epsilon>0\\
\frac32\pi^2&\epsilon<0
\end{cases}+\mathcal O(\epsilon)
\end{aligned}
\end{equation}
The singularity is $-\frac{\pi^2 R}{|\epsilon|}$ as before, but the finite part appears ambiguous: it depends on whether we shrink the second contour or we expand it. This of course makes no sense, the finite part should be scheme-independent. The resolution is that a local counterterm shifts the expectation value by something proportional to the length of the line, but in this case we have two lengths, $2\pi R$ and $2\pi(R+\epsilon)$.  So we split the difference and write the result as $-\frac{\pi}{2|\epsilon|}\times2\pi(R+\epsilon/2)+\pi^2+\mathcal O(\epsilon)$, where we once again read the finite, scheme-independent part as $\pi^2$. (Equivalently, we can also take the symmetric limit $\lim_{\epsilon\to0}=\frac12(\lim_{\epsilon\to0^+}+\lim_{\epsilon\to0^-})$, for the same answer).

As a final comment, we remark that we found the scheme-independent part of $\langle W\rangle$ for a circular geometry to be independent of the radius $R$. This is of course as expected, since a circle has a single scale, and a conformal line has no scales of its own, and therefore there is no dimensionless ratio the expectation value could depend on. Below we will consider a rectangular geometry which has two scales, and we will find that $\langle W\rangle$ is a non-trivial function of their ratio. Another simple geometry with two scales is the ellipse, say with axes $(a,b)$, for which we find $\log\langle W\rangle=-\frac{2 \pi a}{|\epsilon|}E(1-b^2/a^2)+\frac12\pi ^2 (b/a+a/b)+\mathcal O(\epsilon)$, where $E$ is the complete elliptic integral of the second kind. The singularity is once again $-\frac{\pi}{2|\epsilon|}$ times the length of the line, and the finite part $\pi^2(a/b+b/a)$ is a non-trivial function of the dimensionless ratio $a/b$. For $a=b$ we recover the circle answer $\pi^2$.

We now move on to the rectangular geometry that measures the potential energy between two probe charges:
\begin{equation}
\tikz[baseline=1.5cm]{
\draw[thick] (0,0) -- (0,3) -- (2,3) -- (2,0) -- cycle;
\begin{scope}
\clip (-.2,0) rectangle ++(.4,3);
\draw[fermion] (0,0)--(0,3.3);
\end{scope}
\begin{scope}
\clip (0,3-.2) rectangle ++(2,0.5);
\draw[fermion] (0,3)--(2.3,3);
\end{scope}
\begin{scope}
\clip (2-.2,0) rectangle ++(.4,3);
\draw[fermion] (2,3)--(2,-.3);
\end{scope}
\begin{scope}
\clip (0,-.2) rectangle ++(2,0.5);
\draw[fermion] (2,0)--(-.3,0);
\end{scope}
\node[scale=.8,anchor=north east] at (0,.2) {$(0,0)$};
\node[scale=.8,anchor=north west] at (2,.2) {$(0,R)$};
\node[scale=.8,anchor=south east] at (0,2.8) {$(T,0)$};
\node[scale=.8,anchor=south west] at (2,2.8) {$(T,R)$};
}
\end{equation}

We use both regularizations. We begin with $y$ being the straight line between $(0,0)$ and $(T,0)$, while $x$ is the whole rectangle; we will then add the other pieces of the $y$ path. We parametrize $y(\tau')=(\tau' (T+2\epsilon)-\epsilon,-\epsilon)$, with $\tau'\in[0,1]$, as well as
\begin{equation}
x(\tau)=\begin{cases}
(4\tau T,0) & 0<\tau<1/4\\
(T,(4\tau-1)R) & 1/4<\tau<1/2\\
((-4\tau+3)T,R) & 1/2<\tau<3/4\\
(0,(-4\tau+4)R) & 3/4<\tau<1
\end{cases}
\end{equation}
Then, $\mathrm dx^\mu\mathrm dy_\mu=4T(T+2\epsilon)(\Theta(0<\tau<1/4)-\Theta(1/2<\tau<3/4))\mathrm d\tau\mathrm d\tau'$ and so this segment contributes
\begin{equation}
\begin{aligned}
-2T^2&\int_0^{1/4}\mathrm d\tau\int_0^1\mathrm d\tau'
\frac{\mu^2}{(\mu^2T^2(4\tau-\tau')^2)^\alpha}\\
&+2T^2\int_{1/2}^{3/4}\mathrm d\tau\int_0^1\mathrm d\tau'
\frac{\mu^2}{(\mu^2(T^2(-4\tau-\tau'+3)^2+R^2)^\alpha}\\
&=\mu^{2 (1-\alpha )}\biggl(T^2 R^{-2 \alpha } \, _2F_1(\tfrac{1}{2},\alpha ;\tfrac{3}{2};-\tfrac{T^2}{R^2})\\
&+\frac{1}{2 (\alpha -1)}\bigl[-R^{2 (1-\alpha )}+(R^2+T^2)^{1-\alpha }+\tfrac{1}{1-2 \alpha }T^{2(1- \alpha)}\bigr]\biggr)\\
&=-\frac{1}{2 (\alpha -1)}+1+\frac{T }{R}\arctan\frac{T}{R}+\frac{1}{2} \log\frac{\mu ^2 R^2 T^2}{R^2+T^2}+\mathcal O(\alpha-1)
\end{aligned}
\end{equation}
in analytic regularization, and
\begin{equation}
\begin{aligned}
-2T(T+&2\epsilon)\int_0^{1/4}\mathrm d\tau\int_0^1\mathrm d\tau'\frac{1}{(4 \tau  T-\tau' (T+2 \epsilon )+\epsilon )^2+\epsilon ^2}\\
&+\int_{1/2}^{3/4}\mathrm d\tau\int_0^1\mathrm d\tau'\frac{1}{(R+\epsilon )^2+(T (4 \tau +\tau'-3)+2 \tau' \epsilon -\epsilon )^2}\\
&=-\frac{\pi  T}{2 | \epsilon | }+\frac{\pi  T}{4 R}+1+\frac{1}{2} \log\frac{T^2R^2}{2(T^2+R^2)\epsilon^2}\\
&+\frac{T}{2R}(\arctan(T/R)-\arctan(R/T))+\begin{cases}
-\pi/4 & \epsilon>0\\
+3\pi/4 & \epsilon<0
\end{cases}
\end{aligned}
\end{equation}
in point-splitting regularization.

Adding the other three segments amounts adding the same expression with $T\leftrightarrow R$, and multiplying by $2$, so all in all
\begin{equation}\label{eq:W_square}
\log\langle W(\square)\rangle=2\frac{T}{R}\arctan\frac{T}{R}+\log\frac{\mu ^2 R^2 T^2}{R^2+T^2}+(R\leftrightarrow T)\\
\end{equation}
for the analytic regularization approach, after absorbing the $\frac{1}{2(\alpha-1)}$ pole into a corner counterterm. The point-splitting result, after removing the $-\frac{\pi  (T+R)}{| \epsilon | }$ pole into a perimeter counterterm, is in fact identical to this, thanks to standard $\arctan$ identities.

So both approaches yield the same answer, but they both introduce a dependence on the arbitrary scale $\mu$. This dependence is somewhat artificial in that the rescaling $\mu\to\lambda\mu$ can be removed by a corner counterterm $\log(\lambda)$, so the result doesn't depend on $\mu$ in a measurable way. But at the same time it is not possible to remove this dependence, which means that the configuration breaks the $1d$ scale invariance of the line (the more precise statement is that this is a trace anomaly, see e.g.~\cite{Cuomo:2024psk}). This is in fact the cusp anomalous dimension of the line, as we elaborate on below.

Note that in analytic regularization, the circle and rectangle both had no need for a perimeter counterterm. On the other hand, in point-splitting, the circle had a divergence $-\frac{\pi^2 R}{|\epsilon|}$, while the rectangle had a divergence $-\frac{\pi  (T+R)}{| \epsilon | }$; these are both equal to $-\frac{\pi}{2|\epsilon|}\times\text{length}$, so the same counterterm fixes both. This is of course as expected: for a fixed choice of scheme, we should find the same counterterm for any geometry, since counterterms are local and thus do not care about the shape of the line. In other words, if two geometries have different linear divergences, it would not be possible to fix both at the same time, and the observable would be inconsistent. It is reassuring that the circle and rectangle lead to the same counterterm, for both choices of regularization.

Note also that if we take the $T\gg R$ limit of the above, we find $V(R)=-\pi/R+\mathcal O(\log T)$ which, after multiplying by $\frac{e^2}{4\pi^2}$, becomes $V(R)=-\frac{e^2}{4\pi R}$, which is the usual Coulomb potential of free $4d$ Maxwell. On the other hand, for our large $N_f$ QED$_d$ problem,
\begin{equation}
V(R)=-\frac{4\pi g}{(4\pi)^{d/2}\Gamma(d/2-1)}\frac{1}{R}\qquad R\gg e^{2/(d-4)}
\end{equation}
and therefore we learn that these theories are all Coulomb-like. This is of course the only $R$ dependence compatible with scale invariance, so this result is simply telling us that the Wilson line is conformal.

We end this discussion with a quick analysis of the cusp, which we parametrize by two segments $x=(s,0)$ and $x=(s\cos\phi,s\sin\phi)$, where $s\in(0,\infty)$:
\begin{equation}
\tikz[baseline=.5cm]{
\begin{scope}
\clip (-5:3cm) arc(-5:30:3cm) -- (-.3,-.05) -- cycle;
\draw[fermion] (3,0) -- (0,0);
\draw[fermion] (0,0) -- (2.99082, 1.39464);
\end{scope}
\draw  (1,0) arc (0:25:1);
\node[scale=.8] at (1.15,.25) {$\phi$};
}
\end{equation}
We are more cavalier than before, and simply cutoff the $s$ integral at short and long distances; we find the same divergences as before, plus a new type, namely the cusp. This comes from the crossed terms where $x$ is horizontal and $y$ is oblique:
\begin{equation}
-\frac12\times 2\times\int_\infty^0\mathrm ds\int_0^\infty\mathrm ds'\frac{\cos\phi}{(s - s'\cos\phi)^2 + (s' \sin\phi)^2}-(\text{same, with $\phi=\pi$})
\end{equation}
where we subtract the $\phi=\pi$ term so that the cusp dimension of a straight line vanishes.

We interpret $s$ and $s'$ as the proper times of the probe charges, and -- as one often does when using Schwinger's parametrization -- change variables into the ``total proper time'' $\tau=s+s'\in(0,\infty)$, and the ``relative proper time'' $q=\frac{s}{s+s'}\in(0,1)$. The Jacobian is simply $\tau$, so this integral simplifies to
\begin{equation}
\begin{aligned}
&\int_\epsilon^L\frac{\mathrm d\tau}{\tau}\int_0^1\mathrm dq\,\frac{\cos \phi}{(q-(1-q) \cos \phi )^2+(1-q)^2 \sin^2\phi}-(\cdots)\\
&=\bigl(1+2 \cot (\phi ) \arctan(\cot (\phi/2))\bigr)\log\frac{L}{\epsilon}
\end{aligned}
\end{equation}
whence $\Gamma_\text{cusp}(\phi)=-(1+2 \cot (\phi ) \arctan(\cot (\phi/2)))\equiv ([\phi]-\pi)\cot(\phi-\pi)-1$, where $[\phi]=\phi\mod 2\pi$. 

The rectangle has four cusps, each contributing a factor of
\begin{equation}
\bigl(1+2 \cot (\phi ) \arctan(\cot (\phi/2))\bigr)\bigg|_{\phi=\pi/2}\times\log L\equiv +\log(L)
\end{equation}
This is in agreement with our earlier computation~\eqref{eq:W_square}, where we found $\cdots+2\log(R^2)$.

We note that the cusp dimension above satisfies $\Gamma_\text{cusp}(\phi)<0,\Gamma_\text{cusp}'(\phi)>0,\Gamma_\text{cusp}''(\phi)<0$, in agreement with the general analysis of~\cite{Cuomo:2024psk}. For our QED$_d$ problem, we are to multiply this function by $\frac{4g}{(4\pi)^{d/2}\Gamma(d/2-1)}$, which is positive for all $2<d<4$, so these inequalities are still satisfied. This is not entirely obvious, since~\cite{Cuomo:2024psk} explicitly assumes unitarity, which is not present in fractional $d$. Of course, the loss of unitarity does not show up in any of the simple observables we look at in this paper, so it is not entirely surprising that our $\Gamma_\text{cusp}(\phi)$ behaves as if the theory was unitary. Indeed, as pointed out at the beginning of this subsection, to leading order in $1/N_f$ the photon propagator in QED$_d$ has the same $x$ dependence as the propagator of $4d$ free Maxwell, which is of course unitary. So the only non-trivial statement is that the coefficient that connects these two propagators is positive for all $2<d<4$.

In any case, we note that $\frac{4g}{(4\pi)^{d/2}\Gamma(d/2-1)}$ is positive for $2<d<4$, but actually vanishes in $d=2,4$. For $d=4$ this simply reflects the fact that the ``infrared fixed point'' of QED$_4$ is essentially free. For $d=2$, it is due to the fact that, while the Wilson line defines a conformal defect for generic $d$, it becomes topological in $d=2$: it becomes a Verlinde line of $SO(\dim R)_1/G_{T(R)}$, as reviewed in appendix~\ref{sec:2dQCD}.

The factor $\frac{4g}{(4\pi)^{d/2}\Gamma(d/2-1)}$ becomes negative for $d<2$, and the potential $V(R)$ and the cusp $\Gamma_\text{cusp}(\phi)$ become positive. While our general analysis should perhaps not be trusted all that much below $d<2$ (for example, does it even make sense to talk about the cusp angle $\phi$ if we have fewer than two directions?), above we found that the $d\to1$ limit of the general $d$ formulas does give the same answer as  the honest $d=1$ computation (cf.~\ref{sec:1d}), so let us indulge. One possible interpretation for the change of sign in $\frac{4g}{(4\pi)^{d/2}\Gamma(d/2-1)}$ is that this factor is actually the norm of the field strength $f_{\mu\nu}$, namely
\begin{equation}
\begin{aligned}
\langle \partial_{[\rho} a_{\mu]}(x)\partial_{[\sigma} a_{\nu]}(0)\rangle&=-g\int_p  e^{ipx}\frac{p_{[\sigma}\delta_{\nu][\rho}p_{\mu]}}{p^{d-2}}\\
&=\frac{16g}{(4\pi)^{d/2}\Gamma(d/2-1)}\frac{x_{[\sigma}\delta_{\nu][\rho}x_{\mu]}}{x^6}
\end{aligned}
\end{equation}
The fact that $f_{\mu\nu}$ has negative norm below $d=2$ can be understood as the statement that this field carries $d-2$ degrees of freedom, this being the dimension of the vector representation of $SO(d-2)$ (which is itself the little group of massless, helicity 1 particles in $d$ dimensions).

\subsection{Quarks}

In this subsection we sketch the computation of $\langle\bar\psi(x) W\psi(0)\rangle$, with $W$ is a straight Wilson line connecting $x$ to $0$. We again are rather brief, as we shall encounter no surprises.

As reviewed in appendix~\ref{sec:cpt}, our goal is to isolate the log divergence in this correlation function, from whose coefficient we can read off $\Delta^{(1)}$. We collect some useful integration formulas in appendix~\ref{ap:integrals}. In particular, the log divergence is always associated to scale-less integrals, and we introduce the notation
\begin{equation}
\mathscr L:=\int_k\frac{1}{k^d}\equiv \frac{1}{(4\pi)^{d/2}\Gamma(d/2)}\log(\Lambda^2)
\end{equation}
for those.

For this geometry, the eikonal factor (cf.~\eqref{eq:wilson_k_space}) reads
\begin{equation}
V^\mu(x,k)=\int_x^0\mathrm dz^\mu e^{-ikz}=ix^\mu\frac{1-e^{-ikx}}{kx}
\end{equation}
which, diagrammatically, we will denote as the insertion of an external source coupled to the photon: $\tikz[baseline=-2pt]{
\draw[photon] (0,0) -- (.8,0);
\node[scale=.5] at (-.08,0) {$\boldsymbol\otimes$};
}$. Note that, unlike before, here $k_\mu V^\mu$ does not vanish, since the contour is not closed; instead, $k_\mu V^\mu=i(1-e^{-ikx})$. This will cancel the gauge dependent part of the undress correlator $\langle \bar\psi(x)\psi(0)\rangle$.

There are three diagrams that contribute to this two-point function to leading order; we evaluate them in turn:
\begin{equation}\label{eq:psi_self_energy}
\begin{aligned}
\tikz[baseline=-2pt]{
\draw (-2,0) -- (2,0);
\begin{scope}
\clip(-1.2,0) rectangle ++(2.3,1.2);
\draw[photon] (-15:1cm) arc(-15:195:1cm);
\end{scope}
\filldraw (-1.04,0) circle (1pt);
\filldraw (1.04,0) circle (1pt);
\draw[fermion] (2,0) -- (.7,0);
\draw[fermion] (1,0) -- (-1.3,0);
\begin{scope}
\clip (-2,-.2) rectangle ++(1,.4);
\draw[fermion] (-1,0) -- (-2.3,0);
\end{scope}
\node[scale=.8] at (0,1.4) {$k$};
\node[scale=.8] at (-1,-.3) {$\mu$};
\node[scale=.8] at (1,-.3) {$\nu$};
}\ &=-\int_{pk}e^{ipx}S(p)\gamma^\mu S(p+k)\gamma^\nu S(p)\Delta_{\mu\nu}(k)\\
&=ig\int_pe^{ipx}S(p)\gamma^\mu\gamma^\alpha\biggl[\int_k \frac{p_\alpha+k_\alpha}{k^{d-2}(p+k)^2}(\delta_{\mu\nu}+\xi\frac{k_\mu k_\nu}{k^2})\biggr]\gamma^\nu S(p)\\
&=-g\biggl[\frac{(d-2)^2}{d}+\xi\biggr]S(x)\mathscr L
\end{aligned}
\end{equation}

Next,
\begin{equation}
\begin{aligned}
\tikz[baseline=-3pt]{\draw (0,0) -- (2,0);
\draw[photon] (1,0) -- (1,.8);
\node[scale=.5] at (1,.87) {$\boldsymbol\otimes$};
\node[scale=.8] at (1.25,.87) {$\mu$};
\node[scale=.8] at (1,-.25) {$\nu$};
\filldraw (1,0) circle (1pt);
\draw[fermion] (2,0) -- (.7,0);
\begin{scope}
\clip (0,-.2) rectangle ++(1,.4);
\draw[fermion] (1,0) -- (-.3,0);
\end{scope}
}\ &=-\int_{pk}e^{ipx}V^\mu(x,-k)S(p)\gamma^\nu S(p+k)\Delta_{\mu\nu}(k)\\
&=g\int_pe^{ipx}S(p)\gamma^\nu\biggl[\int_kV^\mu(x,-k) S(p+k)\frac{1}{k^{d-2}}(\delta_{\mu\nu}+\xi\frac{k_\mu k_\nu}{k^2})\biggr]
\end{aligned}
\end{equation}
We isolate the log divergence by focusing on the various degeneration limits. For this integral there are two: $p$ large with $p+k$ fixed, and the other way around. These are equal to each other, so we consider one and multiply the answer by  $2$. For the remaining integral, we note that, by Lorentz covariance, we can write
\begin{equation}
\int_kV^\mu(x,-k) S(k)\frac{1}{k^{d-2}}(\delta_{\mu\nu}+\xi\frac{k_\mu k_\nu}{k^2})=-i(\gamma_\nu f(x^2)+x_\nu \slashed x g(x^2))
\end{equation}
for some scalar functions $f,g$ defined by
\begin{equation}
\delta_{\alpha\nu}f(x^2)+x_\alpha x_\nu g(x^2)=\int_kV^\mu(x,-k) \frac{k_\alpha}{k^d}(\delta_{\mu\nu}+\xi\frac{k_\mu k_\nu}{k^2})
\end{equation}
Therefore, the diagram can be simplified into
\begin{equation}
\begin{aligned}
&=2ig\int_pe^{ipx}S(p)\int_kV^\mu(x,-k) \frac{k_\mu}{k^d}(1+\xi)\\
&=2g\int_pe^{ipx}S(p)\int_k (1-e^{ikx})\frac{1}{k^d}(1+\xi)\\
&=2g(1+\xi)S(x)\mathscr L
\end{aligned}
\end{equation}

Finally, the last diagram contributing to this two-point function is
\begin{equation}
\begin{aligned}
\tikz[baseline=5pt]{
\begin{scope}
\clip (-.1,-.2) rectangle ++(2.2,.4);
\draw[fermion] (2.1,0) -- (-.5,0);
\end{scope}
\node[scale=.8] at (0,1) {$\mu$};
\node[scale=.8] at (2,1) {$\nu$};
\draw[photon] (0,.7) -- (2,.7);
\node[scale=.5] at (-.08,.7) {$\boldsymbol\otimes$};
\node[scale=.5] at (2.08,.7) {$\boldsymbol\otimes$};
}\ & =-\frac12S(x)\int_k V^\mu(x,-k)V^\nu(x,k)\Delta_{\mu\nu}(k)\\
&=-\frac12gS(x)\int_k V^\mu(x,-k)V^\nu(x,k) \frac{1}{k^{d-2}}(\delta_{\mu\nu}+\xi\frac{k_\mu k_\nu}{k^2})
\end{aligned}
\end{equation}
where $1/2$ is a symmetry factor. The gauge part is straightforward, since $k_\mu V^\mu\propto 1-e^{ikx}$, which is easily integrated:
\begin{equation}
\begin{aligned}
\int_k V^\mu(x,-k)V^\nu(x,k) \frac{1}{k^{d-2}}\xi\frac{ k_\mu k_\nu}{k^2}&=\xi\int_k (1-e^{ikx})(1-e^{-ikx}) \frac{1}{k^d}\\
&=2\xi\mathscr L
\end{aligned}
\end{equation}

On the other hand, the physical part equals
\begin{equation}
-\frac12gS(x)\int_k \frac{x^2}{(kx)^2}(1-e^{ikx})(1-e^{-ikx}) \frac{1}{k^{d-2}}
\end{equation}
In order to evaluate this integral, we define
\begin{equation}
h(s):=\int_k \frac{1}{(kx)^2}(1-e^{ikxs})(1-e^{-ikxs}) \frac{1}{k^{d-2}}
\end{equation}
and note that the integral we are after is $h(1)$. On the other hand, if we differentiate this twice, we find
\begin{equation}
h''(s)=\int_k (e^{ikxs}+e^{-ikxs}) \frac{1}{k^{d-2}}
\end{equation}
This is just (twice) the position-space scalar propagator, namely $h''(s)=\frac{8}{(4 \pi )^{d/2}\Gamma(d/2-1)}\frac{1}{(sx)^2}$. Therefore, integrating twice, we derive
\begin{equation}
h(s)=-\frac{8\log (s)}{x^2(4\pi)^{d/2} \Gamma(d/2-1)}+c_1(x)+c_2(x)s
\end{equation}
where $c_1,c_2$ are constants of integration. Furthermore, from the definition of $h(s)$, it is clear that a rescaling of $s$ can be compensated by a rescaling of $x$, which means that $\log(s)$ must be accompanied by $\frac12\log(x^2)$, with the same coefficient. Therefore, the physical part evaluates to
\begin{equation}
-gS(x)((2-d)\mathscr L+\# \Lambda|x|)
\end{equation}
where the second term comes from $c_2(x)$. This term, proportional to the length of the line, is a perimeter counterterm, and we simply drop it. So, all in all, this diagram contributes
\begin{equation}
g(d-2-\xi)S(x)\mathscr L
\end{equation}

Adding up all contributions, we see that the $\xi$-dependent part nicely cancels, while the physical parts add up to
\begin{equation}\label{eq:psiWpsi_2pt}
gS(x)\mathscr L\times 4\frac{d-1}{d}
\end{equation}
whence
\begin{equation}\label{eq:Delta_1_quark}
\Delta(\psi)=\frac{d-1}{2}-g\frac{2(d-1)}{(4\pi)^{d/2}\Gamma(d/2+1)}+\mathcal O(g^2)
\end{equation}

We can check this in various special cases. The anomalous dimension vanishes in $d=1$, it equals $-\frac{1}{2N_f}$ in $d=2$, $-\frac{32}{3\pi^2N_f}$ in $d=3$, and vanishes in $d=4$. The first three agree with our low-dimensional review~\eqref{eq:Delta_psi_1},~\eqref{eq:Delta_psi_2_QED},~\eqref{eq:Delta_psi_3}. The fourth simply says that our IR CFT becomes free in $4d$ which, in a rough sense, corresponds to the logarithmic running $e^2\to0$ in QED$_4$.

As another simple check, taking the trace of~\eqref{eq:psiWpsi_2pt} gives the $\mathcal O(N_f^{0})$ correction to the \emph{vev} $\langle \bar\psi\psi\rangle$. This is easily seen to vanish for all $d$, including $d=1$, in agreement with $\langle\bar\psi\psi\rangle=\frac12N_f$ being tree-level exact in $d=1$ (cf.~\eqref{eq:vev_qed}).

Furthermore, the general $d$ answer~\eqref{eq:Delta_1_quark} also agrees with~\cite{Gracey:1993sn} (however, see footnote~\ref{fn:gracey}).

Finally, we describe the generalization to non-abelian gauge groups. As explained around~\eqref{eq:g_QCD}, the first change is that the effective coupling becomes $g_\text{QCD}=g_\text{QED}\times\frac{1}{T(R)}$. Furthermore, the gluon propagator becomes $\Delta_\text{QCD}=\Delta_\text{QED}\delta^{ab}$, where $a,b$ are color indices, and the vertex acquires a factor of $t^a_R$. This gives an extra factor of $(t^a t^a)_{ij}=T(R)\frac{\dim(G)}{\dim(R)}\delta_{ij}$. So all in all, the final result in QCD is equal to the QED one, multiplied by $\frac{\dim(G)}{\dim(R)}$, which agrees with~\eqref{eq:Delta_psi_2_YM}.

\subsection{Mesons}

In this subsection we study bilinear operators of the form $\bar\psi\Gamma\psi$, where $\Gamma$ is some matrix in spinor-flavor space. In particular, we will be interested in the correlation function $\langle\bar\psi\Gamma\psi\,\bar\psi\bar\Gamma\psi\rangle$ where $\Gamma,\bar\Gamma$ are in principle distinct matrices, assumed hermitian for concreteness. We begin with $G=U(1)$, and generalize to QCD afterwards.

To tree level this correlator is given by
\begin{equation}\label{eq:tree_level_bilinear}
\langle\bar\psi\Gamma\psi(p)\,\bar\psi\bar\Gamma\psi(-p)\rangle=\ \tikz[baseline=-3pt]{
\draw[->,>=Stealth,thick] (.01,.5) -- (-.12,.5);
\draw[->,>=Stealth,thick] (-.01,-.5) -- (.12,-.5);
\draw[thick] (0,0) circle (.5cm);
\filldraw (-.5,0) circle (1pt);\filldraw (.5,0) circle (1pt);
}\ + \
\tikz[baseline=-3pt]{
\draw[->,>=Stealth,thick] (.01,.5) -- (-.12,.5);
\draw[->,>=Stealth,thick] (-.01,-.5) -- (.12,-.5);
\draw[thick] (0,0) circle (.5cm);
\filldraw (-.5,0) circle (1pt);

\draw[photon] (.5,0) -- (1,0);
\begin{scope}[shift={(1.5,0)}]
\draw[->,>=Stealth,thick] (.01,.5) -- (-.12,.5);
\draw[->,>=Stealth,thick] (-.01,-.5) -- (.12,-.5);
\draw[thick] (0,0) circle (.5cm);
\filldraw (.5,0) circle (1pt);
\end{scope}
}
\end{equation}

The first diagram will show up quite often, so we give it a name: $M(\Gamma,\bar \Gamma)$. It evaluates to
\begin{equation}
\begin{aligned}
M(\Gamma,\bar\Gamma)&:=-\int_q \tr[S(q)\Gamma S(q+p)\bar\Gamma]\\
&=\frac{1}{g}\frac{1}{2(d-2)}\frac{\tr[\gamma^\alpha\Gamma\gamma^\beta\bar\Gamma]}{\tr[1]}\frac{-\delta_{\alpha\beta}p^2+(2-d)p_\alpha p_\beta}{p^{4-d}}
\end{aligned}
\end{equation}


In this notation, the second diagram is equal to
\begin{equation}
-M(\Gamma,\gamma^\mu)\Delta_{\mu\nu}(p)M(\gamma^\nu,\bar\Gamma)
\end{equation}
We note that
\begin{equation}\label{eq:M_Gamma_gamma}
M(\Gamma,\gamma^\mu)=\frac{1}{g}\frac{1}{\tr[1]}\frac{\tr[\Gamma\gamma^\mu]p^2-p^\mu\tr[\Gamma\slashed p]}{p^{4-d}}
\end{equation}
which vanishes when contracted with $p_\mu$, i.e., this structure is transverse. This implies that the $\xi$-dependent part in $M(\Gamma,\gamma^\mu)\Delta_{\mu\nu}(p)$ drops out, and the correlation function is gauge invariant, as one would expect. So to summarize, to tree level this correlation function is
\begin{equation}\label{eq:meson_tree_general}
\langle\bar\psi\Gamma\psi(p)\,\bar\psi\bar\Gamma\psi(-p)\rangle=M(\Gamma,\bar\Gamma)-\frac{g}{p^{d-2}} M(\Gamma,\gamma^\mu)M(\gamma_\mu,\bar\Gamma)
\end{equation}

Let us evaluate this explicitly for a few special cases. First, when $\Gamma=\gamma^\rho$, the corresponding bilinear is the gauge current $\bar\psi\gamma^\rho\psi$; for this, we find
\begin{equation}
\langle\bar\psi\gamma^\rho\psi\,\bar\psi\bar\Gamma\psi\rangle=M(\gamma^\rho,\bar\Gamma)-\frac{g}{p^{d-2}} M(\gamma^\rho,\gamma^\mu)M(\gamma_\mu,\bar\Gamma)
\end{equation}
which, using~\eqref{eq:M_Gamma_gamma}, becomes
\begin{equation}
\begin{aligned}
&=M(\gamma^\rho,\bar\Gamma)-\frac{\delta^{\rho\mu}p^2-p^\mu p^\rho}{p^2}M(\gamma_\mu,\bar\Gamma)\\
&\equiv0
\end{aligned}
\end{equation}

Therefore, we learn that the correlation function $\langle\bar\psi\gamma^\rho\psi\,\bar\psi\bar\Gamma\psi\rangle$ vanishes for any choice of $\bar\Gamma$. This is in fact the expected result: the gauge current $\bar\psi\gamma^\rho\psi$ is a null operator, since the equations of motion imply
\begin{equation}
\frac{1}{e^2}\partial_\mu f^{\mu\nu}=\bar\psi\gamma^\nu\psi
\end{equation}
which becomes $\bar\psi\gamma^\nu\psi\equiv0$ in the low energy limit $e^2\to\infty$. (If we were to repeat the computation of $\langle\bar\psi\gamma^\rho\psi\,\bar\psi\bar\Gamma\psi\rangle$ using the full photon propagator $\frac{e^2\delta_{\mu\nu}}{p^2(1-e^2\Pi)}$ instead of the low energy one $-\frac{\delta_{\mu\nu}}{p^2\Pi}$, we would find that the correlator decays exponentially $\langle\bar\psi\gamma^\rho\psi(x)(\cdots)\rangle\sim \exp(-\mu|x|)$, where $\mu\sim (N_fe^2)^{1/(4-d)}$ with $e$ the short-distance dimensionful coupling. Therefore, the gauge current creates the higher-$d$ analogue of the Coleman-Schwinger boson; it defines a massive operator and it decouples in the low energy effective theory).

Let us now consider $\Gamma=\bar\Gamma=1$. Then, we find
\begin{equation}
\langle\bar\psi\psi(p)\,\bar\psi\psi(-p)\rangle=M(1,1)-\frac{g}{p^{d-2}} M(1,\gamma^\mu)M(\gamma_\mu,1)
\end{equation}
which, using~\eqref{eq:M_Gamma_gamma}, becomes
\begin{equation}\label{eq:bilinear_tree_general}
\langle\bar\psi\psi(p)\,\bar\psi\psi(-p)\rangle=-\frac{1}{g}\frac{d-1}{d-2}\frac{1}{p^{2-d}}
- \frac{1}{g}\frac{1}{\tr[1]^2}\frac{1}{p^{4-d}}(\tr[\gamma_\mu]\tr[\gamma^\mu]p^2-p^\mu p^\nu\tr[\gamma_\mu]\tr[\gamma_\nu])
\end{equation}
From the $p$ scaling we read off $\Delta^{(0)}=d-1$, as expected. But we also encounter the usual $\tr[\gamma_\mu]$ issue. At this point we could just declare that only $\Delta^{(0)}$ is meaningful, and the overall coefficient is scheme-dependent anyway, so it doesn't really matter if we fail to analytically continue it. But in order to compute $\Delta^{(1)}$ we will have to divide the $\mathcal O(g^0)$ result by this $\mathcal O(g^{-1})$ one, and therefore we cannot really sweep this issue under the rug. Perhaps more importantly, unless we manage to simplify the factor $p^\mu p^\nu$ into $p^2\delta^{\mu\nu}$, the expression above seems to be in tension with Lorentz invariance, which requires the correlator of scalar operators to be a function of $p^2$ only.

The simplest way to deal with this awkward structure is to note that, for integer $d$, the trace $\tr[\gamma_\mu]$ is only non-zero when $d=1$. But when $d=1$, the two terms in $\tr[\gamma_\mu]\tr[\gamma^\mu]p^2-p^\mu p^\nu\tr[\gamma_\mu]\tr[\gamma_\nu]$ are actually equal to each other, and they cancel each other out. Therefore, this structure vanishes for all $d\in\mathbb N$, and the correct analytic continuation is that it vanishes for all $d\in\mathbb C$.

If we insist that $\gamma_\mu$ must make sense at fractional $d$, then we can reach the same conclusion via a more elaborate argument. In this case, we need the formal identity~\eqref{eq:weird_identity} (namely $\tr[\gamma_\rho]\delta_{\mu\nu}=\tr[\gamma_\mu]\delta_{\nu\rho}=\tr[\gamma_\nu]\delta_{\mu\rho}$). Contracting this identity with $\tr[\gamma_\mu]$ yields $\tr[\gamma_\rho]\tr[\gamma_\nu]=\tr[\gamma_\mu]\tr[\gamma^\mu]\delta_{\nu\rho}$, which again implies that the two terms $\tr[\gamma_\mu]\tr[\gamma^\mu]p^2$ and $p^\mu p^\nu\tr[\gamma_\mu]\tr[\gamma_\nu]$ are equal to each other.

Whatever argument one likes best, the end result is very simple:
\begin{equation}\label{eq:psipsi_tree}
\langle\bar\psi\psi(p)\,\bar\psi\psi(-p)\rangle=-\frac{1}{g}\frac{d-1}{d-2}\frac{1}{p^{2-d}}
\end{equation}

Let us now consider the first non-trivial $1/N_f$ correction. This is given by the diagrams
\begin{equation}\label{eq:five_diagrams}
\begin{aligned}
&\tikz[baseline=-3pt]{
\draw[thick] (0,0) circle (.5cm);
\filldraw (-.5,0) circle (1pt);\filldraw (.5,0) circle (1pt);
\begin{scope}
\clip (0,0) circle (.5cm);
\draw[photon] (0,.8) circle (.8cm);
\end{scope}

\draw[thick, postaction={decorate}, decoration={markings, mark=at position 0.5 with {\arrow[rotate=-10]{Stealth}}},rotate=55] (-20:.5cm) arc(-20:120:.5cm);
\draw[thick, postaction={decorate}, decoration={markings, mark=at position 0.5 with {\arrow[rotate=-10]{Stealth}}},rotate=55+180] (-20:.5cm) arc(-20:120:.5cm);

}\ +\ 
\tikz[baseline=-3pt,rotate=180]{
\draw[thick] (0,0) circle (.5cm);
\filldraw (-.5,0) circle (1pt);\filldraw (.5,0) circle (1pt);
\begin{scope}
\clip (0,0) circle (.5cm);
\draw[photon] (0,.8) circle (.8cm);
\end{scope}

\draw[thick, postaction={decorate}, decoration={markings, mark=at position 0.5 with {\arrow[rotate=-10]{Stealth}}},rotate=55] (-20:.5cm) arc(-20:120:.5cm);
\draw[thick, postaction={decorate}, decoration={markings, mark=at position 0.5 with {\arrow[rotate=-10]{Stealth}}},rotate=55+180] (-20:.5cm) arc(-20:120:.5cm);

}\ + \
\tikz[baseline=-3pt]{
\draw[thick] (0,0) circle (.5cm);
\filldraw (-.5,0) circle (1pt);\filldraw (.5,0) circle (1pt);
\begin{scope}
\clip (0,0) circle (.5cm);
\draw[photon] (0,-.55) -- (0,.55);
\end{scope}

\draw[thick, postaction={decorate}, decoration={markings, mark=at position 0.5 with {\arrow[rotate=-10]{Stealth}}},rotate=105] (-20:.5cm) arc(-20:120:.5cm);
\draw[thick, postaction={decorate}, decoration={markings, mark=at position 0.5 with {\arrow[rotate=-10]{Stealth}}},rotate=105+80] (-20:.5cm) arc(-20:120:.5cm);
\draw[thick, postaction={decorate}, decoration={markings, mark=at position 0.5 with {\arrow[rotate=-10]{Stealth}}},rotate=105+180] (-20:.5cm) arc(-20:120:.5cm);
\draw[thick, postaction={decorate}, decoration={markings, mark=at position 0.5 with {\arrow[rotate=-10]{Stealth}}},rotate=105-90] (-20:.5cm) arc(-20:120:.5cm);
}\\[+3ex]
&\qquad+\ 
\tikz[baseline=-3pt]{
\draw[->,>=Stealth,thick] (.01,.5) -- (-.12,.5);
\draw[->,>=Stealth,thick] (-.01,-.5) -- (.12,-.5);
\draw[thick] (0,0) circle (.5cm);
\filldraw (-.5,0) circle (1pt);\filldraw (2,0) circle (1pt);

\begin{scope}
\clip (.36,-.45) rectangle ++(.77,.9);
\draw[photon] (.2,-.35) -- (1.3,-.35);
\draw[photon] (.05,.35) -- (1.3,.35);
\end{scope}
\begin{scope}[shift={(1.5,0)}]
\draw[->,>=Stealth,thick] (.01,.5) -- (-.12,.5);
\draw[->,>=Stealth,thick] (-.01,-.5) -- (.12,-.5);
\draw[thick] (0,0) circle (.5cm);
\end{scope}
}\ +\ 
\tikz[baseline=-3pt]{
\draw[->,>=Stealth,thick] (.01,.5) -- (-.12,.5);
\draw[->,>=Stealth,thick] (-.01,-.5) -- (.12,-.5);
\draw[thick] (0,0) circle (.5cm);
\filldraw (-.5,0) circle (1pt);\filldraw (2,0) circle (1pt);

\begin{scope}
\clip (.35,-.25) rectangle ++(.69,.9);
\draw[photon] (.2,-.55) -- (1.3,.5);
\fill[white] (.76,-.02) circle (2pt);
\end{scope}
\begin{scope}[shift={(0,-.05)}]
\clip (.4,-.32) rectangle ++(.72,.7);
\draw[photon] (.2,.55) -- (1.3,-.5);
\end{scope}

\begin{scope}[shift={(1.5,0)}]
\draw[->,>=Stealth,thick] (.01,.5) -- (-.12,.5);
\draw[->,>=Stealth,thick] (-.01,-.5) -- (.12,-.5);
\draw[thick] (0,0) circle (.5cm);
\end{scope}
}\\[+3ex]
&\hspace{2cm}+\ \tikz[baseline=-3pt]{
\draw[->,>=Stealth,thick] (.01,.5) -- (-.12,.5);
\draw[->,>=Stealth,thick] (-.01,-.5) -- (.12,-.5);
\draw[thick] (0,0) circle (.5cm);
\filldraw (-.5,0) circle (1pt);

\draw[->,>=Stealth,thick] (3.01,.5) -- (3-.12,.5);
\draw[->,>=Stealth,thick] (3-.01,-.5) -- (3.12,-.5);

\draw[thick] (3,0) circle (.5cm);\filldraw (3.5,0) circle (1pt);

\draw[photon] (2,0) -- (2.5,0);

\begin{scope}
\clip (.36,-.45) rectangle ++(.77,.9);
\draw[photon] (.2,-.35) -- (1.3,-.35);
\draw[photon] (.05,.35) -- (1.3,.35);
\end{scope}
\begin{scope}[shift={(1.5,0)}]
\draw[->,>=Stealth,thick] (.01,.5) -- (-.12,.5);
\draw[->,>=Stealth,thick] (-.01,-.5) -- (.12,-.5);
\draw[thick] (0,0) circle (.5cm);
\end{scope}
}\ +\ 
\tikz[baseline=-3pt]{
\draw[->,>=Stealth,thick] (.01,.5) -- (-.12,.5);
\draw[->,>=Stealth,thick] (-.01,-.5) -- (.12,-.5);
\draw[thick] (0,0) circle (.5cm);
\filldraw (-.5,0) circle (1pt);\filldraw (2,0) circle (1pt);

\begin{scope}
\clip (.35,-.25) rectangle ++(.69,.9);
\draw[photon] (.2,-.55) -- (1.3,.5);
\fill[white] (.76,-.02) circle (2pt);
\end{scope}
\begin{scope}[shift={(0,-.05)}]
\clip (.4,-.32) rectangle ++(.72,.7);
\draw[photon] (.2,.55) -- (1.3,-.5);
\end{scope}

\begin{scope}[shift={(1.5,0)}]
\draw[->,>=Stealth,thick] (.01,.5) -- (-.12,.5);
\draw[->,>=Stealth,thick] (-.01,-.5) -- (.12,-.5);
\draw[thick] (0,0) circle (.5cm);
\end{scope}

\draw[->,>=Stealth,thick] (3.01,.5) -- (3-.12,.5);
\draw[->,>=Stealth,thick] (3-.01,-.5) -- (3.12,-.5);
\draw[thick] (3,0) circle (.5cm);\filldraw (3.5,0) circle (1pt);

\draw[photon] (2,0) -- (2.5,0);

}
\end{aligned}
\end{equation}

We will call these three lines the ``one-loop'', ``two-loop'', and ``three-loop'' diagrams, respectively (even though they are all of order $\mathcal O(N_f^0)$).

As reviewed in appendix~\ref{sec:cpt}, our task is to isolate the log divergence in these Feynman integrals. See appendix~\ref{ap:integrals} for some useful integral formulas; in particular, logs are always associated to scale-less integrals, and we denote those by
\begin{equation}
\mathscr L:=\int_k\frac{1}{k^d}\equiv \frac{1}{(4\pi)^{d/2}\Gamma(d/2)}\log(\Lambda^2)
\end{equation}

Let us go ahead and evaluate~\eqref{eq:five_diagrams}. The first and second one-loop diagrams are the simplest: they are equal to each other, and also almost identical to the trace of~\eqref{eq:psi_self_energy}. Specifically, they both equal
\begin{equation}\label{eq:2_loop_1}
\begin{aligned}
-\int_q\tr[\eqref{eq:psi_self_energy}\bar\Gamma S(q+p)\Gamma]&=g\biggl[\frac{(d-2)^2}{d}+\xi\biggr]\mathscr L\int_q\tr[S(q)\bar\Gamma S(q+p)\Gamma]\\
&=-g\biggl[\frac{(d-2)^2}{d}+\xi\biggr]\mathscr LM(\Gamma,\bar\Gamma)
\end{aligned}
\end{equation}

The third one-loop diagram is also relatively straightforward: it equals
\begin{equation}
\int_{qk}\tr[\Gamma S(q)\gamma^\mu S(q+k)\bar\Gamma S(q+k+p)\gamma^\nu S(q+p)]\Delta_{\mu\nu}(k)
\end{equation}
We can extract the log divergence by looking at the limit $p\to0$; we can do this in two ways, either we set $S(q+k+p)\to S(q+k)$ or $S(q+p)\to S(q)$. Adding up these two singularities, we find
\begin{equation}\label{eq:2_loop_2}
\frac{g\mathscr L}{d}M(\gamma^\mu \gamma^\alpha\bar\Gamma\gamma_\alpha\gamma_\mu,\Gamma)+g\mathscr L\xi M(\bar\Gamma ,\Gamma)+(\Gamma\leftrightarrow\bar\Gamma)
\end{equation}


The one-loop diagrams, after cutting open the photon line, determine the tree-level value of the four-point function $\langle a_\mu\, a_\nu\, \bar\psi\Gamma\psi\,\bar\psi\bar\Gamma\psi\rangle$. Therefore, they should be gauge invariant by themselves. And indeed, adding up twice~\eqref{eq:2_loop_1} (for the first two diagrams) plus~\eqref{eq:2_loop_2} (for the third), the $\xi$-dependent factors cancel out, as required.

Next, we evaluate the two-loop diagrams. As a useful first step, let us cut open the photon lines and focus on the left half of these diagrams, i.e., let us compute the three-point function $\langle a_\mu\, a_\nu\,\bar\psi\Gamma\psi\rangle$, as given by the diagrams
\begin{equation}
\tikz[baseline=-.7cm]{

\draw[thick] (0,0) circle (.5cm);
\filldraw (0,.5) circle (1pt);
\begin{scope}
\clip (-.8,.2) to[in=90,out=180] (-1.5,-.6) to[in=180,out=-90] (-1.2,-1.2) to[in=-90,out=0] (-.5,.08) to[in=0,out=90] cycle;
\draw[photon] (-1.3,-1.2) to[in=180,out=90] (-.1,0);
\end{scope}
\begin{scope}[xscale=-1]
\clip (-.8,.2) to[in=90,out=180] (-1.5,-.6) to[in=180,out=-90] (-1.2,-1.2) to[in=-90,out=0] (-.5,.08) to[in=0,out=90] cycle;
\draw[photon] (-1.3,-1.2) to[in=200,out=90] (-.2,.1);
\end{scope}

\draw[thick, postaction={decorate}, decoration={markings, mark=at position 0.5 with {\arrow[rotate=-10]{Stealth}}},rotate=235] (-20:.5cm) arc(-20:120:.5cm);
\draw[thick, postaction={decorate}, decoration={markings, mark=at position 0.5 with {\arrow[rotate=-10]{Stealth}}},rotate=55+50] (-20:.5cm) arc(-20:120:.5cm);
\draw[thick, postaction={decorate}, decoration={markings, mark=at position 0.5 with {\arrow[rotate=-10]{Stealth}}},rotate=55-50] (-20:.5cm) arc(-20:120:.5cm);

}\ +\ 
\tikz[baseline=-.7cm]{

\draw[thick] (0,0) circle (.5cm);
\filldraw (0,.5) circle (1pt);

\begin{scope}
\clip (-.6,.2) to[in=120,out=180] (-.7,-.6) to[in=180,out=-60] (.5,-1.2) to[in=-90,out=0] (-.5,.08) to[in=0,out=90] cycle;
\draw[photon] (-.3,.1) to[in=120,out=180] (-.6,-.4) to[in=160,out=-60] (1.1,-1.3);
\end{scope}
\fill[white] (-.04,-.9) circle (4pt);
\begin{scope}[xscale=-1]
\clip (-.6,.2) to[in=120,out=180] (-.7,-.6) to[in=180,out=-60] (.5,-1.2) to[in=-90,out=0] (-.5,.08) to[in=0,out=90] cycle;
\draw[photon] (-.2,.1) to[in=120,out=180] (-.6,-.4) to[in=160,out=-60] (1.1,-1.3);
\end{scope}

\draw[thick, postaction={decorate}, decoration={markings, mark=at position 0.5 with {\arrow[rotate=-10]{Stealth}}},rotate=235] (-20:.5cm) arc(-20:120:.5cm);
\draw[thick, postaction={decorate}, decoration={markings, mark=at position 0.5 with {\arrow[rotate=-10]{Stealth}}},rotate=55+40] (-20:.5cm) arc(-20:120:.5cm);
\draw[thick, postaction={decorate}, decoration={markings, mark=at position 0.5 with {\arrow[rotate=-10]{Stealth}}},rotate=55-40] (-20:.5cm) arc(-20:120:.5cm);

}
\end{equation}
These evaluate to
\begin{equation}
\int_q\tr[S(q+p)\Gamma S(q)\gamma^\mu S(q+k+p)\gamma^\nu]+\begin{pmatrix}\mu\leftrightarrow\nu\\ k\leftrightarrow -k-p\end{pmatrix}
\end{equation}
This integral is given by some complicated function of the dimensionless ratio $p^2/k^2$; we will consider the limit $p^2\ll k^2$, in which case we can evaluate the integral analytically:
\begin{equation}\label{eq:3pt_a_a_psi}
=-\frac{2i}{g}
\frac{(d-1)(d-4)}{d-2}\frac{1}{k^{4-d}}\frac{\tr[\Gamma\gamma^{[\mu}\slashed k\gamma^{\nu]}]}{\tr[1]}(1+\mathcal O(p^2/k^2))
\end{equation}
Note that this vanishes when contracted with $k_\mu$, as required by gauge invariance. (We also mention that, being a three-point function, the full $p^2/k^2$ dependence is determined from this expression and the general constraints of conformal invariance; this dependence is somewhat obscured in momentum space, and we will not need it anyway, so this limit $p^2\ll k^2$ is enough for our purposes.)

We can now evaluate the two-loop diagrams:
\begin{equation}
\begin{aligned}
\int_{qk\ell}&\tr[S(q+p)\Gamma S(q)\gamma^\mu S(q+p+k)\gamma^\nu]\\
&\qquad\times\tr[S(\ell+p)\gamma^\sigma S(k+p+\ell)\gamma^\rho S(\ell)\bar\Gamma]\\[+1ex]
&\qquad\qquad\times\Delta_{\mu\rho}(k+p)\Delta_{\nu\sigma}(k)+\text{cross-channel}
\end{aligned}
\end{equation}
As before, we isolate the log divergence by looking at the limit $p\to0$; we can do this in either of the first two lines. When we take $p=0$, say, in the first line, we end up with the same integral as in~\eqref{eq:3pt_a_a_psi}. Then, the fact that that  integral is transverse means that these two diagrams add up to something $\xi$-independent as well.

Plugging~\eqref{eq:3pt_a_a_psi} into our two-loop integral, we find
\begin{equation}
\begin{aligned}
-\frac{2ig}{\tr[1]}\frac{(d-1)(d-4)}{d-2}&\int_{k\ell}\frac{\tr[\Gamma\gamma^{[\mu}\slashed k\gamma^{\nu]}]}{k^d}\tr[S(\ell+p)\gamma_\nu S(k+p+\ell)\gamma_\mu S(\ell)\bar\Gamma]+(\Gamma\leftrightarrow\bar\Gamma)
\end{aligned}
\end{equation}
The $k$ integral is now elementary:
\begin{equation}
\begin{aligned}
&=\frac{2g \mathscr L}{\tr[1]}\frac{(d-1)(d-4)}{d(d-2)}(\tr[\Gamma\gamma^\alpha \gamma^{[\mu}\gamma^{\nu]}]-\delta^{\mu\alpha}\tr[\Gamma \gamma^\nu]+\delta^{\nu\alpha}\tr[\Gamma \gamma^\mu])\\
&\qquad\qquad\int_\ell\tr[S(\ell+p)\gamma_\nu \gamma_\alpha \gamma_\mu S(\ell)\bar\Gamma]+(\Gamma\leftrightarrow\bar\Gamma)
\end{aligned}
\end{equation}
which simplifies to
\begin{equation}\label{eq:two_loop}
=-\frac{2g \mathscr L}{\tr[1]}\frac{(d-1)(d-4)}{d(d-2)}\tr[\Gamma\gamma^{\alpha\mu\nu}]M(\bar\Gamma ,\gamma_{\alpha\mu\nu})+(\Gamma\leftrightarrow\bar\Gamma)
\end{equation}

Finally, we evaluate the three-loop diagrams. These are actually the easiest: the left side of these diagrams is identical to the two-loop diagrams we just computed, with $\bar\Gamma\to\gamma^\rho$, and the right side of these diagrams is $\Delta_{\rho\sigma}M(\gamma^\sigma,\bar\Gamma)$. But the two-loop diagrams~\eqref{eq:two_loop} are easily seen to vanish when we replace $\bar\Gamma\to\gamma^\rho$, since both $\tr[\bar\Gamma\gamma^{\alpha\mu\nu}]$ and $M(\bar\Gamma,\gamma_{\alpha\mu\nu})$ vanish when we perform this replacement. Therefore, the three-loop diagrams are zero. (One can reach this same conclusion from charge-conjugation symmetry.)

To summarize, the first non-trivial correction to the $\langle\bar\psi\Gamma\psi\,\bar\psi\bar\Gamma\psi\rangle$ correlation function is
\begin{equation}\label{eq:general_1NF}
\begin{aligned}
\frac{g\mathscr L}{d}\biggl(
-&(d-2)^2M(\Gamma,\bar\Gamma)+
M(\gamma^\mu \gamma^\alpha\bar\Gamma\gamma_\alpha\gamma_\mu,\Gamma)\\
&-\frac{2}{\tr[1]}\frac{(d-1)(d-4)}{(d-2)}\tr[\Gamma\gamma^{\alpha\mu\nu}]M(\bar\Gamma ,\gamma_{\alpha\mu\nu})
\biggr)+(\Gamma\leftrightarrow\bar\Gamma)
\end{aligned}
\end{equation}

As mentioned earlier, this should vanish if any of the insertions is a gauge current, i.e., if, say, $\Gamma=\gamma^\rho$. And indeed, for this choice of $\Gamma$ the first term is easily seen to cancel against the second, while the third vanishes on its own.

Let us now take $\Gamma=\bar\Gamma=1$. Then, we find
\begin{equation}\label{eq:psipsi_1loop}
\begin{aligned}
\frac{4(d-1)g\mathscr L}{d}&\biggl(
2M(1,1)+\frac{1}{\tr[1]}\frac{d-4}{d-2}\tr[\gamma^{\alpha\mu\nu}]M(1,\gamma_{\alpha\nu\mu})
\biggr)\\
&=-\frac{4(d-1)\mathscr L}{d}\frac{d-1}{d-2}\frac{1}{p^{2-d}}\biggl(
2-\frac{d-4}{d-2}\frac{\tr[\gamma^{\alpha\mu\nu}]\tr[\gamma_{\alpha\mu\nu}]}{\tr[1]^2}
\biggr)
\end{aligned}
\end{equation}

Adding the tree-level result~\eqref{eq:psipsi_tree} to the one-loop result~\eqref{eq:psipsi_1loop}, we conclude that
\begin{equation}
\langle\bar\psi\psi(p)\,\bar\psi\psi(-p)\rangle\propto
1
+\frac{4(d-1)g\mathscr L}{d}\biggl(
2-\frac{d-4}{d-2}\frac{\tr[\gamma^{\alpha\mu\nu}]\tr[\gamma_{\alpha\mu\nu}]}{\tr[1]^2}
\biggr)+\mathcal O(g^2)
\end{equation}
from where we read off $\Delta^{(1)}$ as
\begin{equation}\label{eq:Delta_1_bilinear}
\Delta^{(1)}(\bar\psi\psi)=-\frac{g}{(4\pi)^{d/2}\Gamma(d/2)} \frac{4(d-1)}{d}\biggl(
2-\frac{d-4}{d-2}\frac{\tr[\gamma^{\alpha\mu\nu}]\tr[\gamma_{\alpha\mu\nu}]}{\tr[1]^2}
\biggr)
\end{equation}

As a check, let us consider a few special cases of this. If $d=1$, we get $\Delta^{(1)}=0$. If $d=2$, we get $\Delta^{(1)}=-1/N_f$. If $d=3$, we use $\frac{\tr[\gamma^{\alpha\mu\nu}]\tr[\gamma_{\alpha\mu\nu}]}{\tr[1]^2}=-3!$ to get $\Delta^{(1)}=128/3\pi^2N_f$. These agree with the results reviewed in the appendix (cf.~\eqref{eq:Delta_psi_1},~\eqref{eq:Delta_psi_2},~\eqref{eq:Delta_psipsi_3}). And if $d=4$ we once again get $\Delta^{(1)}=0$, as expected.

Before we discuss the result above more thoroughly, let us quickly mention the non-abelian case. As in previous subsections, all diagrams considered here stay the same, and we just multiply by the usual color factor $\dim(G)/\dim(R)$. The only exception is that the gluonic line in~\eqref{eq:tree_level_bilinear} acquires corrections due to ghosts and self-interactions, which contribute to order $\mathcal O(N^0_f)$. But as we argued below~\eqref{eq:bilinear_tree_general}, this diagram does not contribute to $\bar\psi\psi$, so this does not affect our analysis above. And taking our QED result above, and multiplying by $\dim(G)/\dim(R)$, does match our $d=1,2,3$ expectations, see e.g.~\eqref{eq:Delta_psipsi_2_YM}.


How should we think of~\eqref{eq:Delta_1_bilinear} at fractional $d$? If we believe equation~\eqref{eq:tr_gamma3_0} for fractional $d$, and thus also its implication that $(d-3)\tr[\gamma_{\alpha\mu\nu}]\equiv0$, then we would conclude that the second term is exactly zero for all $d\neq3$, and this scaling dimension is discontinuous there.

This conclusion is not actually precise enough. The correct interpretation of~\eqref{eq:Delta_1_bilinear} for $d\in\mathbb C$ requires us to think about (a regularized version of) $\operatorname{Rep}(Pin(\infty))$ more carefully. Before we do that, we would like to mention the more naive approach which, while ultimately not well-defined, yields some amusing results.

For the naive approach, we assume that $\gamma^\mu$ makes sense for $d\in\mathbb C$, and that its various traces are non-zero, analytic functions of $d$. If this is the case, we must reckon with the extraneous operators in~\eqref{eq:extraneous}; these are formally scalars, and have the same tree-level scaling dimension as $\bar\psi\psi$, and therefore we must solve a (infinite-dimensional) degenerate perturbation theory problem. We don't think this is the correct perspective, and therefore we won't dwell on this too much. But as an illustration, let us truncate the extraneous operators to the subset $\{\bar\psi\psi,\tr[\gamma_{\rho\sigma\tau}]\bar\psi\gamma^{\rho\sigma\tau}\psi\}$. Now, take the general formula~\eqref{eq:general_1NF} and plug in $\Gamma=\bar\Gamma=1+\chi\tr[\gamma_{\rho\sigma\tau}]\gamma^{\rho\sigma\tau}$ for some coefficient $\chi$, to be chosen below. Then, one finds
\begin{equation}\label{eq:chi_general}
\begin{aligned}
\langle\bar\psi\Gamma\psi(p)\,\bar\psi\bar\Gamma\psi(-p)\rangle
&=M(\Gamma,\bar\Gamma)-
\frac{4g\mathscr L}{d}\biggl(
(d-4)\biggl\{4+\frac{\chi^{-1}-6\tr[1]}{\tr[1]}\frac{d-1}{d-2}\biggr\}M(\bar\Gamma ,\Gamma)\\
&-\biggl\{6 (d-3)+\frac{\chi^{-1}-6\tr[1]}{\tr[1]}\frac{(d-1)(d-4)}{(d-2)}\biggr\}M(1,\Gamma)
\biggr)
\end{aligned}
\end{equation}

The point of~\eqref{eq:chi_general} is the following: if we choose the coefficient $\chi$ such that the term proportional to $M(1,\Gamma)$ vanishes, namely
\begin{equation}
\chi^{-1}\equiv -\frac{12}{(d-4) (d-1)}\tr[1]
\end{equation}
then the two-point function becomes
\begin{equation}
\begin{aligned}
\langle\bar\psi\Gamma\psi(p)\,\bar\psi\bar\Gamma\psi(-p)\rangle=M(\Gamma,\bar\Gamma)
\biggl(1+\frac{8(d-1)g\mathscr L}{d}+\mathcal O(g^2)\biggr)
\end{aligned}
\end{equation}
from which we can read off $\Delta^{(1)}$ as
\begin{equation}\label{eq:Delta_1_analytic}
\Delta^{(1)}\bigl(\bar\psi(1+\chi\tr[\gamma_{\rho\sigma\tau}]\gamma^{\rho\sigma\tau})\psi\bigr)=-\frac{8(d-1)g}{d(4\pi)^{d/2}\Gamma(d/2)}
\end{equation}

This is now analytic in $d$. But if we take $d=3$, we get $-64/3\pi^2N_f$, which is not the correct answer~\eqref{eq:Delta_psipsi_3}. The reason for this is that, if we take $d=3$, the operator in parentheses in~\eqref{eq:Delta_1_analytic} actually vanishes. Indeed, in $d=3$ one has $\tr[\gamma_{\rho\sigma\tau}]\gamma^{\rho\sigma\tau}=-6\tr[1]$, and therefore
\begin{equation}
\biggl(-\frac{12}{(d-4) (d-1)}\tr[1]\biggr)^{-1}\tr[\gamma_{\rho\sigma\tau}]\gamma^{\rho\sigma\tau}\xrightarrow{d\to3}-1
\end{equation}

So in this approach, we would conclude that we either work with the operator $\bar\psi\psi$, whose scaling dimension is discontinuous at $d=3$, or with an operator $\bar\psi(1+\chi\tr[\gamma_{\rho\sigma\tau}]\gamma^{\rho\sigma\tau})\psi$ whose dimension is analytic in $d$, but the operator itself is discontinuous, and it becomes null at $d=3$. Furthermore, there is no scalar operator in QED$_3$ with dimension $-64/3\pi^2N_f$, so the formula~\eqref{eq:Delta_1_analytic} does not match anything in the physical theory\footnote{Intriguingly, there is one bilinear in QED$_3$ with this scaling dimension: the adjoint scalar $\bar\psi_i\psi^j-\text{trace}$, which has $\Delta^{(1)}=-64/3\pi^2$~\cite{PhysRevB.76.149906,Chester:2016ref}. This is not a flavor singlet, so this operator is distinct from our $\bar\psi\psi$. It is not clear to us if this is just a coincidence, but we do mention that, as explained around~\eqref{eq:gamma_d_dict}, the notion of flavor quantum numbers is somewhat ambiguous in fractional dimension, so it is conceivable that the singlet and the adjoint get mixed up somehow. It would be interesting to study the interplay between flavor symmetries and analytic continuation of $d$ more carefully in the future.}. This is reminiscent of various results in the literature, where one observes that continuous families of CFTs, when particularized to concrete values of the parameter, appear to contain operators that are not present if one began with that choice of the parameter, see e.g.~\cite{DeCesare:2025ukl,DeCesare:2026dwm,Zan:2026oyb} for recent examples.

We now explain what we believe is a more fruitful perspective -- this being that one should be working in ``$\operatorname{Rep}(Pin(\infty))$''. In short, we will try to work with \emph{stable} quantities, namely we will ignore low $d$ exceptions and only use formulas that are correct for \emph{all} sufficiently large $d\in\mathbb N$. In particular, all traces over the even part of the Clifford algebra are stable, while the traces over the odd part will be replaced by their stable limit, i.e., they will be declared to be zero. This last prescription is correct at large enough $d$, but wrong at small $d\in\mathbb N$, and therefore we will have to fix the low rank exceptions by hand, in a systematic way.

Note that declaring that odd traces are zero is formally equivalent, as we did in the discussion of $\bar\psi(1+\chi\tr[\gamma_{\mu\nu\sigma}]\gamma^{\mu\nu\sigma})\psi$, to adding suitable extraneous operators with coefficients chosen so that these traces cancel out. The notion of a direct limit makes this precise.

In any event, once we declare that all traces with an odd number of gammas vanish, we can easily write down expressions for scaling dimensions that are analytic in $d$. The issue is that these formulas will yield the wrong answer for small $d$, as these expressions are missing the contribution of these traces that we set to zero by hand. The obvious question is whether we can harmonize the two approaches and write down observables that are both analytic in $d$ and correct for all $d\in\mathbb N$, and not just for large enough $d$.

A hint toward realizing this goal is to notice that $\bar\psi\gamma_{\mu_1\cdots\mu_n}\psi$ is a rank-$n$ anti-symmetric tensor only \emph{generically}, but not always. For example, $\bar\psi\gamma_{\mu_1\mu_2\mu_3}\psi$ is actually a scalar when $d=3$. More generally, the low-rank exceptions where $\tr[\gamma_{\mu_1\cdots\mu_n}]$ is non-zero are precisely the situations where these fermion bilinars are actually scalars.

This means that if we want to understand the scalar operator $\bar\psi\psi$ at low $d$, we must also consider the higher-rank forms $\bar\psi\gamma_{\mu_1\cdots\mu_n}\psi$ as well, since the latter are sometimes scalars too. In a formal sense, in fractional $d$ we must allow for something like mixing of Lorentz tensors of different spin, since at exceptional values of $d$, these representations become isomorphic.

So, with this insight in mind, let us compute $\Delta(\bar\psi\gamma_{\mu_1\cdots\mu_n}\psi)$ to leading order in $1/N_f$. For this we can just reuse our general formulas~\eqref{eq:meson_tree_general},~\eqref{eq:general_1NF} and simply plug in $\Gamma=\bar\Gamma=\gamma_{\mu_1\cdots\mu_n}$. For generic $d$ these formulas will involve, much like in the case of $\bar\psi\psi$, traces of an odd number of gammas. The stable limit of these dimensions is easily worked out, the answer being
\begin{equation}\label{eq:naive_Delta}
\begin{aligned}
\widehat\Delta(\bar\psi\psi)&=1-\frac{8(d-1)}{d}\frac{g}{(4\pi)^{d/2}\Gamma(d/2)}\\
\widehat\Delta(\bar\psi\gamma_\mu\psi)&=0\\
\widehat\Delta(\bar\psi\gamma_{\mu_1\mu_2}\psi)&=1+\frac{8(d-3)}{d} \frac{g}{(4\pi)^{d/2}\Gamma(d/2)}\\
\widehat\Delta(\bar\psi\gamma_{\mu_1\mu_2\mu_3}\psi)&=1-\frac{8(d-4) (d+1)}{d(d-2)}\frac{g}{(4\pi)^{d/2}\Gamma(d/2)}
\end{aligned}
\end{equation}
etc. Here, the hat means that this scaling dimension refers to the direct limit $d\to\infty$, i.e., these expressions are only correct for large enough $d$, and they fail at small $d\in\mathbb N$, since they ignore traces that are actually non-zero sometimes. The vanishing of $\widehat\Delta(\bar\psi\gamma_\mu\psi)$ simply reflects the fact that this bilinear is the gauge current and hence a null operator. We will forget about this bilinear from now on.

As a quick check of the result above note that, if we take $d=4-2\epsilon$ and expand at small $\epsilon$, we find
\begin{equation}
\begin{aligned}
\widehat\Delta(\bar\psi\psi)&= 3-2\epsilon+\frac{1}{\tr[1]}\biggl(-18 \epsilon+15 \epsilon ^2+\tfrac{35}{2} \epsilon ^3+(\tfrac{83}{4}-36 \zeta(3)) \epsilon ^4+\cdots\biggr)\\
\widehat\Delta(\bar\psi\gamma_{\mu_1\mu_2}\psi)&= 3-2\epsilon+\frac{1}{\tr[1]}\biggl(
6\epsilon - 13\epsilon^2 - \tfrac92 \epsilon^3 + (\tfrac74 + 12\zeta(3)) \epsilon^4+\cdots
\biggr)
\\
\widehat\Delta(\bar\psi\gamma_{\mu_1\mu_2\mu_3}\psi)&= 3-2\epsilon+\frac{1}{\tr[1]}\biggl(30\epsilon^2 + 13\epsilon^3 -\tfrac{35}{2}\epsilon^4+\cdots\biggr)
\end{aligned}
\end{equation}
in perfect agreement with~\eqref{eq:Delta_epsilon}. The reason these two computations match is that the $\epsilon$ expansion of~\cite{DiPietro:2017kcd} uses the 't Hooft-Veltman prescription that is adapted to a neighborhood of $d=4$, in which traces with an odd number of gammas do vanish. 

The key point of~\eqref{eq:naive_Delta} is the following. If we take $\widehat\Delta(\bar\psi\psi)$ and plug in $d=3$, we find $1-\frac{64}{3\pi^2N_f}$, which differs from the correct answer as computed directly in $d=3$, namely $1+\frac{128}{3 \pi ^2 N_f}$ (cf.~\eqref{eq:Delta_psipsi_3}). On the other hand, in $d=3$ the operator $\bar\psi\gamma_{\mu_1\mu_2\mu_3}\psi$ is a scalar and, if we take $\widehat\Delta(\bar\psi\gamma_{\mu_1\mu_2\mu_3}\psi)$ and plug in $d=3$, we do get the correct answer, i.e., $1+\frac{128}{3 \pi ^2 N_f}$. That being said, the formula for $\widehat\Delta(\bar\psi\gamma_{\mu_1\mu_2\mu_3}\psi)$ gives the wrong answer for $d=2$, while the formula for $\widehat\Delta(\bar\psi\psi)$ gives the correct answer in this dimension. So neither $\widehat\Delta(\bar\psi\psi)$ nor $\widehat\Delta(\bar\psi\gamma_{\mu_1\mu_2\mu_3}\psi)$ matches the honest $\Delta(\bar\psi\psi)$ for all $d\in\mathbb N$, but they do each give the correct answer for some specific $d\in\mathbb N$. (The reader might find the graph on page~\pageref{fig:plot_delta}, which plots the functions~\eqref{eq:naive_Delta}, useful.)


To leading order in $1/N_f$ the operator $\bar\psi\psi$ only picks up the trace $\tr[\gamma_{\mu_1\mu_2\mu_3}]$, which is why we find one extra branch adapted to $d=3$. To higher orders in $1/N_f$ one finds higher odd traces $\tr[\gamma_{\mu_1\cdots\mu_n}]$, and therefore there are infinitely many branches for $\Delta(\bar\psi\psi)$, each adapted to a neighborhood of a given odd $d$. The analytic continuation of $d$ for a generic QFT does not live on $\mathbb C$, but rather on some infinite cover thereof; only on such cover are observables truly analytic. It would be interesting to try and make this statement mathematically rigorous, especially in relation to analytic continuation of other discrete parameters, like the number of colors in large $N_c$ gauge theories.

An important aspect that should be considered more carefully is that the factor of $\frac{g}{(4\pi)^{d/2}\Gamma(d/2)}$ in~\eqref{eq:naive_Delta} is $\sim \frac{2^d}{\tr[1]}\sin(\pi d/2)\mathcal O(1/\sqrt d)$ at large $d$. This grows with $d$ and therefore the $\Delta^{(1)}$ correction is larger than the tree-level result $\Delta^{(0)}$ for sufficiently large $d$, and the large-$N_f$ expansion breaks down. In other words, our result for $\Delta(\bar\psi\psi)$ is only reliable in the limit $2^{\lfloor d/2\rfloor} N_f\gg 2^d$. This puts any notion of uniqueness into question since, as we argued, analytic continuation can only possibly be unique if we impose ``nice behavior'' at $d\to\infty$. For fixed $N_f$, the inequality $2^{\lfloor d/2\rfloor} N_f\gg 2^d$ is always disobeyed for sufficiently large $d$, and therefore it is impossible to establish whether our result is unique or not. It appears to us that, ultimately, the topic of uniqueness in analytic continuation of QFTs can only be addressed in non-perturbative settings.

\section{QED$_d$ Gross--Neveu--Yukawa model at large $N_f$}
\label{section_QEDGNY}
We now add a Gross--Neveu--Yukawa interaction to large-$N_f$ QED$_d$, as studied in the previous section. The purpose is to understand how the
branch structure discussed above is modified by the additional interaction. We consider the Euclidean action \cite{Boyack,Janssen}
\begin{equation}
\mathcal{L}
=
\sum_{i=1}^{N_f}
\left[
-\bar\psi_i\gamma^\mu(\partial_\mu+iA_\mu)\psi_i
+
\phi\,\bar\psi_i\psi_i
\right]
+
\frac{1}{4e^2}F_{\mu\nu}^2
+
\frac{N_f}{2u^2} (\partial_\mu \phi)^2
+
N_f\lambda_0^2\phi^4 
\end{equation}
and take the large-$N_f$ limit with $e^2N_f$, $u$, and $\lambda_0$
held fixed. 

The scalar field is normalized so that its bare propagator is
\begin{equation}
G^{\rm bare}_\phi(p)
=
\frac{u^2}{N_f p^2} 
\end{equation}
At large $N_f$, the leading correction to the scalar two-point function
comes from the fermion bubble
\begin{equation}
\Sigma(p)
=
\int_\ell
\tr\left[
\frac{\slashed p+\slashed \ell}{(p+l)^2}
\frac{\slashed \ell}{\ell^2}
\right]
=
-\frac{1}{2}\tr[1]\,
\kappa_d(1,1)|p|^{d-2} 
\end{equation}
In $d=3$, this gives $\Sigma(p)=-N_f|p|/8$ when $\tr[1]=2N_f$.
The effective scalar propagator is therefore
\begin{equation}
\begin{aligned}
G^{\rm eff}_\phi(p)
&=
\left[
\left(G^{\rm bare}_\phi(p)\right)^{-1}
-
\Sigma(p)
\right]^{-1} \\
&=
\left[
\frac{N_f p^2}{u^2}
+
\frac{1}{2}\tr[1]\,
\kappa_d(1,1)|p|^{d-2}
\right]^{-1} 
\end{aligned}
\end{equation}
For $2<d<4$, the fermion bubble dominates in the infrared, and hence
\begin{equation}
G_\phi(p)
=
\frac{h}{|p|^{d-2}}+\cdots,
\qquad
h
:=
\frac{2}{\tr[1] \kappa_d(1,1)} 
\end{equation}
Here, the dots denote corrections subleading in the large-$N_f$ infrared
expansion.
Thus, in the infrared, the effective scalar propagator has the same 
large-$N_f$ scaling as the gauge-field propagator.

The Feynman rules relevant for the leading $1/N_f$ computation are
\begin{equation}
\begin{aligned}
\tikz[baseline=-3pt]{
\begin{scope}
\clip (0,-.2) rectangle ++(1.5,0.5);
\draw[fermion] (0,0)--(1.8,0);
\end{scope}
\node[scale=.8] at (.1,-.3) {\phantom{$\mu$}};
\node[scale=.8] at (1.4,-.3) {\phantom{$\nu$}};
}
&=S(p)=
-i\frac{\slashed p}{p^2}
\\[+1ex]
\tikz[baseline=-3pt]{
\begin{scope}
\clip (0,-.2) rectangle ++(1.5,0.5);
\draw[photon] (0,0)--(1.8,0);
\end{scope}
\node[scale=.8] at (.1,-.3) {$\mu$};
\node[scale=.8] at (1.4,-.3) {$\nu$};
}
&=\Delta_{\mu\nu}(p)= 
\frac{g}{p^{d-2}}
\left(
\delta^{\mu\nu}
+
\xi\frac{p^\mu p^\nu}{p^2}
\right)
\\[+1ex]
\tikz[baseline=-3pt]{
\begin{scope}
\clip (0,-.2) rectangle ++(1.5,0.5);
\draw[scalar] (0,0)--(1.8,0);
\end{scope}
\node[scale=.8] at (.1,-.3) {\phantom{$\mu$}};
\node[scale=.8] at (1.4,-.3) {\phantom{$\nu$}};
}
&=G(p)=
\frac{h}{p^{d-2}} 
\end{aligned}
\end{equation}
where $g$ is the same effective coupling as defined in \eqref{eq:eff_coupling}. The gauge and scalar vertices are\footnote{The quartic interaction is irrelevant for
$d<4$, and we will not include it in the leading large-$N_f$ analysis.}
\begin{equation}
\begin{aligned}
\tikz[baseline=-3pt]{
\begin{scope}
\draw[photon](0,0)--(-1, 0);
\draw[fermion] (0,0)--(1, 0.6);
\draw[fermion] (1, -0.6) -- (0,0);
\end{scope}
\node[scale=.8] at (-0.1,-.3) {$\mu$};
}
&=
-i\gamma^\mu,
\qquad
\tikz[baseline=-3pt]{
\begin{scope}
\draw[scalar](0,0)--(-1, 0);
\draw[fermion] (0,0)--(1, 0.6);
\draw[fermion] (1, -0.6) -- (0,0);
\end{scope}
}
=
1 
\end{aligned}
\end{equation}

\subsection{Fermion bilinears}

We now consider bilinears $\bar\psi\Gamma\psi$ and use the
notation $M(\Gamma,\bar\Gamma)$ of Sec.~\ref{sec:results} for the corresponding
fermion bubble.  At leading order, the connected two-point function receives one new contribution, namely scalar exchange. Thus, the connected two-point function is given by
\begin{equation}\label{eq:conn_2pt_gny}
\langle\bar\psi\Gamma\psi(p)\,\bar\psi\bar\Gamma\psi(-p)\rangle
=
\tikz[baseline=-3pt]{
\draw[->,>=Stealth,thick] (.01,.5) -- (-.12,.5);
\draw[->,>=Stealth,thick] (-.01,-.5) -- (.12,-.5);
\draw[thick] (0,0) circle (.5cm);
\filldraw (-.5,0) circle (1pt);\filldraw (.5,0) circle (1pt);
}
+
\tikz[baseline=-3pt]{
\draw[->,>=Stealth,thick] (.01,.5) -- (-.12,.5);
\draw[->,>=Stealth,thick] (-.01,-.5) -- (.12,-.5);
\draw[thick] (0,0) circle (.5cm);
\filldraw (-.5,0) circle (1pt);
\draw[photon] (.5,0) -- (1,0);
\begin{scope}[shift={(1.5,0)}]
\draw[->,>=Stealth,thick] (.01,.5) -- (-.12,.5);
\draw[->,>=Stealth,thick] (-.01,-.5) -- (.12,-.5);
\draw[thick] (0,0) circle (.5cm);
\filldraw (.5,0) circle (1pt);
\end{scope}
}
+
\tikz[baseline=-3pt]{
\draw[->,>=Stealth,thick] (.01,.5) -- (-.12,.5);
\draw[->,>=Stealth,thick] (-.01,-.5) -- (.12,-.5);
\draw[thick] (0,0) circle (.5cm);
\filldraw (-.5,0) circle (1pt);
\draw[scalar] (.5,0) -- (1,0);
\begin{scope}[shift={(1.5,0)}]
\draw[->,>=Stealth,thick] (.01,.5) -- (-.12,.5);
\draw[->,>=Stealth,thick] (-.01,-.5) -- (.12,-.5);
\draw[thick] (0,0) circle (.5cm);
\filldraw (.5,0) circle (1pt);
\end{scope}
} 
\end{equation}
which combine to give
\begin{equation}
\langle
\bar\psi\Gamma\psi(p)\,
\bar\psi\bar\Gamma\psi(-p)
\rangle_{\rm tree}
=
M(\Gamma,\bar\Gamma)
-
\frac{g}{p^{d-2}}
M(\Gamma,\gamma^\mu)
M(\gamma_\mu,\bar\Gamma)
+
\frac{h}{p^{d-2}}
M(\Gamma,1)M(1,\bar\Gamma) 
\end{equation}
As before, $M(\Gamma,\gamma^\mu)$ is transverse, so the $\xi$-dependent part of the photon propagator drops out. 

For the scalar bilinear $\Gamma=\bar\Gamma=1$, one has
\begin{align}
M(1,1)
=
-\frac{1}{2}\tr[1] \,\kappa_d(1,1)|p|^{d-2},
\qquad
M(1,\gamma^\mu)=0 .
\end{align}
The first and the third terms in \eqref{eq:conn_2pt_gny} therefore cancel,
\begin{equation}
\langle
\bar\psi\psi(p)\,
\bar\psi\psi(-p)
\rangle_{\rm tree}
=
0 
\end{equation}
Thus, the leading connected correlator of the scalar bilinear vanishes.
This is just the statement that, at the critical point, the inverse effective scalar propagator is generated by the same fermion bubble that appears in $M(1,1)$.

In what follows, this vanishing connected correlator will not be used
as the starting point for extracting an anomalous dimension.
Rather, to compare with the pure QED result for
$\Gamma=\bar\Gamma=1$,
we focus on the logarithmic corrections to the one-particle-irreducible
bilinear two-point function, namely the fermion bubble.\footnote{Since the leading connected correlator in this channel
vanishes after scalar exchange is included, the $\mathscr L$ term in
the 1PI bubble is not a correction to a nonzero leading two-point
function of $\bar\psi\psi$.  Therefore, the usual extraction of an
anomalous dimension from a logarithmic correction to a two-point
function does not directly apply.}


At the next order in the large-$N_f$ expansion,
the logarithmic correction to the 1PI bilinear two-point function
receives contributions from three classes of diagrams:
pure gauge-exchange diagrams, pure scalar-exchange diagrams,
and mixed gauge-scalar diagrams.
The pure gauge-exchange contribution is the same as in the
pure QED$_d$ calculation above.
We discuss the last two classes in turn.

We first consider the scalar-exchange contribution to the logarithmic
part of the 1PI bilinear two-point function.  Equivalently, this is the
Gross--Neveu--Yukawa part of the calculation with the gauge field
turned off.  The relevant diagrams are the scalar analogues of the
diagrams considered in pure QED$_d$
\begin{equation}\label{eq:five_diagrams_GN}
\tikz[baseline=-3pt]{
\draw[thick] (0,0) circle (.5cm);
\filldraw (-.5,0) circle (1pt);\filldraw (.5,0) circle (1pt);
\begin{scope}
\clip (0,0) circle (.5cm);
\draw[scalar] (0,.8) circle (.8cm);
\end{scope}
\draw[thick, postaction={decorate}, decoration={markings, mark=at position 0.5 with {\arrow[rotate=-10]{Stealth}}},rotate=55] (-20:.5cm) arc(-20:120:.5cm);
\draw[thick, postaction={decorate}, decoration={markings, mark=at position 0.5 with {\arrow[rotate=-10]{Stealth}}},rotate=55+180] (-20:.5cm) arc(-20:120:.5cm);
}
+
\tikz[baseline=-3pt,rotate=180]{
\draw[thick] (0,0) circle (.5cm);
\filldraw (-.5,0) circle (1pt);\filldraw (.5,0) circle (1pt);
\begin{scope}
\clip (0,0) circle (.5cm);
\draw[scalar] (0,.8) circle (.8cm);
\end{scope}
\draw[thick, postaction={decorate}, decoration={markings, mark=at position 0.5 with {\arrow[rotate=-10]{Stealth}}},rotate=55] (-20:.5cm) arc(-20:120:.5cm);
\draw[thick, postaction={decorate}, decoration={markings, mark=at position 0.5 with {\arrow[rotate=-10]{Stealth}}},rotate=55+180] (-20:.5cm) arc(-20:120:.5cm);
}
+
\tikz[baseline=-3pt]{
\draw[thick] (0,0) circle (.5cm);
\filldraw (-.5,0) circle (1pt);\filldraw (.5,0) circle (1pt);
\begin{scope}
\clip (0,0) circle (.5cm);
\draw[scalar] (0,-.55) -- (0,.55);
\end{scope}
\draw[thick, postaction={decorate}, decoration={markings, mark=at position 0.5 with {\arrow[rotate=-10]{Stealth}}},rotate=105] (-20:.5cm) arc(-20:120:.5cm);
\draw[thick, postaction={decorate}, decoration={markings, mark=at position 0.5 with {\arrow[rotate=-10]{Stealth}}},rotate=105+80] (-20:.5cm) arc(-20:120:.5cm);
\draw[thick, postaction={decorate}, decoration={markings, mark=at position 0.5 with {\arrow[rotate=-10]{Stealth}}},rotate=105+180] (-20:.5cm) arc(-20:120:.5cm);
\draw[thick, postaction={decorate}, decoration={markings, mark=at position 0.5 with {\arrow[rotate=-10]{Stealth}}},rotate=105-90] (-20:.5cm) arc(-20:120:.5cm);
}
+
\tikz[baseline=-3pt]{
\draw[->,>=Stealth,thick] (.01,.5) -- (-.12,.5);
\draw[->,>=Stealth,thick] (-.01,-.5) -- (.12,-.5);
\draw[thick] (0,0) circle (.5cm);
\filldraw (-.5,0) circle (1pt);\filldraw (2,0) circle (1pt);
\begin{scope}
\clip (.36,-.45) rectangle ++(.77,.9);
\draw[scalar] (.2,-.35) -- (1.3,-.35);
\draw[scalar] (.05,.35) -- (1.3,.35);
\end{scope}
\begin{scope}[shift={(1.5,0)}]
\draw[->,>=Stealth,thick] (.01,.5) -- (-.12,.5);
\draw[->,>=Stealth,thick] (-.01,-.5) -- (.12,-.5);
\draw[thick] (0,0) circle (.5cm);
\end{scope}
}
+
\tikz[baseline=-3pt]{
\draw[->,>=Stealth,thick] (.01,.5) -- (-.12,.5);
\draw[->,>=Stealth,thick] (-.01,-.5) -- (.12,-.5);
\draw[thick] (0,0) circle (.5cm);
\filldraw (-.5,0) circle (1pt);\filldraw (2,0) circle (1pt);
\begin{scope}
\clip (.35,-.25) rectangle ++(.69,.9);
\draw[scalar] (.2,-.55) -- (1.3,.5);
\fill[white] (.76,-.02) circle (2pt);
\end{scope}
\begin{scope}[shift={(0,-.05)}]
\clip (.4,-.32) rectangle ++(.72,.7);
\draw[scalar] (.2,.55) -- (1.3,-.5);
\end{scope}
\begin{scope}[shift={(1.5,0)}]
\draw[->,>=Stealth,thick] (.01,.5) -- (-.12,.5);
\draw[->,>=Stealth,thick] (-.01,-.5) -- (.12,-.5);
\draw[thick] (0,0) circle (.5cm);
\end{scope}
} 
\end{equation}
The first two diagrams are self-energy insertions on the fermion lines,
the third diagram is the vertex correction, and the last two diagrams
contain two scalar exchanges.
Similarly, the logarithmic terms can be extracted using the same
momentum-region argument as in the pure QED$_d$ calculation.
Keeping only the relevant large momentum regions gives the
logarithmic contribution to the 1PI bilinear two-point function.

The scalar correction to the fermion propagator gives
\begin{equation}
\int_k S(p)S(k+p)S(p)G(k)
=
-h\frac{d-2}{d}\mathscr{L}\,S(p) 
\end{equation}
Thus, the first two diagrams contribute
\begin{equation}
-2h\frac{d-2}{d}\mathscr{L}\,M(\Gamma,\bar\Gamma) 
\end{equation}
The third diagram gives
\begin{equation}
-\frac{h}{d}\mathscr{L}\,
M(\Gamma,\gamma^\alpha\bar\Gamma\gamma_\alpha)
+
(\Gamma\leftrightarrow\bar\Gamma) 
\end{equation}
The logarithmic contributions from the last two diagrams cancel. The scalar contribution to the logarithmic correction is therefore
\begin{equation}
\left(
-h\frac{d-2}{d}\mathscr{L}\,M(\Gamma,\bar\Gamma)
-
\frac{h}{d}\mathscr{L}\,
M(\Gamma,\gamma^\alpha\bar\Gamma\gamma_\alpha)
\right)
+
(\Gamma\leftrightarrow\bar\Gamma) 
\end{equation}
For the scalar bilinear $\Gamma=\bar\Gamma=1$, this becomes
\begin{equation}
-\frac{4h(d-1)}{d}\,\mathscr{L}\,M(1,1) 
\end{equation}

There are also mixed diagrams, with one gauge exchange and one scalar
exchange
\begin{equation}\label{eq:five_diagrams_QEDGNY}
\tikz[baseline=-3pt]{
\draw[<-,>=Stealth,thick] (.04,.5) -- (-.13,.5);
\draw[->,>=Stealth,thick] (.13,-.5) -- (-.04,-.5);
\draw[thick] (0,0) circle (.5cm);
\filldraw (-.5,0) circle (1pt);\filldraw (2,0) circle (1pt);
\begin{scope}
\clip (.36,-.45) rectangle ++(.77,.9);
\draw[scalar] (.2,-.35) -- (1.3,-.35);
\draw[photon] (.05,.35) -- (1.3,.35);
\end{scope}
\begin{scope}[shift={(1.5,0)}]
\draw[<-,>=Stealth,thick] (.04,.5) -- (-.13,.5);
\draw[->,>=Stealth,thick] (.13,-.5) -- (-.04,-.5);
\draw[thick] (0,0) circle (.5cm);
\end{scope}
}
+
\tikz[baseline=-3pt]{
\draw[<-,>=Stealth,thick] (.04,.5) -- (-.13,.5);
\draw[->,>=Stealth,thick] (.13,-.5) -- (-.04,-.5);
\draw[thick] (0,0) circle (.5cm);
\filldraw (-.5,0) circle (1pt);\filldraw (2,0) circle (1pt);
\begin{scope}
\clip (.36,-.45) rectangle ++(.77,.9);
\draw[photon] (.2,-.35) -- (1.3,-.35);
\draw[scalar] (.05,.35) -- (1.3,.35);
\end{scope}
\begin{scope}[shift={(1.5,0)}]
\draw[<-,>=Stealth,thick] (.04,.5) -- (-.13,.5);
\draw[->,>=Stealth,thick] (.13,-.5) -- (-.04,-.5);
\draw[thick] (0,0) circle (.5cm);
\end{scope}
}
+
\tikz[baseline=-3pt]{
\draw[<-,>=Stealth,thick] (.04,.5) -- (-.13,.5);
\draw[->,>=Stealth,thick] (.13,-.5) -- (-.04,-.5);
\draw[thick] (0,0) circle (.5cm);
\filldraw (-.5,0) circle (1pt);\filldraw (2,0) circle (1pt);
\begin{scope}
\clip (.35,-.25) rectangle ++(.69,.9);
\draw[scalar] (.2,-.55) -- (1.3,.5);
\fill[white] (.76,-.02) circle (2pt);
\end{scope}
\begin{scope}[shift={(0,-.05)}]
\clip (.4,-.32) rectangle ++(.72,.7);
\draw[photon] (.2,.55) -- (1.3,-.5);
\end{scope}
\begin{scope}[shift={(1.5,0)}]
\draw[<-,>=Stealth,thick] (.04,.5) -- (-.13,.5);
\draw[->,>=Stealth,thick] (.13,-.5) -- (-.04,-.5);
\draw[thick] (0,0) circle (.5cm);
\end{scope}
}
+
\tikz[baseline=-3pt]{
\draw[<-,>=Stealth,thick] (.04,.5) -- (-.13,.5);
\draw[->,>=Stealth,thick] (.13,-.5) -- (-.04,-.5);
\draw[thick] (0,0) circle (.5cm);
\filldraw (-.5,0) circle (1pt);\filldraw (2,0) circle (1pt);
\begin{scope}
\clip (.35,-.25) rectangle ++(.69,.9);
\draw[photon] (.2,-.55) -- (1.3,.5);
\fill[white] (.76,-.02) circle (2pt);
\end{scope}
\begin{scope}[shift={(0,-.05)}]
\clip (.4,-.32) rectangle ++(.72,.7);
\draw[scalar] (.2,.55) -- (1.3,-.5);
\end{scope}
\begin{scope}[shift={(1.5,0)}]
\draw[<-,>=Stealth,thick] (.04,.5) -- (-.13,.5);
\draw[->,>=Stealth,thick] (.13,-.5) -- (-.04,-.5);
\draw[thick] (0,0) circle (.5cm);
\end{scope}
} 
\end{equation}
For general bilinear insertions $\Gamma$ and $\bar\Gamma$, the mixed
diagrams are gauge invariant by themselves. Indeed, the $\xi$-dependent
part of the photon propagator is proportional to
\begin{equation}
k_\mu k_\alpha
\left[
M(\Gamma,\gamma^\mu\gamma^\alpha)
-
M(\Gamma,\gamma^\alpha\gamma^\mu)
\right]
\end{equation}
which vanishes. We may therefore keep only the transverse part of the
photon propagator.

After extracting the logarithmic contribution, the mixed diagrams give
\begin{equation}
\mathcal{M}_{\rm mix}
=
\frac{4(d-3)gh\,\kappa_d(1,1)\mathscr{L}}{d}\,
M(\Gamma,\gamma^{\mu\alpha})\,
\tr[\bar\Gamma\gamma_{\alpha\mu}]
+
(\Gamma\leftrightarrow\bar\Gamma) 
\end{equation}
For $\Gamma=\bar\Gamma=1$, this term vanishes by using $\tr[\gamma_{\alpha\mu}]=0$.
Thus, the mixed diagrams do not contribute to the scalar bilinear at
this order.

The gauge contribution is the same one found in the pure QED$_d$
calculation. Combining the scalar, gauge, and mixed contributions, the logarithmic correction to the 1PI bilinear two-point function is
\begin{equation}\label{eq:gny_full_kernel}
\begin{aligned}
&\left(
-h\frac{d-2}{d}\mathscr L\,M(\Gamma,\bar\Gamma)
-\frac{h}{d}\mathscr L\,
M(\Gamma,\gamma^\alpha\bar\Gamma\gamma_\alpha)
\right)\\
&\quad
+\frac{g\mathscr L}{d}\biggl(
-(d-2)^2M(\Gamma,\bar\Gamma)
+
M(\gamma^\mu\gamma^\alpha\bar\Gamma\gamma_\alpha\gamma_\mu,\Gamma)\\
&\qquad\qquad
-\frac{2}{\tr[1]}
\frac{(d-1)(d-4)}{d-2}
\tr[\Gamma\gamma^{\alpha\mu\nu}]
M(\bar\Gamma,\gamma_{\alpha\mu\nu})
\biggr)\\
&\quad
+\frac{4(d-3)gh\,\kappa_d(1,1)\mathscr L}{d}
M(\Gamma,\gamma^{\mu\alpha})
\tr[\bar\Gamma\gamma_{\alpha\mu}]
+(\Gamma\leftrightarrow\bar\Gamma).
\end{aligned}
\end{equation}
This formula is the QED$_d$--GNY analogue of the logarithmic 1PI
meson result in the pure gauge theory.  

For the scalar bilinear $\Gamma=\bar\Gamma=1$, 
the mixed term vanishes, and one finds
\begin{align}
    M(1,1) \mathscr{L}\left[-\frac{4 h(d-1)}{d}+\frac{4 g(d-1)}{d}\left(2-\frac{d-4}{d-2} \frac{\operatorname{tr}\left(\gamma^{\alpha \mu \nu}\right) \operatorname{tr}\left(\gamma_{\alpha \mu \nu}\right)}{\operatorname{tr}(1)^2}\right)\right]
\end{align}
Thus, the scalar channel in the QED$_d$--GNY model inherits the same
low-rank obstruction as in pure QED$_d$.  For generic $d$ the odd trace
is absent in the stable limit, while at $d=3$ it is nonzero and changes
the logarithmic coefficient.  The Yukawa interaction shifts the scalar
part of the 1PI correction, but it does not remove the branch structure
associated with the exceptional identity $\gamma_{\mu\nu\rho}\sim\epsilon_{\mu\nu\rho}$
in three dimensions.

We close with a caution about the interpretation of this result.  The
combination of a vanishing leading connected correlator and a nonzero
logarithm in the 1PI kernel may look like a logarithmic
CFT. This is not the right conclusion yet.  In the GNY theory,
$\bar\psi\psi$ belongs to a scalar-singlet sector containing operators
such as $\phi$, $\phi^3$, and $\partial^2\phi$.  These operators are not
all independent, since the scalar equation of motion relates them.  Thus,
the correct question is not whether the single object $\bar\psi\psi$ has
a logarithmic two-point function.  One must first quotient by the
equation-of-motion relation and then separate descendant directions from
primary directions.  Only on this reduced space of independent scalar
primaries does it make sense to ask whether the dilatation operator is
diagonalizable, or instead has a Jordan block.  We will not carry out
this operator-mixing problem here.

\section{Conclusions and future directions}\label{sec:conclusions}

In this paper we have argued that analytic continuation in the number of spacetime dimensions $d$ is a meaningful operation, but one that is also remarkably subtle.

We showed that, if we think of $d$ as living in $\mathbb C$, then correlation functions, OPEs, and scaling dimensions of generic operators are all discontinuous, even in theories as simple as the free fermion CFT. On the other hand, we provided evidence that analyticity is recovered if we consider some infinite cover of $\mathbb C$, although this last perspective definitely warrants a much more thorough study. 

A simple application of our results is the following. The QCD theory we studied in section~\ref{sec:results} is known to be dual to a WZW model when $d=2$. Then, the analytic continuation in $d$ we constructed can be interpreted as one specific choice for analytic continuation of this $2d$ theory (see e.g.~\cite{Ma:2019ysf,Nahum:2019fjw} for somewhat recent efforts for such problem). The level of the WZW model is $N_f$, so we only have a continuation for large level. The question of how to construct a non-perturbative continuation is still open, of course.

As stressed in the introduction, analytic continuation in $d$ requires a choice of target for all $d\in\mathbb N$, i.e., it is not an operation that is intrinsically defined for a QFT that lives in one specific $d$. That being said, some families might not have a meaningful continuation if we insist on having a target for \emph{all} $d\in\mathbb N$, but the continuation might exist if we content ourselves with having a target for some congruence class in $\mathbb N$. As a simple illustration, the integer sequence $a_n=(-1)^n$ does not have a nice continuation to the complex $n\in\mathbb C$ plane, but if we unzip $\mathbb N$ into the even and odd integers, then we do get two healthy continuations, namely $a_n=1$ and $a_n=-1$. In this sense, it might be the case that some QFT that is naturally defined for $d=3$ might fail to have a nice continuation if we insist that we want all $d\in\mathbb N$, but the continuation might be well-defined if we only require a target at $d\in 2\mathbb N+1$.

For example, because of Bott periodicity, fermions might have a nicer continuation in $d$ if we consider the eight cases $d\in 8\mathbb N+k$, $k\in\{0,\dots,7\}$ separately. This would allow us, for example, to analytically continue Majorana fermions.

An interesting family of QFTs one might want to analytically continue in $d$ is Chern-Simons(-matter) theories. In the same way we set up the analytic continuation of QED/QCD via the $1/N_f$ expansion, in CS one can perform a $1/k$ expansion, which one could use to try and define a continuation perturbatively. This continuation in $d$ should present all the same subtleties we explored here\footnote{As we mentioned before, generic operators in a chiral theory should behave similarly to parity-odd operators in a vector-like theory. For example, we saw that in regular QCD the parity-odd operator $\bar\psi\psi$ was rather subtle at fractional $d$, but the gauge sector (Wilson lines) were perfectly well-behaved. Once we give up reflection symmetry, the gauge sector should become subtle as well. For instance, without this symmetry, the most general photon two-point function compatible with gauge invariance is $\Pi_{\mu\nu}(p)=(p^2\delta_{\mu\nu}-p_\mu p_\nu)\Pi^\text{even}(p^2)+i p^\sigma\tr[\gamma_{\mu\nu\sigma}]\Pi^\text{odd}(p^2)$. Therefore, if CS theories admit continuation in $d$ at all, the gauge sector will acquire a branched structure in $d$ similar to the one in $\bar\psi\psi$ discussed in this paper.}. As before, defining this continuation requires a choice of target for an infinite family of $d\in\mathbb N$; here one natural choice is to take the $d$-dimensional Chern-Simons form for all $d\in2\mathbb N+1$. 

Although we do not have anything sharp to say about this, it should be useful to think about the complications induced by $\tr[\gamma_{\mu_1\cdots\mu_n}]$ in the language of von Neumann algebras. Indeed, if $Pin(d)$ for fractional $d$ is morally $Pin(\infty)$, then we should perhaps think of the $\gamma^\mu$'s for fractional $d$ as operators acting on some infinite-dimensional space, and their trace should be regularized somehow (while preserving all the properties required by QFT, namely Lorentz invariance, linearity and cyclicity). Looking at our various expressions for $\Delta(\bar\psi\psi)$, we see that observables always depend on these traces via the quotient $\tr[\gamma_{\mu_1\cdots\mu_n}]/\tr[1]$, i.e., these are indeed regularized traces, if only formally.

Another aspect that invites further analysis is the connection between analytic continuation of $d$ and of $N$. One simple remark in this context is that the difficulty in defining traces of $\gamma^\mu$ in fractional $d$ can be reformulated entirely in the language of analytic continuation of flavor symmetries. Indeed, the gamma matrices can equivalently be thought of as Majorana fermions in an auxiliary $0+1d$ system, in terms of which $\tr[\gamma_\mu\gamma_\nu\cdots]=\langle \Psi_\mu\Psi_\nu\cdots\rangle$.\footnote{The fact that $\tr[\gamma_{\mu_1\cdots\mu_n}]$ is non-zero only when $n=d$ has a simple interpretation in this language: the correlation function vanishes due to the $d$ Ramond zero-modes, unless we insert the same number of $\Psi$'s.} As far as these Majorana fermions is concerned, the group $Pin(d)$ is a flavor symmetry.\footnote{More precisely, the flavor symmetry is $O(d)$; the double-cover $Pin(d)$ appears because the vacua of these fermions realize the symmetry projectively. This is just the statement that this flavor symmetry has a 't Hooft anomaly. Note that the correlation functions of $\Psi^\mu$ present unstable behavior even though this operator transforms under the vector representation of $O(d)$, which is stable; this happens because the ground state of the theory is itself unstable.}

Let us also emphasize that the obstructions to analyticity in $d$ or $N$ are always due to low rank exceptions. Indeed, the direct limit of a representation is always analytic, by construction, since this limit is defined by considering only the asymptotic behavior in $d,N\to\infty$, which is smooth. In this sense, the non-analyticity reflects features that are non-perturbative in $1/d$ or $1/N$. It would be nice to understand this better, specifically by drawing insights developed in the context of large $N_c$ gauge theories.\footnote{We would like to thank Justin Kulp for useful discussions regarding this point.} (There, unstable representations are usually called ``heavy'' or ``large''.)

Let us finally go back to one of the original motivations we mentioned in the introduction, namely dim-reg in the context of supersymmetry. Our main conclusion is that analytic continuation in $d$ is very subtle; therefore, if we only care about this operation as a simplifying device for loop computations, then we have nothing to offer: the rigorous way to analytically continue requires more work than dim-reg saves. Conversely, if one is interested in this continuation for its own sake (e.g., can we find an interpolation between $d=3$ and $d=4$ Seiberg-like dualities?), then unfortunately we also have reasons to be pessimistic: supersymmetry itself is, in a sense, a low $d$ exception. It is not clear to us how to set up a healthy SUSY target for all $d\in\mathbb N$. We leave this possibility open for now.

\section*{Acknowledgments}

We would like to thank Sriram Bharadwaj, Gabriel Cuomo, Thomas Dumitrescu, Justin Kulp, Adar Sharon, and Anna Wolz for helpful conversations, and again Justin Kulp for useful comments on a first draft. D.R.~would like to thank Michael Gutperle for helpful discussions. The work of D.D.~was supported in part by U.S. Department of Energy award DE-SC0009937. Y.L.~was supported in part by U.S. Department of Energy award DE-SC0009937 and by the Simons Collaboration on Global Categorical Symmetries. We are grateful to the Bhaumik Institute for Theoretical Physics for support.

\appendix

\section{Background}\label{sec:background}

In this appendix we lay down our conventions and review some background. 

\subsection{Useful integrals}\label{ap:integrals}

We will encounter various expressions like
\begin{equation}
\int_k \frac{1}{k^{2a}(k+p)^{2b}}
\end{equation}
where $\int_k$ is a formal operation, which agrees with $\int\frac{\mathrm d^dk}{(2\pi)^d}$ when $d$ is an integer. In this discussion we will assume that $a$ and $b$ are such that the integral is absolutely convergent.

We can evaluate our integral using, for example, Feynman's parametrization
\begin{equation}
\frac{1}{k^{2a}(k+p)^{2b}}=\frac{\Gamma(a+b)}{\Gamma(a)\Gamma(b)}\int_u\frac{u^{a-1}(1-u)^{b-1}}{(u k^2+(1-u)(k+p)^2)^{a+b}}
\end{equation}
where $\int_u=\int_0^1\mathrm du$. We now shift $k\to k-(1-u)p$ so that the denominator becomes $k^2+u(1-u)p^2$, which is rotationally invariant, and write
\begin{equation}
\begin{aligned}
\int_k \frac{1}{k^{2a}(k+p)^{2b}}&=\frac{\Gamma(a+b)}{\Gamma(a)\Gamma(b)}\frac{2\pi^{d/2}}{\Gamma(d/2)}\frac{1}{(2\pi)^d}\\
&\qquad\times\int_uu^{a-1}(1-u)^{b-1}\int_0^\infty \frac{k^{d-1}\mathrm dk}{(k^2+u(1-u)p^2)^{a+b}}
\end{aligned}
\end{equation}
where the last expression is a regular one-dimensional integral. This integral is just a beta function:
\begin{equation}
\begin{aligned}
&=\frac{\Gamma(a+b)}{\Gamma(a)\Gamma(b)}\frac{\Gamma(d/2) \Gamma(a+b-d/2)}{2 \Gamma (a+b)}\frac{2\pi^{d/2}}{\Gamma(d/2)}\\
&\qquad \frac{1}{(2\pi)^d}\frac{1}{p^{2(a+b)-d}}\int_uu^{a-1}(1-u)^{b-1}(u (1-u))^{-a-b+d/2}
\end{aligned}
\end{equation}
and the $u$ integral is also straightforward:
\begin{equation}
=\kappa_d(a,b)\frac{1}{p^{2(a+b)-d}}
\end{equation}
where
\begin{equation}\label{eq:kappa}
\kappa_d(a,b):=\frac{1}{(4\pi)^{d/2}}\frac{\Gamma(a+b-d/2)\Gamma(d/2-a) \Gamma(d/2-b)}{\Gamma (d-a-b)\Gamma(a)\Gamma(b)}
\end{equation}

Tensor integrals can be evaluated using the same technique, and we end up with
\begin{equation}\label{eq:master_integrals}
\begin{aligned}
\int_k \frac{1}{k^{2a}(k+p)^{2b}}&=\kappa_d(a,b)\frac{1}{p^{2(a+b)-d}}\\
\int_k \frac{k^\alpha}{k^{2a}(k+p)^{2b}}&=\kappa_d(a,b)\frac{d-2a}{2(d-a-b)}\frac{-p^\alpha}{p^{2(a+b)-d}}\\
\int_k \frac{k^\alpha k^\beta}{k^{2a}(k+p)^{2b}}&=\kappa_d(a,b)\frac{d-2 a}{4 (d-a-b)(d+1-a-b) }\\
&\qquad\frac{-\frac{d-2 b}{d+2 (1-a-b)}\delta^{\alpha\beta}p^2+(d+2-2 a) p^\alpha p^\beta}{p^{2(a+b)-d}}\\
\int_k \frac{k^\alpha k^\beta k^\gamma}{k^{2a}(k+p)^{2b}}&=\kappa_d(a,b)\frac{(d-2 a)(d+2-2 a) }{8(d-a-b) (d+1-a-b)(d+2-a-b)  }\\
&\qquad\frac{\frac{d-2 b}{d+2 (1-a-b)}(p^2\delta^{\alpha\beta}p^\gamma +\text{perm.})-(d+4-2 a) p^\alpha p^\beta p^\gamma}{p^{2 (a+b)-d}}
\end{aligned}
\end{equation}
etc.

An important limiting case is $a+b\to d/2$, since in this case the integral loses its scale, and it becomes logarithmically divergent. Either by explicitly cutting off the integral $k\le\Lambda$, or by taking the limit $a+b\to d/2$ in the previous expressions and identifying the pole $\frac{1}{a+b-d/2}\to\log(\Lambda^2)$, we find
\begin{equation}
\int_k\frac{1}{k^{2a}(k+p)^{d-2a}}=\mathscr L
\end{equation}
where
\begin{equation}
\mathscr L:=\frac{1}{(4\pi)^{d/2}\Gamma(d/2)}\log(\Lambda^2)=\text{``}\kappa_d(a,b)|_{a+b=d/2}\text{''}
\end{equation}

For some purposes it is also useful to perform a multipole (or, more precisely, Gegenbauer) expansion
\begin{equation}
\begin{aligned}
\frac{1}{(p+k)^\alpha}&=\frac{1}{k^\alpha}-\alpha\frac{k\cdot p}{k^{\alpha+2}}+\frac12\alpha\frac{(\alpha+2)(k\cdot p)^2-k^2p^2}{k^{\alpha+4}}\\
&+\frac12\alpha(\alpha+2)\frac{-\frac13(\alpha+4)(k\cdot p)^3+k^2p^2 k\cdot p}{k^{\alpha+6}}+\cdots
\end{aligned}
\end{equation}

\subsection{QCD$_d$ at large $N_f$}\label{sec:review_nf}

Here we review the basics of the $1/N_f$ expansion, following~\cite{Chester:2016ref}. We begin with the abelian case $G=U(1)$, and then point out the differences that arise when dealing with more general gauge groups at the end.

The lagrangian of the theory in Euclidean signature is given by
\begin{equation}
L=\frac{1}{4e^2} f^{\mu \nu}f_{\mu \nu} - \bar{\psi} \gamma^\mu (\partial_\mu + i a_\mu) \psi
\end{equation}
with $\gamma^\mu$ satisfying~\eqref{eq:cliff}.

The coupling $e^2$ is dimensionful, it has mass-dimension $[e^2]=4-d$. Therefore, at least as far as the $1/N_f$ expansion is concerned, this theory has two fixed-points: $e^2\to0$ and $e^2\to\infty$. The direction of the flow depends on the sign of $4-d$. Although we define these CFTs for all $d\in\mathbb N$, we will call $e^2\to0$ the short-distance CFT, and $e^2\to\infty$ the long-distance CFT, for all $d$, even though this terminology is correct for $d<4$ only.

From the lagrangian we read off the tree-level propagators. In momentum space, the fermion propagator is $S(p)=-\frac{i\slashed p}{p^2}$. Up to gauge-fixing terms, the free photon propagator is $\Delta^0_{\mu\nu}=\frac{e^2}{p^2}\delta_{\mu\nu}$. The interaction vertex is $-i\gamma^\mu$. We will tune the theory to its non-trivial ``low-energy'' CFT; this fixed point is non-perturbative in $e^2$ but weakly coupled in $1/N_f$.

If we denote the photon self-energy by $\Pi_{\mu\nu}(p)=\Pi(p^2)(p^2\delta^{\mu\nu}-p^\mu p^\nu)$, then the exact photon two-point function reads
\begin{equation}
\begin{aligned}
\Delta_{\mu\nu}&=\frac{e^2}{p^2}\delta_{\mu\nu}+\frac{e^4}{p^4}\Pi_{\mu\rho}+\frac{e^6}{p^6}\Pi_{\mu\rho}\Pi_{\rho\nu}+\cdots\\
&=\frac{e^2}{p^2}\delta_{\mu\nu}(1+e^2\Pi+e^4\Pi^2+\cdots)+\text{gauge}\\
&=\frac{e^2\delta_{\mu\nu}}{p^2(1-e^2\Pi)}+\text{gauge}
\end{aligned}
\end{equation}

At the fixed point $e^2\to\infty$, the photon two-point function equals $-\frac{\delta_{\mu\nu}}{p^2\Pi(p^2)}$, again up to gauge-fixing terms. We can solve for $\Pi(p^2)$ analytically in the limit $N_f\gg1$. To leading order in $1/N_f$, the self-energy corresponds to the fermion bubble
\begin{equation}\label{eq:amp_Pi}
\begin{aligned}
\tikz[baseline=-3pt]{
\draw[->,>=Stealth,thick] (.01,.5) -- (-.12,.5);
\draw[->,>=Stealth,thick] (-.01,-.5) -- (.12,-.5);
\draw[thick] (0,0) circle (.5cm);
\filldraw (-.5,0) circle (1pt);\filldraw (.5,0) circle (1pt);
\node[scale=.8] at (-1.2,0) {$\mu$};
\node[scale=.8] at (1.2,0) {$\nu$};
\begin{scope}
\clip (-1,-.2) rectangle ++(0.5,0.5) (.5,-.2) rectangle ++(0.5,0.5);
\draw[photon] (-1.15,0) -- (-.2,0);
\draw[photon] (1.15,0) -- (.3,0);
\end{scope}
\node[scale=.8] at (0,.8) {$q$};
}\biggr|_\text{amp}\ &=-\int_q \tr[S(q)(-i\gamma_\nu)S(q+p)(-i\gamma_\mu)]\\
&=-\tr[\gamma^\alpha\gamma^\nu\gamma^\beta\gamma^\mu]\int_q\frac{q_\alpha(q_\beta+p_\beta)}{q^2(q+p)^2}
\end{aligned}
\end{equation}
Next, we use the trace identity~\eqref{eq:master_trace} together with the standard $d$-dimensional integrals (collected in appendix~\ref{ap:integrals} for reference), and find (see~\eqref{eq:kappa} for the definition of $\kappa_d(a,b)$)
\begin{equation}
\Pi_{\mu\nu}=\tr[1]\frac{\kappa_d(1,1)}{p^{4-d}}\frac{2-d}{2(d-1)}(p^2\delta^{\mu\nu}-p^\mu p^\nu)
\end{equation}
which implies that the one-loop corrected photon propagator is
\begin{equation}
-\frac{\delta_{\mu\nu}}{p^2\Pi(p^2)}=\frac{g\delta_{\mu\nu}}{p^{d-2}}
\end{equation}
where the effective coupling is $g:=\frac{1}{\tr[1]}\frac{2}{\kappa_d(1,1)}\frac{d-1}{d-2}$. More explicitly, using~\eqref{eq:kappa} for $\kappa_d(1,1)$, the effective coupling reads
\begin{equation}
g=\frac{(4\pi)^{d/2}}{\tr[1]}\frac{2\Gamma (d-2)}{\Gamma(2-d/2)\Gamma(d/2-1)^2}\frac{d-1}{d-2}
\end{equation}
Echoing our discussion in subsection~\ref{sec:gamma}, we see that the counting parameter is not really $N_f$, but $\tr[1]\sim N_f\times 2^{\lfloor d/2\rfloor}$ instead.

To the one-loop resummed propagator above we add by hand a (non-local) gauge fixing term $S_{\mathrm{gf}}=\frac{1}{2 g(1+\xi)} \int_p |p|^{d-4} p^\mu  p^\nu a_\mu(p) a_\nu(-p)$, so our propagators read
\begin{equation}
\begin{aligned}
\tikz[baseline=-3pt]{
\begin{scope}
\clip (0,-.2) rectangle ++(1.5,0.5);
\draw[fermion] (0,0)--(1.8,0);
\end{scope}
\node[scale=.8] at (.1,-.3) {\phantom{$\mu$}};
\node[scale=.8] at (1.4,-.3) {\phantom{$\nu$}};
}\ &=-i\frac{\slashed p}{p^2}\\[+1ex]
\tikz[baseline=-3pt]{
\begin{scope}
\clip (0,-.2) rectangle ++(1.5,0.5);
\draw[photon] (0,0)--(1.8,0);
\end{scope}

\node[scale=.8] at (.1,-.3) {$\mu$};
\node[scale=.8] at (1.4,-.3) {$\nu$};
}\ &=\frac{g}{p^{d-2}}(\delta^{\mu\nu}+\xi p^\mu p^\nu/p^2)
\end{aligned}
\end{equation}

Then, the $1/N_f$ expansion looks like standard perturbation theory, but using an effective photon propagator with non-standard, conformal scaling $1/p^{d-2}$ (as opposed to the short distance scaling $1/p^2$). But do note that this is not a \emph{loop} expansion: the counting parameter is not just photon lines, as is usually the case, but rather photon lines minus fermion loops. Moreover, we never add corrections on top of a photon line, as those are already accounted for by using the exact photon propagator. Furthermore, to higher orders in $1/N_f$, there are new effective vertices one must consider; for example, the ``light-by-light'' box diagram generates a quartic coupling with four photon legs.


The non-abelian case is rather similar, but it requires the introduction of ghosts, and cubic and quartic gluonic vertices. Here we will be working to first non-trivial order in $1/N_f$, where these complications will not play a role, and we can just recycle the abelian result by multiplying by suitable color factors. For example, the gluonic self-energy~\eqref{eq:amp_Pi} acquires a factor $\tr_R[t^at^b]$, where $R$ is the representation under which the fermions transform, and $a,b$ are adjoint indices. This equals $T(R)\delta^{ab}$, where $T(R)$ is the Dynkin index of $R$; therefore, the QCD effective coupling is
\begin{equation}\label{eq:g_QCD}
g_\text{QCD}=g_\text{QED}\times \frac{1}{T(R)}
\end{equation}
Other similar color factors will be pointed out as they arise. Note that the factor of $1/T(R)$ implies that the perturbative expansion should be reliable even if $N_f$ is small, as long as $R$ is large. Indeed, fermion loops are a trace over flavor, but also over color and spin, and therefore the counting parameter is $N_f\times T(R)\times 2^{\lfloor d/2\rfloor}$.

We end this quick review with some final remarks:
\begin{itemize}
\item We expect the limit $d\to2$ to be subtle because the four-fermi operator $(\bar\psi\psi)^2$ hits tree-level marginality in this dimension, so one might expect BKT-like behavior (see~\cite{Cherman:2019hbq,Komargodski:2020mxz,Delmastro:2022prj,Cherman:2024onj} for some recent studies of four-fermi interactions in two-dimensional gauge theories). The limit $d\to1$ is even more complicated since in this case we have an infinite tower of operators $(\bar\psi\psi)^n$ which all have $\Delta=0$. (This tower of course truncates at finite $N_f$).
\item We restrict ourselves to $\psi$ and $\bar\psi\psi$ because these are the lightest operators (not counting $f_{\mu\nu}$). For quartics and higher-order composites we would have to worry about operator mixing, which makes the analysis more cumbersome. For example, the operators schematically of the form $(\bar\psi\psi)^n \partial^{2k}$ are all tree-level degenerate with $(\bar\psi\psi)^nf^k$.
\item Furthermore, there are operators that are not tree-level degenerate for generic $d$, but they do become degenerate at special values of this parameter. For example, $\Delta^{(0)}((\bar\psi\psi)^{2n})=2n(d-1)$ and $\Delta^{(0)}(f^{2n})=4n$. These are different for $d\neq3$, but there is level crossing in three dimensions. Naive perturbation theory, where these are treated as separate operators, breaks down around $d=3$~\cite{Korchemsky:2015cyx}.
\item Of course, these matters are complicated even further by the presence of evanescent operators, as we already explained in the introduction. These are key in solving for quartics and higher order composites, but will not play an important role for the low lying $\psi$ and $\bar\psi\psi$.
\end{itemize}

\subsection{Conformal perturbation theory}\label{sec:cpt}

Here we recall some basic concepts about conformal perturbation theory that we shall need in our $1/N_f$ expansion, again following~\cite{Chester:2016ref}.

Recall that the position-space two-point function of a given (primary) operator $\mathcal O(x)$ is
\begin{equation}
\langle \mathcal O(x)\mathcal O(0)\rangle\propto x^{-2\Delta}
\end{equation}
The Fourier transform of this is
\begin{equation}
\int \mathrm d^d x\ e^{ipx}\frac{1}{x^{2\Delta}}\propto \frac{1}{p^\alpha}
\end{equation}
where $\alpha$ is determined by dimensional analysis, namely $\alpha=d-2\Delta$.

In conformal perturbation theory, where $\Delta=\Delta^{(0)}+\Delta^{(1)}+\cdots$, we have
\begin{equation}
\frac{1}{p^{d-2\Delta}}=\frac{1}{p^{d-2\Delta^{(0)}}}\biggl(1+\Delta^{(1)}\log(p^2)+\cdots\biggr)
\end{equation}
In other words, we can read off $\Delta^{(1)}(\mathcal O)$ from the coefficient of $\log(p^2)\in \langle \mathcal O(p)\mathcal O(-p)\rangle$. Of course, this is the same as the coefficient of $-\log(\Lambda^2)$, where $\Lambda$ is the dimensionful scale introduced by the regularization scheme.

Things become more complicated when there are multiple operators with the same tree level dimension $\Delta^{(0)}$, in which case one must perform degenerate perturbation theory instead. The basic recipe is as follows. Say we have scalar operators $\mathcal O_i$, all of which have the same $\Delta^{(0)}$, and we wish to compute the first non-trivial correction $\Delta^{(1)}$. Then, we compute the matrix of two-point functions
\begin{equation}
\langle\mathcal O_i(p)\mathcal O_j(-p)\rangle=p^{-d+2\Delta^{(0)}}(\mathcal M^{(0)}_{ij}+\log(p^2)\mathcal M^{(1)}_{ij}+\cdots)
\end{equation}
The $\Delta^{(1)}$'s are given by the generalized eigenvalues of $\mathcal M^{(1)}$ with respect to $\mathcal M^{(0)}$. If $\mathcal M^{(0)}$ is invertible, these are just the regular eigenvalues of $(\mathcal M^{(0)})^{-1}\mathcal M^{(1)}$. If the set $\{\mathcal O_i\}$ is not a basis (for example, if we accidentally included the same operator twice, or if there are evanescent operators and we tune $d$ to be an integer), then $\mathcal M^{(0)}$ will not be full rank, but the generalized eigenvalue problem is still well-defined.

\subsection{QCD in $d=1$}\label{sec:1d}

In the main text we perform various computations in QCD for arbitrary $d$. Here we review the known answer in $d=1$ so as to have something to compare our general results with. We begin with abelian gauge group, namely $G=U(1)$, and generalize afterwards.

We take $N_f$ Dirac fermions, and write down a Lagrangian
\begin{equation}
L=\bar\psi(\partial_t+ia_t)\psi-ika_t
\end{equation}
where $k$ is the CS level -- imaginary in euclidean signature, and assumed non-negative for definiteness. Whether it is an integer or a half-integer depends on the parity of $N_f$.

The gauge field has no kinetic term so it is just a Lagrange multiplier. The extended Hilbert space is the usual $2^{N_f}$ dimensional Fock space, and the physical space is given by the Gauss/filling constraint ${:}\bar\psi\psi{:}=k$. Given that ${:}\bar\psi\psi{:}=\bar\psi\psi-N_f/2$, this becomes $\bar\psi\psi=k+N_f/2$, explaining the correlation between the parity of $N_f$ and the fractional part of $k$. The dimension of the physical Hilbert space is simply the number of operators of the form $\psi_{i_1}\psi_{i_2}\cdots\psi_{i_n}$ with $i_1<i_2<\cdots<i_n$, where $n=k+N_f/2$. In other words, the partition function is $Z_k=\binom{N_f}{k+N_f/2}$.

Therefore, the expectation value of a Wilson loop is
\begin{equation}\label{eq:W_qed_1d}
\begin{aligned}
\langle W(\bigcirc)\rangle&=\frac{Z_{k+1}}{Z_k}\\
&=\frac{\frac12N_f-k}{\frac12 N_f+k+1}\\
&\overset{k=0}=1-\frac{2}{N_f}+\mathcal O(N_f^{-2})
\end{aligned}
\end{equation}
where in the last line we restrict to the parity symmetric case $k=0$, which is relevant to our more general QCD$_d$ problem. As a perhaps interesting thing to note, the large $N_f$ expansion of this expectation value converges for all $N_f>2(k+1)$, vaguely reminiscent of similar bounds in $d=3$ Chern-Simons-matter dualities.

One can also compute various correlation functions, such as $\langle\bar\psi(t)W\psi(0)\rangle$, with $W$ a Wilson line connecting $\bar\psi(t)$ to $\psi(0)$. But in this theory the Hamiltonian vanishes, and therefore correlation functions are independent of $t$:
\begin{equation}\label{eq:Delta_psi_1}
\langle\bar\psi(t)W\psi(0)\rangle=\text{const.}
\end{equation}
up to contact terms. The same is of course true for any operator in this theory, such as $\bar\psi\psi$.

We can also discuss condensates. The simplest one is the \emph{vev} of $\bar\psi\psi$, since by definition all states have $\bar\psi\psi=k+N_f/2$. So $\langle i|\bar\psi\psi|j\rangle=(k+\frac12N_f)\delta_{ij}$ where $i,j$ run over the $\binom{N_f}{k+N_f/2}$ vacua of the theory. Focusing on the parity-symmetric case $k=0$, and averaging over the degenerate vacua, we simply find
\begin{equation}\label{eq:vev_qed}
\langle\bar\psi\psi\rangle=\tfrac12N_f
\end{equation}

We now sketch the generalization to non-abelian gauge groups. We take this group $G$ to be semi-simple, and the fermions to transform under some complex representation $R\in\operatorname{Rep}(G)$. As above, the problem can be reduced to counting gauge singlets in $\wedge^\bullet R$ or, more generally in presence of a Wilson line $W_\rho$, counting singlets in $\rho\otimes\wedge^\bullet R$. In other words,\footnote{This is the thermal expectation value. One can also consider the Ramond expectation value, which amounts to the replacement $\det(1+g)\to\det(1-g)$. If $(-1)^F\in G$, the NS and R partition functions agree.
}
\begin{equation}
Z_\rho=\int_G\mathrm dg\, \tr_\rho(g)\det{}_{\!R}(1+g)
\end{equation}
where $\mathrm dg$ is the normalized Haar measure on $G$. 

We note in passing that, if $\operatorname{ker}(R)$ is non-zero (i.e., if the theory has a one-form symmetry), then $Z_\rho$ vanishes unless the charge of $R$ divides the charge of $\rho$~\cite{Delmastro:2022pfo}.

We can expand $Z_\rho$ at large $N_f$ using the Laplace method of steepest descent. To this end, we write
\begin{equation}
Z_\rho=\int_G\tr_\rho(g)e^{N_f S(g)}\mathrm dg,\qquad S(g):=\log \det{}_{\!R}(1+g)
\end{equation}
The ``action'' $S$ is maximized at $g=1$, so we expand around this point, i.e., around the Lie algebra of $G$. We write $g=e^{i x}$ and note that
\begin{equation}
\begin{aligned}
\log\det{}_{\!R}(1+g)&=\dim(R)\log 2-\frac18\tr_R(x^2)+\cdots\\
\tr_\rho(g)&=\dim(\rho)-\frac12\tr_\rho(x^2)+\cdots\\
\frac{\mathrm dg}{\mathrm dx}&=1+\frac{1}{12}\tr_\text{adj}(x^2)+\cdots
\end{aligned}
\end{equation}
and therefore
\begin{equation}
Z_\rho\approx\int_{\mathfrak g}(\dim(\rho)-\tfrac12\tr_\rho(x^2))(1+\tfrac{1}{12}\tr_\text{adj}(x^2))e^{-N_f\frac18\tr_R(x^2)}\mathrm dx
\end{equation}
up to $\rho$-independent factors (which drop out in normalized expectation values). Finally, noting that $\tr_R(x^2)=T(R)\vec x\cdot\vec x$, this becomes a standard Gaussian integral with result
\begin{equation}
Z_\rho=\dim(\rho)+\frac{2\dim(G)}{N_fT(R)}(\tfrac{1}{6}\dim(\rho)T(\text{adj})-T(\rho))+\mathcal O(N_f^{-2})
\end{equation}
again up to $\rho$-independent factors. Therefore, the expectation value of the Wilson line is
\begin{equation}\label{eq:W_1d_QCD}
\langle W_\rho\rangle=\frac{Z_\rho}{Z_1}=\dim(\rho)-\frac{2T(\rho)\dim(G)}{T(R)N_f}+\mathcal O(N_f^{-2})
\end{equation}
It would be interesting to analyze more carefully the radius of convergence of this expansion. Similarly, it could also be interesting to study a double-scaling limit in which $\rho$ grows with $N_f$ (such as $\rho=\rho_0^{\otimes N_f}$); this would involve a non-trivial saddle.

\subsection{QCD in $d=2$}\label{sec:2dQCD}

In the main text we perform various computations in QCD for arbitrary $d$. Here we review the known answer in $d=2$ so as to have something to compare our general results with. We begin with an Abelian gauge group, namely $G=U(1)$, and generalize afterwards.

The solution of QED$_2$ is textbook material by now; here we follow~\cite{Kim:1999wn}\footnote{D.D.~would like to thank Yin-Chen He for explaining this derivation to him.}. We work in a basis where $\gamma^1=\sigma_x$, $\gamma^2=\sigma_y$, so that $\gamma^\star=i\gamma^1\gamma^2=-\sigma_z$. Therefore, we can write $\psi=\frac{1}{\sqrt2}\begin{pmatrix} \psi_L\\\psi_R \end{pmatrix}$ where $\psi_L$ is left-handed and $\psi_R$ is right-handed.

We take $N_f$ Dirac fermions, and write down the Lagrangian
\begin{equation}
L=\psi_L^\dagger(\partial_z+ia_z)\psi_R+\psi_R^\dagger(\partial_{\bar z}+ia_{\bar z})\psi_L-\frac{2}{e^2}(\partial_z a_{\bar z}-\partial_{\bar z}a_z)^2
\end{equation}
where $z=x^1+ix^2$ and $a_z=\frac12(a_1-ia_2)$.
Next, we use the Hodge decomposition to write  $a=\star\mathrm d\phi+\mathrm d\chi$ for a pair of scalars $\phi,\chi$ (the expression for $a$ is valid up to harmonic forms, which do not matter in $\mathbb R^2$). We fix the gauge by setting $\chi=0$, namely $a_z=-i\partial_z\phi$, in terms of which
\begin{equation}
L=\psi_L^\dagger(\partial_z-\partial_z\phi)\psi_R+\psi_R^\dagger(\partial_{\bar z}+ \partial_{\bar z}\phi)\psi_L
+\frac{1}{2e^2}(\partial^2\phi)^2
\end{equation}

The point of this discussion is that, if we now define a new pair of fermions $\xi_L=e^{\phi}\psi_L$ and $\xi_R=e^{-\phi}\psi_R$, then the fermion kinetic term becomes
\begin{equation}
\xi_L^\dagger \partial_z\xi_R+\xi^\dagger_R\partial_{\bar z}\xi_L
\end{equation}
which is just the standard free-fermion kinetic term -- the field redefinition decouples the fermion from the photon. That being said, this change of variables has another important effect: due to the chiral anomaly, it generates a theta-like term for $a$, except that the angle is $\phi$ rather than a constant. So the photon kinetic term reads
\begin{equation}
\frac{1}{2e^2}(\partial^2\phi)^2-\frac{N_f}{2\pi} \phi\partial^2\phi
\end{equation}

The final conclusion is that the original, interacting gauge theory is equivalent to the free, decoupled theory of a fermion $\xi$, plus a scalar $\phi$, the latter having a higher derivative kinetic term. (One could worry that this new presentation has ghosts; these are, in fact, removed by the leftover gauge invariance that allows shifting $a$ by harmonic forms.) The propagators in the decoupled frame are
\begin{equation}
\langle \xi^i_L(z)\bar\xi^j_R(0)\rangle=\frac{\delta^{ij}}{\pi z},\qquad 
\end{equation}
as well as
\begin{equation}
\langle \phi(x)\phi(0)\rangle=\int_k e^{ikx}\frac{1}{\frac{1}{e^2}k^4+\frac{N_f}{\pi} k^2}=-\frac{1}{2N_f}K_0\bigl(ex\sqrt{\tfrac{N_f}{2\pi}}\bigr)-\frac{1}{2N_f}\log (\mu  x)
\end{equation}
We are interested in the short-distance limit $x\to0$, in which $\langle\phi(x)\phi(0)\rangle\to \mathcal O(1)$, and the long-distance limit $x\to\infty$, in which $\langle\phi(x)\phi(0)\rangle\to -\frac{1}{2N_f}\log(\mu x)$. Note that the vertex operator $e^{q\phi}$ has two-point function $\langle e^{q\phi(x)}e^{-q\phi(0)}\rangle = \exp(\frac{q^2}{2}\langle(\phi(x)-\phi(0))^2\rangle)$, which is $\mathcal O(1)$ at short distances, and $x^{\frac{q^2}{2N_f}}$ at long distances.

With these considerations in mind, we find
\begin{equation}
\begin{aligned}
\langle\bar\psi\psi(x)\bar\psi\psi(0)\rangle&=\frac14\langle (\bar\psi_L\psi_L+\bar\psi_R\psi_R)(x)(\bar\psi_L\psi_L+\bar\psi_R\psi_R)(0)\rangle\\
&=\frac14\langle e^{-2\phi(x)}e^{2\phi(0)}\rangle\langle (\bar\xi_L\xi_L)(x)(\bar\xi_R\xi_R)(0)\rangle\\
&+\frac14\langle e^{2\phi(x)}e^{-2\phi(0)}\rangle\langle(\bar\xi_R\xi_R)(x)(\bar\xi_L\xi_L)(0)\rangle
\end{aligned}
\end{equation}
which can be evaluated using Wick's theorem, since in these variables the theory is quadratic. At short distances we find
\begin{equation}
\langle\bar\psi\psi(x)\bar\psi\psi(0)\rangle\propto x^{-2}
\end{equation}
which implies $\Delta_\text{UV}(\bar\psi\psi)=1$, consistent with asymptotic freedom. On the other hand, at long distances we find
\begin{equation}\label{eq:Delta_psi_2}
\langle\bar\psi\psi(x)\bar\psi\psi(0)\rangle\propto x^{-2(1-1/N_f)}
\end{equation}
which implies $\Delta_\text{IR}(\bar\psi\psi)=1-1/N_f$.

We now consider a circular Wilson loop. Writing the gauge field in terms of $\phi(x)$ we find
\begin{equation}
\begin{aligned}
\langle a_\mu(x)a_\nu(0)\rangle&\sim \partial^2\log(x^2)\\
\end{aligned}
\end{equation}
This is a total derivative, and therefore, up to contact terms (i.e., up to a perimeter counterterm) the expectation value of the Wilson loop equals
\begin{equation}\label{eq:W_qed_2d}
\langle W\rangle\equiv 1
\end{equation}


Consider now a straight Wilson line from $0$ to $x=(R,0)$ parameterized by $y^\mu(s) = (s,0)$ with $0\leq s \leq R.$ We have
\begin{equation}
\begin{aligned}
W(R,0) &= \exp \biggl( -i\int a_\mu \mathrm dy^\mu \biggr) \\
&= \exp \biggl( -i\int_0^R a_1(s,0) \mathrm ds \biggr)\\
&= \exp \biggl( i\int_0^R \partial_2 \phi(s,0) \mathrm ds \biggr)
\end{aligned}
\end{equation}
using the gauge field parameterization $a_1 = -\partial_2 \phi$, $a_2 = \partial_1 \phi.$ 
We can now compute the correlation function of the Wilson line connecting the fermion insertions by making use of 
\begin{equation}
\begin{aligned}
\langle \psi^i_L(R) W(R,0) \psi_R^{\dagger j}(0) \rangle & = \langle \xi^i_L (R) \xi_R^{\dagger j}(0)\rangle \; \langle e^{B[\phi]} \rangle \\
    &= \frac{\delta^{ij}}{\pi R} \langle e^{B[\phi]} \rangle 
\end{aligned}
\end{equation}
where 
\begin{equation}
B[\phi]
=
\underbrace{
-\phi(R)+\phi(0)
}_{B_\text{end}}
+
\underbrace{
i\int_0^R \mathrm ds\,\partial_2\phi(s)
}_{ B_\text{line}} 
\end{equation}
Since the Lagrangian is quadratic in $\phi$ and $B[\phi]$ is linear in $\phi,$ we have that
\begin{equation}
\begin{aligned}
\langle  e^{B[\phi]} \rangle &= \exp \biggl(\frac{1}{2}{\langle B[\phi]^2 \rangle }\biggr)\\
    &= \exp \biggl(  \frac{1}{2}\langle B^2_\mathrm{end} \rangle + \langle B_\mathrm{end} B_\mathrm{line} \rangle  + \frac{1}{2} \langle B^2_\mathrm{line} \rangle  \biggr)
\end{aligned}
\end{equation}
Defining $\langle \phi(x) \phi(y) \rangle \equiv D_\phi(x-y), $ we see that
\begin{align}
\frac{1}{2}\langle B^2_\mathrm{end} \rangle
= D_\phi(0) - D_\phi(R)
= \frac{1}{2N_f}\biggl[\gamma_E+\log\biggl(\frac{m_\gamma R}{2}\biggr)+K_0(m_\gamma R)\biggr].
\end{align}
In the UV limit with $R \rightarrow 0$, we get $\frac{1}{2}\langle B^2_\mathrm{end} \rangle = \mathcal{O}(1).$ In the IR limit with $R \rightarrow \infty$, the Bessel function dies exponentially and the $R$-dependent piece is
\begin{equation}
\frac{1}{2}\langle B^2_\mathrm{end} \rangle_{\mathrm{IR}} = \frac{1}{2N_f} \log R .
\end{equation}
The cross term is
\begin{equation}
\langle B_\mathrm{end} B_\mathrm{line} \rangle = i \int_0^R \mathrm ds \left( \langle \phi(0) \partial_2 \phi(s) \rangle - \langle \phi(R) \partial_2 \phi(s) \rangle  \right) .
\end{equation}
This vanishes by reflection symmetry $x_2\to -x_2.$ The final term that we need is
\begin{equation}
\begin{aligned}
\frac{1}{2} \langle B^2_\mathrm{line} \rangle 
&= - \frac{1}{2} \int_0^R \mathrm ds \int_0^R \mathrm ds' \; \langle \partial_2 \phi(s) \partial_2 \phi(s') \rangle \\
&= \frac{1}{2} \int_0^R \mathrm ds \int_0^R \mathrm ds' \; \partial^2_y D_\phi(s-s',y)|_{y=0} \\
&= -\frac{1}{2N_f} \int_0^R \mathrm du \; (R-u) \biggl( \frac{1}{u^2} - \frac{m_\gamma K_1(m_\gamma u)}{u} \biggr)
\end{aligned}
\end{equation}
where $u \equiv |s-s'|,$ and we have used
\begin{equation}
D_\phi(u,y) = -\frac{1}{2N_f}\biggl[\log(\mu \sqrt{u^2+y^2})+K_0(m_\gamma \sqrt{u^2+y^2})\biggr],
\qquad 
m_\gamma^2 \equiv \frac{N_f e^2}{\pi}.
\end{equation}
For $u \ll 1/m_\gamma,$ we get
\begin{equation}
\frac{1}{u^2} - \frac{m_\gamma K_1(m_\gamma u)}{u}= -\frac{m_\gamma^2}{2} \biggl( \log\biggl( \frac{m_\gamma u}{2} \biggr) + \gamma_E - \frac{1}{2} \biggr) + \dots
\end{equation}
so the $1/u^2$ singularity cancels, and in the UV limit we simply get $\frac{1}{2} \langle B^2_\mathrm{line} \rangle_\mathrm{UV} \rightarrow 0$. In the IR, the Bessel term is exponentially suppressed for $u\gg 1/m_\gamma.$ Therefore,
\begin{equation}
\frac{1}{2} \langle B^2_\mathrm{line} \rangle_\mathrm{IR} = -\frac{1}{2N_f} ( c'R - \log R + \mathcal{O}(1)).
\end{equation}
The linear term is the Wilson-line perimeter term and is removed by Wilson-line renormalization. The universal $R$-dependent term is therefore
\begin{equation}
\biggl[\frac{1}{2} \langle B^2_\mathrm{line} \rangle_\mathrm{IR}\biggr]_\text{ren}
    = \frac{1}{2N_f} \log R .
\end{equation}
Thus, 
\begin{equation}\label{eq:Delta_psi_2_QED}
\langle 
\psi_L^i(R) W(R,0) \psi_R^{\dagger j}(0)
\rangle
\sim
\begin{cases}
\dfrac{\delta^{ij}}{\pi R}, 
& \text{UV: } R\to 0, \\[2mm]
\dfrac{\delta^{ij}}{\pi R^{1-1/N_f}}, 
& \text{IR: } R\to \infty .
\end{cases}
\end{equation}

For non-abelian gauge theories we do not have an exact solution for the full RG flow, but the IR fixed point is understood anyway~\cite{Dalley:1992yy,KUTASOV1995447,Gross:1997mx,Isachenkov:2014zua,Komargodski:2020mxz,10.21468/SciPostPhys.8.5.072,Dempsey:2021xpf,Delmastro:2021otj,Delmastro:2022prj}. For example, the exact scaling dimension of $\psi$ is
\begin{equation}\label{eq:Delta_psi_2d}
\Delta(\psi)=\frac{c_\text{IR}}{2c_\text{UV}}
\end{equation}
where
\begin{equation}
\begin{aligned}
c_\text{UV}&=N_f\dim(R)\\
c_\text{IR}&=c_\text{UV}-\frac{N_f T(R)}{N_f T(R)+T(\text{adj})}\dim(G)
\end{aligned}
\end{equation}
are the short- and long-distance central charges, respectively. Thus, in the large $N_f$ limit, this becomes
\begin{equation}\label{eq:Delta_psi_2_YM}
\Delta(\psi)=\frac12-\frac{1}{N_f}\frac{\dim(G)}{2\dim(R)}+\mathcal O(N_f^{-2})
\end{equation}
Given that $\bar\psi\psi$ factorizes into chiral and anti-chiral components, each half being $\psi$ itself, its scaling dimension is just twice~\eqref{eq:Delta_psi_2d}:
\begin{equation}\label{eq:Delta_psipsi_2_YM}
\begin{aligned}
\Delta(\bar\psi\psi)&=2\Delta(\psi)\\
&=1-\frac{1}{N_f}\frac{\dim(G)}{\dim(R)}+\mathcal O(N_f^{-2})
\end{aligned}
\end{equation}
Intuitively, this reflects the fact that the Wilson line connecting $\bar\psi$ to $\psi$ is topological in $d=2$, it is a Verlinde line of the infrared CFT $SO(\dim R)_1/G_{T(R)}$.

We note in passing that, from the exact solution, the large $N_f$ expansion of this observable converges for all $N_f$ larger than $T(\text{adj})/T(R)$.

Although we will not need this for the rest of this paper, we mention for completeness that one can also write down general formulas for $\Delta$ for more complicated composite operators. For example, let $G$ be simple and $R$ complex. Then, there are three types of quadratic defect operators, namely $\psi^2_S,\psi^2_A,\psi^2_\text{adj}$, where the index refers to the symmetry structure under the $\mathfrak{su}(N_f)$ symmetry; these operators live at the end of the Wilson line with gauge quantum numbers opposite to those of the flavor symmetry. Using the data of $SO(\dim R)_1/G_{T(R)},$ we find
\begin{equation}
\begin{aligned}
\Delta(\psi^2_S)&=\frac{(N_f\dim(R)-2)c_\text{IR}+c_\text{UV}}{(N_f\dim(R)-1)c_\text{UV}}\\
&=2\Delta(\psi)+\frac{1}{N_f^2}\frac{\dim(G)}{\dim(R)^2}+\mathcal O(N_f^{-3})\\
\Delta(\psi^2_A)&=\frac{(N_f\dim(R)+2)c_\text{IR}-c_\text{UV}}{(N_f\dim(R)+1)c_\text{UV}}\\
&=2\Delta(\psi)-\frac{1}{N_f^2}\frac{\dim(G)}{\dim(R)^2}+\mathcal O(N_f^{-3})\\
\Delta(\psi^2_\text{adj})&=\frac{(N_f\dim(R))^2c_\text{IR}-c_\text{UV}}{((N_f\dim(R))^2-1)c_\text{UV}}\\
&=2\Delta(\psi)+\mathcal O(N_f^{-3})
\end{aligned}
\end{equation}

Finally, the expectation value $\langle W_\rho(\bigcirc)\rangle$ is given by the quantum dimension of $\rho\in\operatorname{Rep}(G_{T(R)})$ (assuming $\rho$ is integrable, i.e., it doesn't scale with $N_f$). In other words,
\begin{equation}\label{eq:W_2d_QCD}
\begin{aligned}
\langle W_\rho(\bigcirc)\rangle&=\prod_{\alpha>0}\frac{\sin\pi\frac{(\rho+\rho_w,\alpha)}{N_fT(R)+T(\mathrm{adj})}}{\sin\pi\frac{(\rho_w,\alpha)}{N_fT(R)+T(\mathrm{adj})}}\\[+2ex]
&=\dim(\rho)\biggl(1-\frac{\pi^2}{6}\frac{T(\mathrm{adj})\dim(G)}{N^2_fT(R)^2}\frac{T(\rho)}{\dim(\rho)}+\mathcal O(N_f^{-3})\biggr)
\end{aligned}
\end{equation}
where $\rho_w=\frac12\sum_{\alpha>0}\alpha$ is the Weyl vector of $\mathfrak g$ and $(\cdot,\cdot)$ denotes the scalar product. Perhaps surprisingly, the expectation value vanishes to leading order in $1/N_f$. As before, this expansion also converges for all $N_f$ larger than $T(\text{adj})/T(R)$.

\subsection{QCD in $d=3$}

In the main text we perform various computations in QCD for arbitrary $d$. Here we review some known answers in $d=3$ so as to have something to compare our general results with. Unlike $d=1,2$, three-dimensional QCD is not exactly solvable; we stick to leading order in $1/N_f$.

We only quote results in the case of abelian gauge group, namely $G=U(1)$. To leading order in $1/N_f$, the non-abelian result simply requires an additional color factor $\dim(G)/\dim(R)$~\cite{Gracey:1993ua,Ye:2002ep}. The scaling dimension of $\psi$ has been computed in, for example,~\cite{Ye:2002fn,Gracey:1993sn}; it reads\footnote{Reference~\cite{Gracey:1993sn} claims to compute $\Delta(\bar\psi\psi)$ for any $d$, where they find $\Delta^{(1)}=-\frac{(d-1) \Gamma (d)}{\Gamma(2-d/2) \Gamma(d/2+1) \Gamma(d/2)^2}$ (see equation 15 therein). Taken literally, this cannot be correct since, upon plugging in $d=3$, one gets $-64/3\pi^2$ (see equation 27 therein), which disagrees with the more careful evaluation of~\cite{PhysRevB.76.149906}, namely $128/3\pi^2$. The mistake in~\cite{Gracey:1993sn} is that they missed a Feynman diagram in their analysis (namely, the two-loop diagram in eqs.~3,4 in~\cite{PhysRevB.76.149906}; this is of course a subtle mistake and in fact~\cite{PhysRevB.76.149906} also missed this same diagram in their initial paper, the correct answer only appearing in an appended erratum). This diagram does not contribute to $\Delta(\psi)$, which is why the result in~\cite{Gracey:1993sn} is correct if we interpret it as computing $\Delta(\psi)$ instead of $\Delta(\bar\psi\psi)$. That being said, if we take their $1/N_f^2$ result and plug in $d=2$, we get $\Delta^{(2)}=-1/4$, which is different from the value of both $\Delta(\psi)$ and $\Delta(\bar\psi\psi)$ that we found in the previous subsection, where we learned that $\Delta$ is one-loop exact for both of these operators.\label{fn:gracey}}
\begin{equation}\label{eq:Delta_psi_3}
\Delta(\psi)=1-\frac{32}{3\pi^2N_f}+\mathcal O(N_f^{-2})
\end{equation}
Similarly, the scaling dimension of $\bar\psi\psi$ is given by~\cite{PhysRevB.76.149906} (see also~\cite{Chester:2016ref})
\begin{equation}\label{eq:Delta_psipsi_3}
\Delta(\bar\psi\psi)=2+\frac{128}{3\pi^2 N_f}+\mathcal O(N_f^{-2})
\end{equation}

\subsection{QCD in $d=4^-$}

Continuing with our low $d$ review, the next logical step would be $d=4$. But in four dimensions, gauge theories are infrared free for large enough $N_f$, and therefore there is nothing interesting we could compare with. Instead, let us discuss the limit $d\to4^-$. We denote $d=4-2\epsilon$, and report the scaling dimensions of bilinear operators in the limit of small $\epsilon$, as computed in~\cite{DiPietro:2017kcd}. These authors define
\begin{equation}
B^{(n)}_\text{sing}=\bar\psi\gamma_{\mu_1\cdots\mu_n}\psi
\end{equation}
and they compute $\Delta(B^{(n)})$ to order $\epsilon^3$ and exactly in $N_f$. For our purposes, the $1/N_f$ correction is enough; writing
\begin{equation}
\Delta(B^{(n)}_\text{sing})=d-1+\frac{1}{\tr[1]}\Delta^{(1)}_n(\epsilon)+\mathcal O(1/N_f^2)
\end{equation}
the paper~\cite{DiPietro:2017kcd} finds
\begin{equation}\label{eq:Delta_epsilon}
\begin{aligned}
\Delta^{(1)}_0(\epsilon)&=-18\epsilon+15 \epsilon^2+\tfrac{35}{2} \epsilon^3+\cdots \\
\Delta^{(1)}_1(\epsilon)&=0\\
\Delta^{(1)}_2(\epsilon)&=6\epsilon-13 \epsilon^2-\tfrac92 \epsilon^3+\cdots \\
\Delta^{(1)}_3(\epsilon)&=30\epsilon ^2+13 \epsilon^3 +\cdots
\end{aligned}
\end{equation}


\section{Chiral anomaly}\label{app:chiral}

Although in this paper we have focused on discontinuities at odd $d$ resulting from traces with an odd number of gammas, an almost identical story holds for discontinuities at even $d$ resulting from traces with an even number of gammas, with an extra insertion of $\gamma^\star$. One can make this connection very concrete: if we denote by $\gamma$ the gamma matrices in odd $d$, and by $\tilde\gamma$ the gamma matrices in even $d-1$, then one can choose a basis such that $\gamma^{\tilde\mu}=\tilde\gamma^{\tilde\mu}$ for $\tilde\mu=1,2,\dots,d-1$, and $\gamma^d=\tilde\gamma^\star$. Therefore, we can write
\begin{equation}
\tr[\gamma_{\rho_1\cdots\rho_n}]\tr[\gamma^{\rho_1\cdots\rho_n}]=\tr[\tilde\gamma_{\tilde\rho_1\cdots\tilde\rho_n}]\tr[\tilde\gamma^{\tilde\rho_1\cdots\tilde\rho_n}]+n\tr[\tilde\gamma^\star\tilde\gamma_{\tilde\rho_1\cdots\tilde\rho_{n-1}}]\tr[\tilde\gamma^\star\tilde\gamma^{\tilde\rho_1\cdots\tilde\rho_{n-1}}]
\end{equation}
and thus the problem of defining $\tr[\gamma_{\rho_1\cdots\rho_n}]$ is related to the problem of defining $\tr[\tilde\gamma^\star\tilde\gamma_{\tilde\rho_1\cdots\tilde\rho_{n-1}}]$ in one lower dimension. Of course, the former is related to the existence of Chern-Simons terms in odd $d$, and the latter to the existence of chiral anomalies in even $d-1$, and these are naturally connected via inflow~\cite{Callan:1984sa}.

Consider for example free fermions in presence of a background $A$ for its vector-like $U(1)$ symmetry. Then, one has
\begin{equation}
\begin{aligned}
\langle\bar\psi\Gamma\psi\rangle&=i\int_k\tr[S(k)\slashed A S(k+p)\Gamma]+\mathcal O(A^2)\\
&=-iA^\sigma M(\gamma_\sigma,\Gamma)+\mathcal O(A^2)
\end{aligned}
\end{equation}
If we take $\Gamma=\gamma^\star\gamma^\mu$ we find
\begin{equation}\label{eq:chiral_2d}
\begin{aligned}
p_\mu\langle\bar\psi\gamma^\star\gamma^\mu\psi\rangle&=-iA^\sigma \frac{1}{g}\frac{1}{\tr[1]}\frac{p^\alpha\tr[\gamma^\star\gamma_{\alpha\sigma}]}{p^{2-d}}+\mathcal O(A^2)
\end{aligned}
\end{equation}
The trace $\tr[\gamma^\star\gamma_{\alpha\sigma}]$ vanishes for all integers $d$ except for $d=2$, where it equals $-i\tr[1]\epsilon_{\alpha\sigma}$. Then, in $d=2$ we recover the standard chiral anomaly
\begin{equation}
\partial_\mu(\bar\psi\gamma^\star\gamma^\mu\psi)=- \frac{N_f}{2\pi}\epsilon_{\alpha\sigma}F^{\alpha\sigma}
\end{equation}
On the other hand, for $d\in\mathbb C$, it is not clear to us what should be done with~\eqref{eq:chiral_2d}. As above, the problem of analytically continuing $\tr[\gamma^\star\gamma_{\alpha\sigma}]$ away from $d=2$ is more or less equivalent to the problem of continuing $\tr[\gamma^{\mu\nu\rho}]$ away from $d=3$. So the discontinuity in $\Delta(\bar\psi\psi)$ at $d=3$ is related to the discontinuity of $\Delta(\bar\psi\gamma^\star\gamma^\mu\psi)$ at $d=2$. It would be interesting to explore the branched structure of $\bar\psi\gamma^\star\gamma^\mu\psi$ more carefully.

\clearpage
\bibliographystyle{unsrt}
\bibliography{ref.bib}

\end{document}